\documentclass{aa}

\usepackage{graphicx}
\usepackage{txfonts}
\usepackage{natbib,twoopt}
\usepackage[breaklinks=true]{hyperref} 
\bibpunct{(}{)}{;}{a}{}{,}             
\makeatletter
  \newcommandtwoopt{\citeads}[3][][]{\href{http://adsabs.harvard.edu/abs/#3}%
    {\def\hyper@linkstart##1##2{}%
     \let\hyper@linkend\@empty\citealp[#1][#2]{#3}}}
  \newcommandtwoopt{\citepads}[3][][]{\href{http://adsabs.harvard.edu/abs/#3}%
    {\def\hyper@linkstart##1##2{}%
     \let\hyper@linkend\@empty\citep[#1][#2]{#3}}}
  \newcommandtwoopt{\citetads}[3][][]{\href{http://adsabs.harvard.edu/abs/#3}%
    {\def\hyper@linkstart##1##2{}%
     \let\hyper@linkend\@empty\citet[#1][#2]{#3}}}
  \newcommandtwoopt{\citeyearads}[3][][]%
    {\href{http://adsabs.harvard.edu/abs/#3}
    {\def\hyper@linkstart##1##2{}%
     \let\hyper@linkend\@empty\citeyear[#1][#2]{#3}}}
\makeatother
\usepackage{amstext}

\begin{document}

   \title{Hypervelocity star candidates from {\it Gaia} DR2 and DR3
          proper motions and parallaxes}


\titlerunning{HVS candidates from {\it Gaia} astrometry}

   \author{R.-D. Scholz\inst{1}
          }

   \institute{Leibniz-Institut f\"ur Astrophysik Potsdam,
              An der Sternwarte 16, D--14482 Potsdam, Germany\\
              \email{rdscholz@aip.de}
             }

   \date{Received 2023 October 30; accepted 2024 February 12}


  \abstract
   {Hypervelocity stars (HVSs) unbound to the Galaxy can
   be formed with extreme stellar interactions, e.g. close encounters
   with supermassive black holes or in massive star clusters, supernova
   explosions in binary systems, or the stripping of dwarf galaxies.
   Observational evidence comes from measurements of radial velocities (RVs)
   of objects crossing the outer Galactic halo and of tangential
   velocities based on high proper motions (HPMs) and distances 
   of relatively nearby stars.}
   {I searched for new nearby HVS candidates and reviewed 
   known objects using their {\it Gaia} astrometric measurements.}
   {Candidates were selected with significant {\it Gaia} parallaxes 
   of $>$0.1\,mas, proper motions of $>$20\,mas/yr, 
   and computed Galactocentric tangential velocities $vtan\_g$$>$500\,km/s.
   The DR2 and DR3 samples of several thousand HVS candidates 
   were studied with respect to their proper
   motions, sky distribution, number of observations, location in crowded 
   fields, colour-magnitude diagrams, selection effects with magnitude,
   and RVs in DR3. The most extreme 
   ($vtan\_g$$>$700\,km/s) and nearest (within 4\,kpc)
   72 DR3 HVS candidates were investigated with respect to
   detected close neighbours, flags and astrometric 
   quality parameters of objects of similar magnitudes in DR3. The quality
   checks involved HPM objects in a global comparison
   and all objects in the vicinity of each target.}
   {Spurious HPMs in the Galactic centre region led to false HVS 
   interpretations in {\it Gaia} DR2 and are still present in DR3,
   although to a lesser extent. Otherwise there is good agreement between
   the HPMs of HVS candidates in DR2 and DR3.
   However, HVS candidates selected from DR2 tend to have larger 
   parallaxes hence lower tangential velocities in DR3. Most DR3 
   RVs are much lower than the tangential 
   velocities, indicating that the DR3 HVS candidates are still 
   affected by underestimated parallaxes.
   None of the 72 extreme nearby DR3 HVS candidates,
   including three D$^6$ stars, passed all quality checks.
   Their tangential velocities may turn out to be smaller,
   but at least some of them still appear unbound to the Galaxy.
} 
   {}

   \keywords{
Parallaxes --
Proper motions --
Stars: distances --
Hertzsprung-Russell and C-M diagrams --
Stars: kinematics and dynamics --
Galaxy: halo
               }

   \maketitle


\section{Introduction}
\label{Sect_intro}

The high-speed stellar objects at issue in this study occur only rarely, 
and their formation has been explained by exotic astrophysical processes.
However, their fascinating discoveries and interpretations can only be 
described incompletely here due to their abundance. Many individual objects 
were discussed repeatedly in the light of different scenarios and measured 
velocities. The underlying data are, on the one hand extreme stellar radial 
velocities (RVs) from spectroscopic measurements, and on the other hand very 
high tangential velocities of stars computed from astrometric measurements
of their high proper motions (HPMs) and distance estimates, which may rely on 
astrometry or photometry. Whereas extreme RVs are formally much more
precisely measured than high tangential velocities, they should in principle 
also be verified and monitored by repeated spectroscopic measurements with 
different instruments. But this could be the topic of another special
investigation. The focus of this study lies on a critical review of 
astrometric measurements (of both HPMs and parallaxes) 
that can lead to very high tangential velocities of stars, 
indicating that they may escape the Galaxy.

Early investigations on the local Galactic escape velocity
\citepads{1981ApJ...251...61C} 
already indicated a range of 550-650\,km/s in agreement with modern values
\citepads[see][and references therein]{2023AJ....166...12L}. 
As illustrated in
\citetads[][Fig.4]{2023AJ....166...12L}. 
the upper limit of five recent studies corresponds to 580$\pm$63\,km/s found by
\citetads{2018A&A...616L...9M}, 
and this value turns out to be only 50\,km/s higher or lower
at Galactocentric distances of about 4\,kpc or 12\,kpc, respectively.
In the outer Galactic halo, the escape velocity is further reduced.
Using a slightly modified three-component bulge–disk–halo potential model of
\citetads{2008ApJ...680..312K}, 
\citetads{2014ApJ...787...89B} 
found 367\,km/s at the Galacocentric distance of 50\,kpc 
and 578\,km/s at 8\,kpc.
The latter value is in very good agreement with the upper limit of the
local Galactic escape velocity given by
\citetads{2018A&A...616L...9M}. 

A hypervelocity star 
\citepads[HVS;][]{2015ARA&A..53...15B} 
is moving so fast through the Galaxy that it is
unbound to the Galactic potential. 
This definition, which is independent of the ejection mechanism
and relies only on a very high velocity, is used in this study.
A possible HVS formation scenario was 
first predicted theoretically by 
\citetads{1988Natur.331..687H}, 
namely the tidal disruption of a tight binary by the supermassive 
black hole (SMBH) in the Galactic centre (GC). 
\citetads{2003ApJ...599.1129Y} 
considered three scenarios of HVS ejection by the (binary) SMBH in the GC,
including the possible ejection of single stars.
Shortly thereafter, 
\citetads{2005ApJ...622L..33B} 
reported the first HVS observation in the outer Milky Way halo. This and the
following HVS discoveries by Brown et al.
(e.g. \citeyearads{2006ApJ...640L..35B}; 
\citeyearads{2009ApJ...690.1639B}; 
\citeyearads{2012ApJ...751...55B}; 
\citeyearads{2014ApJ...787...89B}) 
were based on RV measurements alone and consisted of 
late B-type stars on the main sequence (MS) more than 50\,kpc away 
from the GC but all consistent with a GC origin.

Other early HVS discoveries included blue stars of different types, e.g.
a subluminous O star
\citepads{2005A&A...444L..61H}, 
the B-type giant \object{HD 271791}
\citepads{2008A&A...483L..21H}, 
an A-type star
\citepads{2009A&A...507L..37T}, 
and an sdB star
\citepads{2011A&A...527A.137T}. 
A growing number of HVSs, e.g. those first observed by
\citetads{2005ApJ...634L.181E}, 
\citetads{2008A&A...483L..21H}, 
Tillich et al.
(\citeyearads{2009A&A...507L..37T}; 
\citeyearads{2011A&A...527A.137T}) 
began to throw discredit on their generally accepted interaction 
with the SMBH in the GC. 
The B-type HVS \object{HE 0437-5439}, located much closer to
the Large Magellanic Cloud (LMC) than to the GC
\citepads{2005ApJ...634L.181E}, 
was considered to originate from the LMC by several studies, e.g. of
\citetads{2007MNRAS.376L..29G}, 
\citetads{2008A&A...480L..37P}, and 
\citetads{2008ApJ...675L..77B}. 
However, based on Hubble Space Telescope (HST) proper motion measurements,
which were not very significant for this outer halo star,
\citetads{2010ApJ...719L..23B} 
preferred a GC origin.
An alternative origin for some of the known HVSs from a recently disrupted 
dwarf galaxy that passed near the GC was suggested by 
\citetads{2009ApJ...691L..63A}. 
A possible HVS ejection in a supernova explosion was first proposed by
\citetads{2008ApJ...684L.103P} 
for \object{HD 271791}. A massive star that could be ejected as a HVS
from an extremely dense young star cluster and later become a pulsar was
discussed by
\citetads{2008MNRAS.385..929G} 
in connection with \object{PSR B1508+55}. This pulsar was previously
found to be an unusual HVS, because of its high tangential velocity of 
about 1000\,km/s from parallax and HPM measurements, and because
of its birth site in the Galactic plane
\citepads{2005ApJ...630L..61C}. 

The question about the existence of an older population of late-type HVSs
was raised by
\citetads{2007ApJ...664..343K}. 
Using data from the Sloan Digital Sky Survey (SDSS),
\citetads{2012ApJ...744L..24L} 
and
\citetads{2014ApJ...780....7P} 
found F- and G/K-type HVS candidates, where the latter selected their 
candidates by proper motion.
\citetads{2015MNRAS.448L...6T} 
compared the observed velocity distributions of late B-type 
\citepads{2014ApJ...787...89B} 
and G/K-type HVS candidates
\citepads{2014ApJ...780....7P} 
mentioning that the tangential velocity errors of the latter are ten times
larger than the RV errors of the former and that the 
typically higher tangential velocities of the relatively nearby G/K-type
HVS candidates may be affected by distance errors. From his simulations
of HVSs ejected from disrupted binaries in asymmetric supernova explosions,
\citetads{2015MNRAS.448L...6T} 
found maximum possible velocities of 770\,km/s and 1280\,km/s, respectively
for late B-type and G/K-type stars. He concluded that the binary disruption
scenario could explain all observed G/K-type HVS candidates and is, ''under
the most extreme favourable conditions'', even possible for late B-type HVSs.
However, the  G/K-type HVS candidates of 
\citetads{2014ApJ...780....7P} 
were not confirmed by
\citetads{2015A&A...576L..14Z} 
who measured smaller ground-based proper motions than claimed before.
A general revision of pre-{\it Gaia} proper motion measurements of late-type 
HVS candidates towards smaller values and consequently lower tangential 
velocities using {\it Gaia} DR2 
\citepads{2018A&A...616A...1G} 
was presented by
\citepads{2018MNRAS.479.2789B}.  
The latest HVS search based on the combination of astrometric data from
{\it Gaia} DR3 
\citepads{2023A&A...674A...1G} 
and spectroscopic data from various large-scale Galactic surveys
\citepads{2023AJ....166...12L} 
revealed
52 metal-poor late-type candidates within 5\,kpc from the Sun, 
none of which was definitely ejected from the GC.

Several HVS studies based on {\it Gaia} DR2, e.g.
\citetads{2019MNRAS.490..157M} 
and
\citetads{2018ApJ...866..121H}, 
made use of both astrometric and RV measurements, which were
available for the relatively small subsample of 7 million stars. 
Searching among the about 34 million stars with 6D information in
{\it Gaia} DR3,
\citetads{2022MNRAS.515..767M} 
did ''not identify any HVS candidates with a velocity higher
than 700\,km/s and robustly observed kinematics'', whereas
\citetads{2023ApJ...944L..39L} 
found two HVS candidates with a potential GC origin.
\citetads{2023MNRAS.525..561E} 
considered only one object, \object{S5-HVS1}, for which 
\citetads{2020MNRAS.491.2465K} 
measured a precise and constant RV of 1017$\pm$2.7\,km/s
and {\it Gaia} DR3 confirmed its DR2 proper motion of about 35\,mas/yr but
also a still non-significant parallax, as an HVS with a clear GC origin.
On the other hand,
\citetads{2018ApJ...865...15S} 
and
\citetads{2018ApJ...868...25B}. 
demonstrated that an effective HVS search
could already be carried out with {\it Gaia} DR2
HPMs and parallaxes alone. 

The 19 HVS candidates found by 
\citetads{2018ApJ...868...25B} 
applying relatively strong astrometric and photometric quality criteria,
including 5$\sigma$ parallax measurements ($Plx/e\_Plx$$>$5), fall in
the {\it Gaia} colour-magnitude diagram (CMD) near to the MS 
and in the late-type giants region. Their heliocentric distances
determined from the {\it Gaia} DR2 parallaxes are about 2-15\,kpc, whereas 
the Galactocentric tangential velocities (Sect.~\ref{subS_vtan_g}) are 
in the range 670-920\,km/s. The three extreme HVS candidates discovered by
\citetads{2018ApJ...865...15S} 
using less stringent quality criteria
have DR2-based $vtan\_g$$\approx$1300-2400\,km/s and distances from the Sun
between only 1\,kpc and 2.5\,kpc. Their spectroscopic observations allowed
these authors to classify these objects as unusual white dwarfs (WDs), 
located between the normal WD sequence and the MS in the {\it Gaia} CMD, 
and measure their RVs. Surprisingly, for two of these three unusual WDs the 
RV was consistent with 0, and only one turned out to have a high RV of 
about 1200\,km/s, comparable with its tangential velocity. Nevertheless, 
for the responsible ejection mechanism of all three objects,
\citetads{2018ApJ...865...15S} 
suggested a ''dynamically driven double-degenerate double-detonation type Ia
supernova (D$^6$)'' scenario, in which the D$^6$ objects are the surviving
former close companions of their exploded primaries. 
The position of the three D$^6$ objects in the {\it Gaia} CMD appears to 
be close to the well-measured HVS \object{LP 40-365}, which is another 
unusual WD that may have been exploded itself but only partially 
destroyed in a subluminous type Iax supernova
\citepads{2017Sci...357..680V}, 
and similar objects (called ''LP 40-365 stars'') subsequently found by 
\citetads{2019MNRAS.489.1489R}. 
The accurate {\it Gaia} DR2 parallax of the 
prototype \object{LP 40-365} was for the first time crucial to estimate the 
size and, thus, to confirm its nature as a supernova Ia survivor
\citepads{2018MNRAS.479L..96R}. 
Selecting blue HPM stars with non-significant and even negative
parallaxes in {\it Gaia} DR3,
\citetads{2023OJAp....6E..28E} 
recently confirmed four ''hot D$^6$ stars'' with heliocentric 
tangential velocities of the order of 1000-5000\,km/s 
by measuring comparably large but (formally) precise RVs. 
With these discoveries, more and more alternative HVS ejection scenarios, 
beyond the classical SMBH interaction in the GC, come into focus, and
\citetads{2023MNRAS.518.6223I} 
discussed already eight (!) different possible HVS formation channels.

A first brief critical review of the astrometric reliability of the fastest
nearby HVS candidates selected based on {\it Gaia} DR2 Galactocentric 
tangential velocities was presented by
\citetads{2018RNAAS...2..211S}. 
Only six candidates passed the quality checks applied therein 
using some of the most important astrometric parameters. Among those six 
candidates were two of the 19 HVS candidates previously presented by
\citetads{2018ApJ...868...25B} 
and only one of the three D$^6$ objects of
\citetads{2018ApJ...865...15S}. 
The latter object, \object{D6-2}, was however already considered a suspicious
HVS because of its zero RV measured by these authors, and it was also
mentioned as doubtful by
\citetads{2018RNAAS...2..211S}. 
because of its DR2 quality parameters just slightly
below the allowed limits. 

This extended and more detailed study of HVS candidates with high tangential
velocities first involves all objects of moderately HPMs, 
significant parallaxes, and parallactic distances up to 10\,kpc in {\it Gaia} 
DR2 and DR3 (Sect.~\ref{Sect_vtan}). 
In Sect.~\ref{Sect_nearextrHVS}, the nearest (within 4\,kpc)
and most extreme HVS candidates selected from {\it Gaia} DR3 are investigated 
with respect to a large number of astrometric quality parameters available
in and derived from DR3.
Conclusions on the reliability of the HVS status derived from
{\it Gaia} astrometric measurements and the location of HVS candidates
in the CMD are drawn in Sect.~\ref{Sect_concl}.

\section{{\it Gaia} HVS selection from tangential velocities}
\label{Sect_vtan}

The initial selection of relatively nearby HVS candidates in the last two
{\it Gaia} data releases investigated here was made by a conservative HPM
cut $PM$$>$20\,mas/yr, with a minimum parallax $Plx$$>$0.1\,mas, and by only
one astrometric quality criterion requiring a 5$\sigma$ significance of the 
parallax measurement $Plx/e\_Plx$$>$5 in DR2 (the ratio $Plx/e\_Plx$
is also 
listed as $RPlx$ (\textsf{parallax\_over\_error}) 
in DR2 and DR3). However, in DR3,
objects with 3$<$$Plx/e\_Plx$$<$5 were included as low-priority candidates. 
This selection was sensitive to heliocentric tangential velocities of roughly 
$>$1000\,km/s at 10\,kpc distance, $>$500\,km/s at 5\,kpc, and
$>$200\,km/s at 2\,kpc, respectively. 
The final step in the astrometric selection of HVS candidates was
carried out after transforming the tangential velocities to the Galactic
rest frame.

\subsection{Galactocentric tangential velocities}
\label{subS_vtan_g}

For a comparison with the estimated upper limit of the local Galactic 
escape velocity of 580$\pm$63\,km/s
\citepads{2018A&A...616L...9M} 
heliocentric tangential velocities have to be corrected for solar motion.
Galactocentric tangential velocities $vtan\_g$
were computed in the same way as in
\citetads{2018RNAAS...2..211S} 
by following Eq.\,1 of
\citetads{2018ApJ...866..121H} 
but using a local standard of rest velocity of 235\,km/s as applied by
\citetads{2018ApJ...868...25B}. 

In the first part of the study described in the following subsections, 
a conservative minimum of $vtan\_g$$>$500\,km/s was used
in view of the relatively large tangential velocity errors of the order of
$\pm$100\,km/s (even about $\pm$200\,km/s for low-priority candidates).
These errors will be discussed for {\it Gaia} DR3 HVS candidates
in Sects.~\ref{subS_DR3magsel} and \ref{subS_compRV} and 
shown in Fig.~\ref{Fig_DR3evtanG}.
Note that the tangential velocity errors are completely dominated by the 
parallax errors, and the proper motion errors can be neglected.
Later, as shown in Sect.~\ref{Sect_nearextrHVS}, the limit was increased 
to $vtan\_g$$>$700\,km/s to select the most promising extreme HVS candidates, 
and in particular to verify the nearest extreme HVS candidates within 4\,kpc
by checking additional {\it Gaia} DR3 astrometric quality parameters.

\subsection{{\it Gaia} DR2 candidates}
\label{subS_DR2}

The preceding DR2-based selection of HVS candidates by
\citetads{2018RNAAS...2..211S} 
was restricted to the nearest candidates with the highest velocities and
yielded only six apparently well-measured stars with $PM$$>$60\,mas/yr 
and $vtan\_g$$>$700\,km/s. These six stars are marked in 
Fig.~\ref{Fig_dr2pm} and in many of the following 
figures for comparison. The only three of these six candidates, 
\object{D6-2}, \object{Gaia DR2 3841458366321558656}, 
and \object{Gaia DR2 6097052289696317952}, which have 
still $vtan\_g$$>$700\,km/s, when their new {\it Gaia} DR3 data are used
(Sect.~\ref{subS_DR3}), are labelled in the figures with their D$^6$
name or abbreviated DR2 designations, respectively.

   \begin{figure}[h!]
   \resizebox{\hsize}{!}{\includegraphics[angle=270]{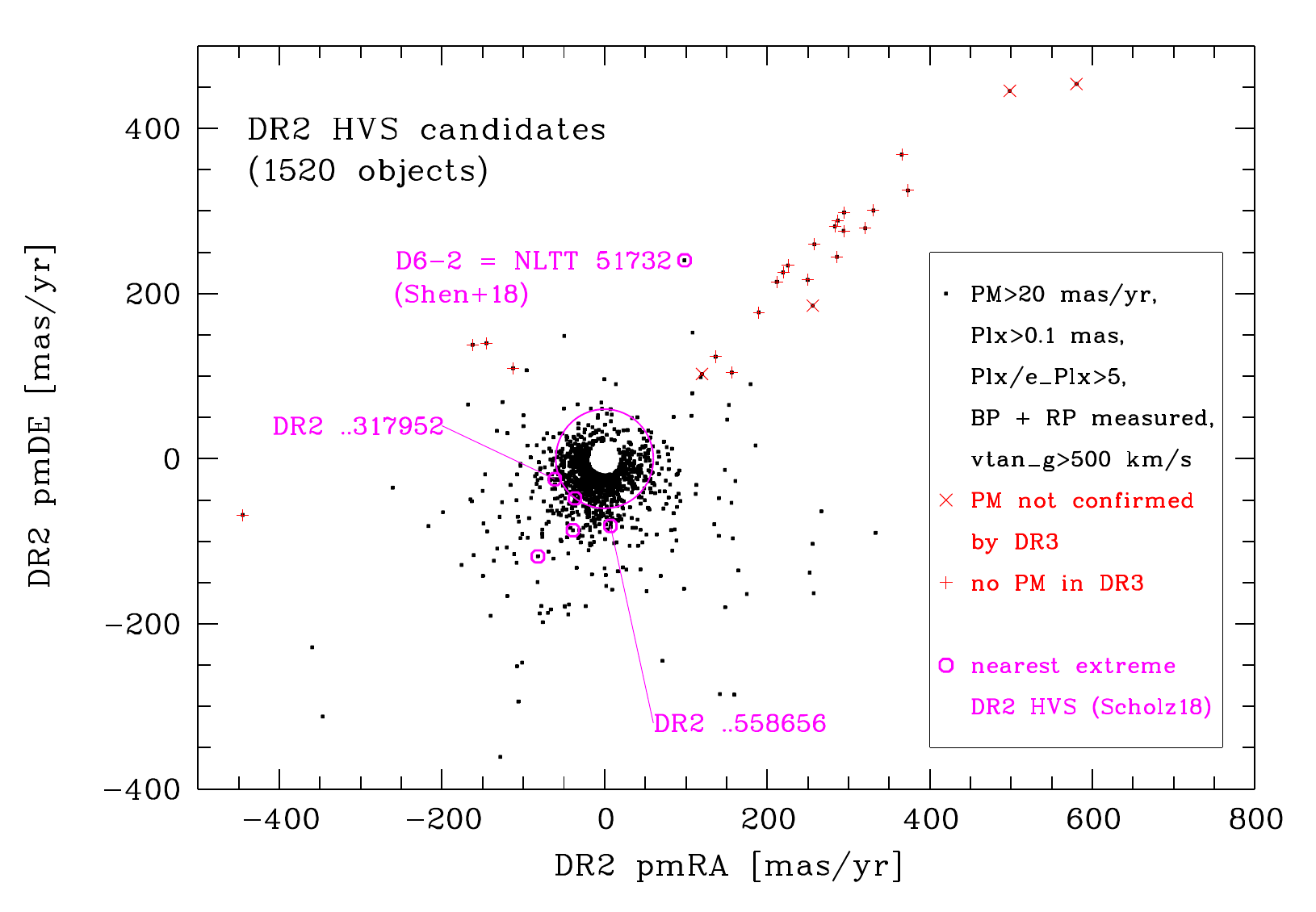}}
      \caption{Proper motions of HVS candidates selected from {\it Gaia} DR2
           with $PM$$>$20\,mas/yr, $Plx$$>$0.1\,mas, available $BP,RP$
           photometry, and $vtan\_g$$>$500\,km/s.
           All 1520 high-priority candidates with $Plx/e\_Plx$$>$5 are shown.
           Red symbols mark objects with unconfirmed (crosses) or not measured
           (plus signs) proper motions in DR3. The magenta circle and symbols
           indicate the minimum HPM of 60\,mas/yr and the six
           extreme ($vtan\_g$$>$700\,km/s) nearby HVS
           candidates selected with additional DR2 quality criteria by
\citetads{2018RNAAS...2..211S}, 
           respectively. The three labelled objects have based on DR3 still
           $vtan\_g$$>$700\,km/s.
              }
      \label{Fig_dr2pm}
   \end{figure}

\begin{figure}[h!]
   \resizebox{\hsize}{!}{\includegraphics[angle=270]{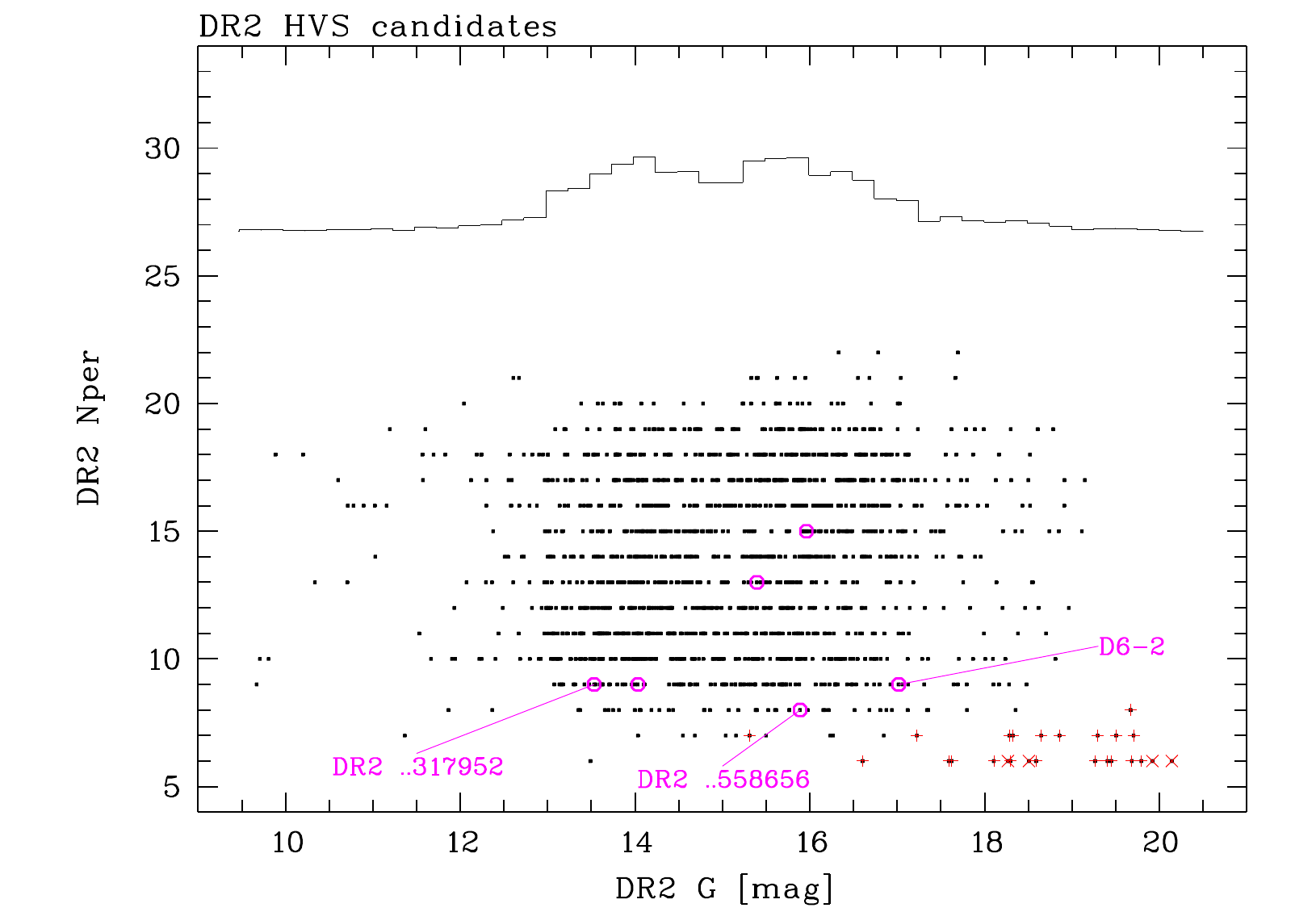}}
      \caption{{\it Gaia} DR2 $G$ magnitudes and numbers of visibility periods
           $Nper$ of 1520 high-priority HVS candidates selected from DR2.
           Overplotted coloured symbols and labelled objects as in
           Fig.~\ref{Fig_dr2pm}. The black histogram illustrates the
           magnitude distribution.
              }
      \label{Fig_dr2NperG}
   \end{figure}

Some of the HPMs of 1520 DR2 HVS candidates, selected here with 
only one astrometric quality criterion ($Plx/e\_Plx$$>$5) but also by limiting
the sample to objects with measured $BP$ and $RP$ photometry in DR2,
were not confirmed (4) or not measured (21) by DR3 (Fig.~\ref{Fig_dr2pm}). 
Most of them show an unusual distribution with almost equally large positive
components $pmRA$ and $pmDE$, while a few have similar large $pmRA$ and $pmDE$
values but with different 
signs. As one can see in 
Fig.~\ref{Fig_dr2NperG}, 
all these doubtful candidates are faint and have obviously not enough epochs
of observation, as indicated by small numbers of visibility periods $Nper$
(= \textsf{visibility\_periods\_used}).
They are also located at low Galactic latitudes, 
where the majority is concentrated
in the GC region (two are close to the Galactic anti-centre). The three
labelled extreme HVS candidates with confirmed $vtan\_g$$>$700\,km/s in DR3
had also small $Nper$ values in DR2 but are generally brighter and at higher
Galactic latitudes.

   \begin{figure*}
   \centering
   \includegraphics[angle=270,width=15cm]{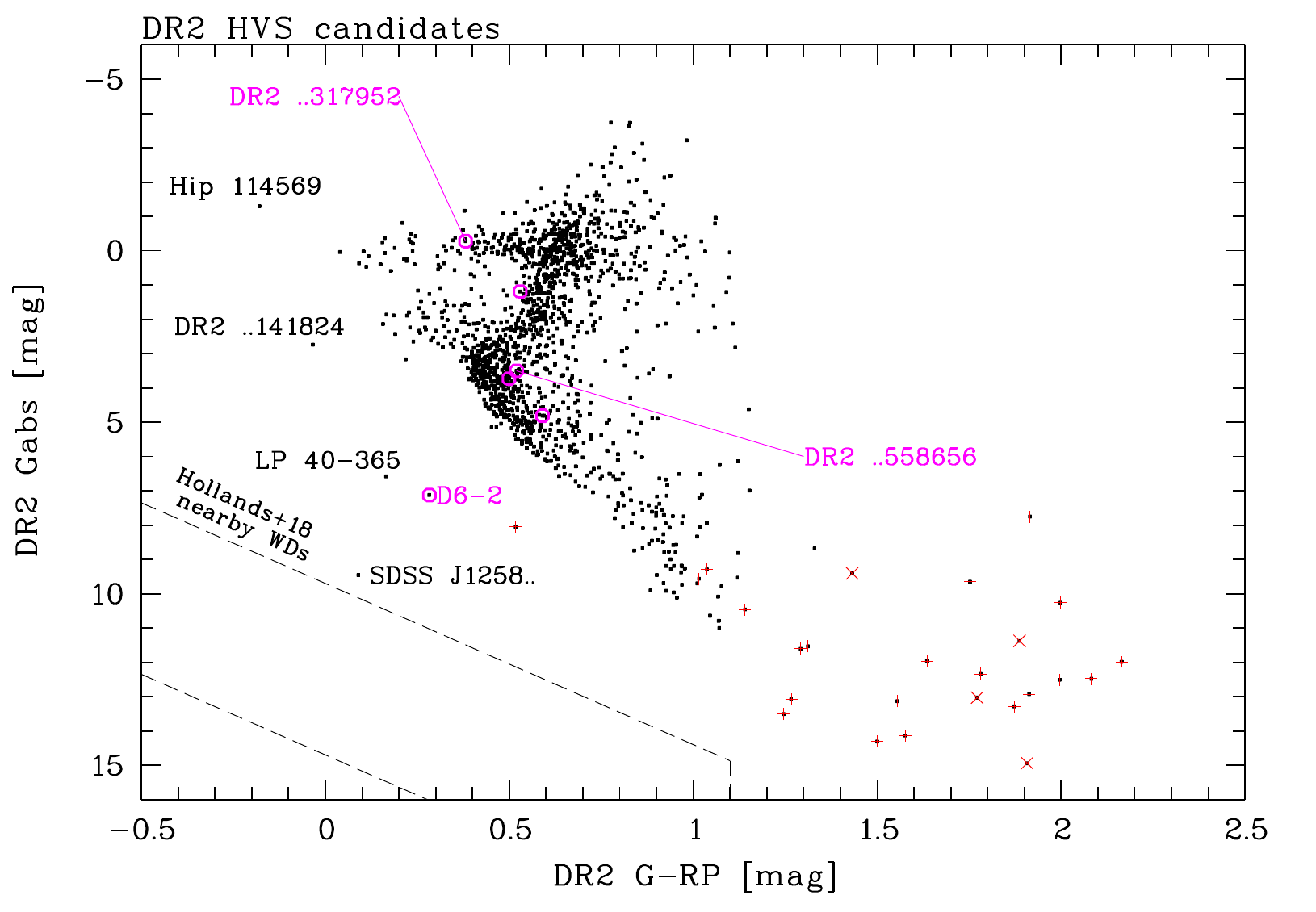}
      \caption{{\it Gaia} DR2 colour-magnitude diagram of 1520 high-priority
           DR2 HVS candidates. The region occupied by nearby WDs
\citepads{2018MNRAS.480.3942H} 
           is bounded by dashed lines.
           Overplotted coloured symbols and labelled objects
           as in Fig.~\ref{Fig_dr2pm}. For other labelled stars see text.
              }
      \label{Fig_dr2cmd}
   \end{figure*}

Out of the shown 25 suspicious HPM objects, 7 were included in the
DR2 tangential velocity-selected sample of 28 apparent late-type HVSs of
\citetads[][their Table 2]{2019ApJS..244....4D}. 
Another 7 of their candidates have no $BP$ and $RP$ measurements in DR2
and are not shown here. Together, these 14 stars, which in a visual 
inspection of available multi-epoch imaging 
data\footnote{https://irsa.ipac.caltech.edu/applications/finderchart/} 
did also not display any HPM, represent half of their sample.
The problem of selecting false {\it Gaia} HPM stars in crowded 
fields, when the significance of the parallaxes is used as the only
quality criterion, will be further illuminated in Sect.~\ref{subS_falsepm}.

The bimodal magnitude distribution seen in Fig.~\ref{Fig_dr2NperG} can be
roughly attributed to distant giants and nearby dwarfs among the HVS
candidates. The median parallax of 711 candidates with $G$$<$15\,mag
is 0.176\,mas, whereas for the 809 candidates with $G$$>$15\,mag
it is 0.368\,mas. The CMD showing the absolute magnitudes
$G_{abs}=G+5\times\log{(Plx/100)}$ (where the parallax $Plx$ is in
units of mas) vs. the $G$$-$$RP$ colour, all 1520 high-priority DR2 candidates
(Fig.~\ref{Fig_dr2cmd}) also includes nearly equal numbers of giants and
dwarfs, if one simply counts all objects with absolute magnitudes
brighter (797) or fainter (723) than the turnoff point 
at $G_{abs}$$\approx$2.5\,mag. The 25 candidates with lacking or unconfirmed
proper motions in DR3 mostly appear as a diffuse cloud of unusual red 
and faint stars in this CMD. This is another hint to their overestimated 
parallaxes, in addition to their strongly overestimated proper motions 
and likely problematic photometric measurements, in DR2. 
The bluest (at $G$$-$$RP$$\approx$0.5\,mag) of these 
25 objects (\object{Gaia DR2 3444106859889780608})
is with $G$$=$15.3\,mag also the brightest in Fig.~\ref{Fig_dr2NperG},
has according to other catalogues
\citepads[e.g.][]{2015AJ....150..101Z} 
a proper motion below 10\,mas/yr, and was apparently resolved in DR3 as
two nearly equally bright stars, each with poor astrometric quality, 
separated by about 0.2\,arcsec.

In addition to three labelled extreme HVS candidates from
\citetads{2018RNAAS...2..211S} 
(see Fig.~\ref{Fig_dr2pm}), the two bluest objects and two luminous WDs
are also labelled in the DR2 CMD (Fig.~\ref{Fig_dr2cmd}). All these four
stars, have DR2-based Galactocentric tangential velocities just 
slightly 
above the chosen limit of 500\,km/s.
In order of absolute magnitude, these are
\object{Hip 114569}, an early-type runaway candidate from an open cluster
\citepads{2022A&A...663A..39B}, 
\object{Gaia DR2 2139774976276141824}, a hot subdwarf found in DR2
\citepads{2019A&A...621A..38G}, 
\object{LP 40-365}, the already mentioned (in Sect.~\ref{Sect_intro}) 
well-measured HVS whose RV is also of the order of 500\,km/s
\citepads{2017Sci...357..680V}, 
and
\object{SDSS J125834.93-005948.4}, a WD found in SDSS DR10
\citepads{2015MNRAS.446.4078K}. 
The last one lies close to the colour box of nearby WDs
\citepads{2018MNRAS.480.3942H}. 
Interestingly, all of those four stars have slightly larger parallaxes 
measured in DR3, following a general trend that will be described in more
detail in Sect.~\ref{subS_trends}, so that only \object{LP 40-365} remains a
HVS in the DR3-based selection (Sect.~\ref{subS_DR3}).

   \begin{figure}
   \resizebox{\hsize}{!}{\includegraphics[angle=270]{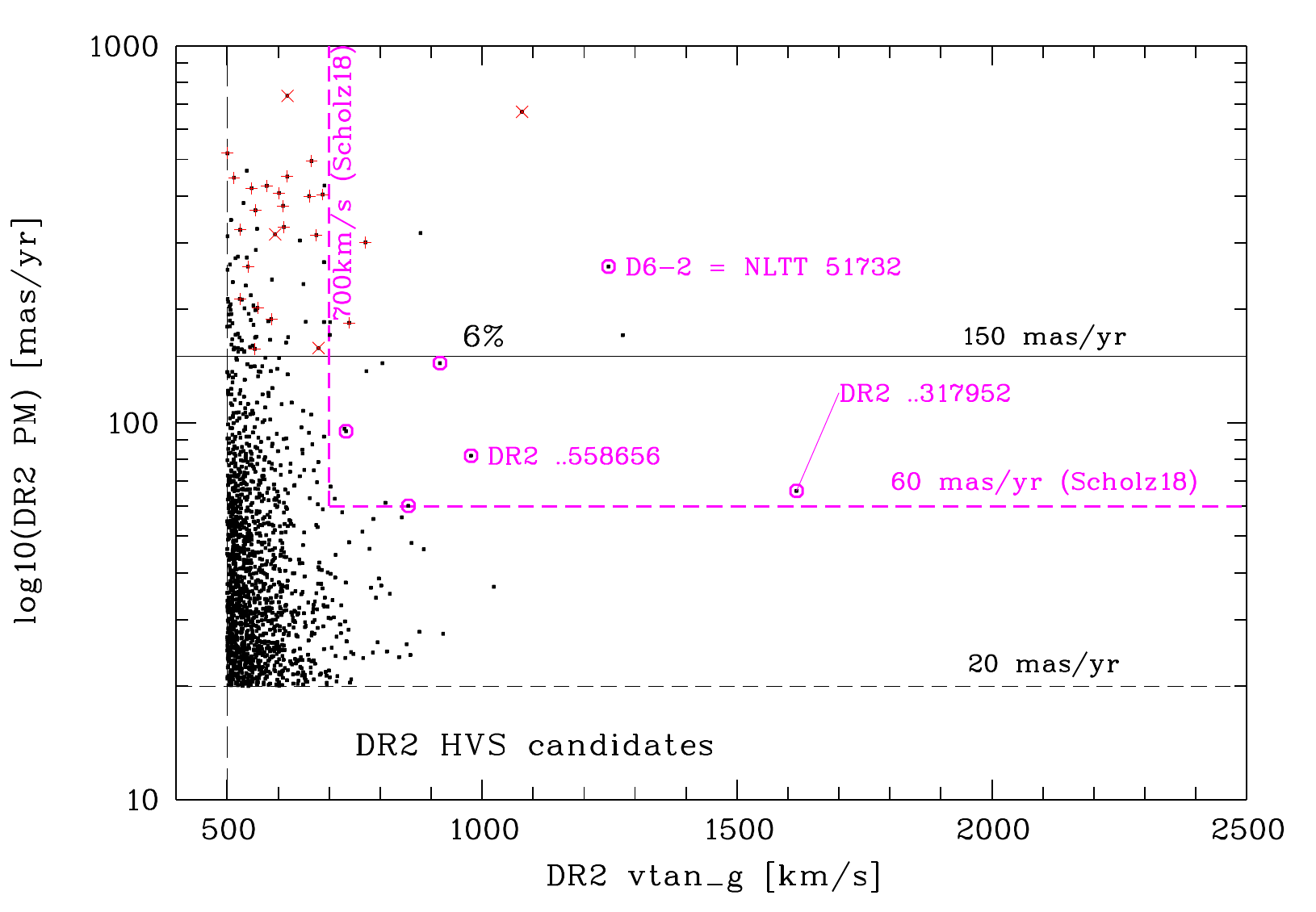}}
      \caption{Galactocentric tangential velocities $vtan\_g$
           and total proper motions $PM$
           of 1520 high-priority {\it Gaia} DR2 HVS candidates.
           Overplotted coloured symbols and labelled objects
           as in Fig.~\ref{Fig_dr2pm}. The magenta dashed lines show the search
           limits of
\citetads{2018RNAAS...2..211S}. 
              }
      \label{Fig_dr2vtangPM}
   \end{figure}

The measured total proper motions and computed $vtan\_g$ of the
DR2 HVS candidates are shown in Fig.~\ref{Fig_dr2vtangPM}. Due to the
25 spurious HPM stars (red symbols), 
the fraction of 'classical' HPM stars with $PM$$>$150\,mas/yr 
\citepads[see e.g.][]{2005AJ....129.1483L} 
falling above the black solid line
appears with about 6\% relatively high. Concerning the marked six candidates of
\citetads{2018RNAAS...2..211S}, 
the magenta dashed lines indicate the more extreme HPM and $vtan\_g$
limits used for them in comparison to the moderate 20\,mas/yr 
and $vtan\_g$$=$500\,km/s (black dashed lines) applied here.
It is remarkable that only the three of them with the highest $vtan\_g$
(labelled) 'survived' with $vtan\_g$$>$700\,km/s in the later DR3-based HVS 
selection. However, as will be demonstrated later (Sect.~\ref{subS_DR3}
and \ref{subS_trends}), 
\object{Gaia DR2 6097052289696317952} appearing as the fastest HVS
in Fig.~\ref{Fig_dr2vtangPM} suffered from a dramatic decrease in velocity
because of its much larger parallax in DR3.

\subsection{{\it Gaia} DR3 candidates}
\label{subS_DR3}

\begin{figure}
   \resizebox{\hsize}{!}{\includegraphics[angle=270]{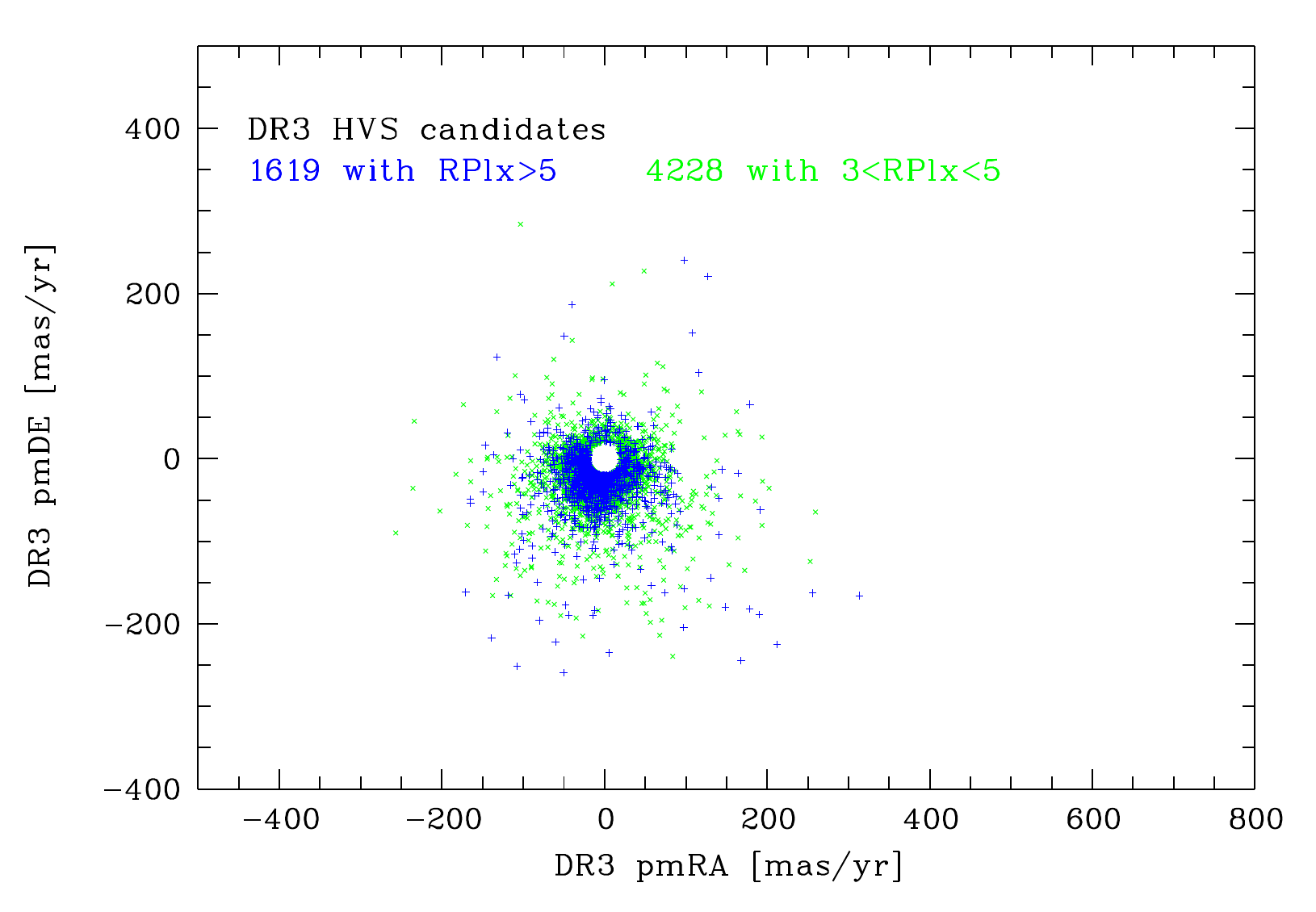}}
           \caption{Proper motions of {\it Gaia} DR3
           high-priority (blue plus signs) and low-priority (green crosses)
           HVS candidates.
              }
      \label{Fig_dr3pm}
   \end{figure}

\begin{figure}
   \resizebox{\hsize}{!}{\includegraphics[angle=270]{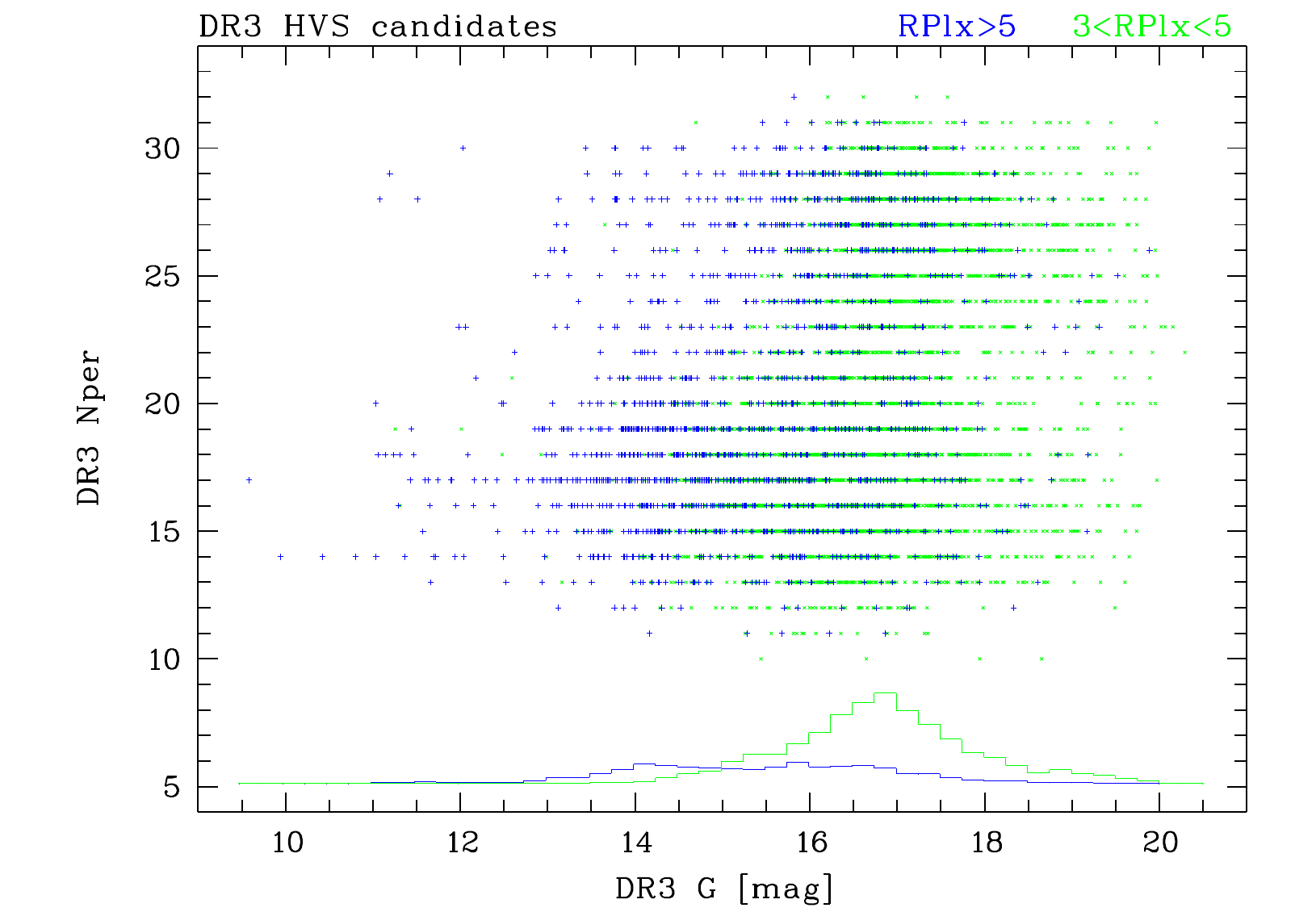}}
           \caption{{\it Gaia} DR3 $G$ magnitudes and numbers of visibility
           periods $Nper$ of 1619 high-priority (blue plus signs)
           and 4228 low-priority (green crosses) HVS candidates selected
           from DR3 (Sect.~\ref{subS_DR3}).
           The blue and green histograms illustrate their corresponding
           magnitude distributions.
              }
      \label{Fig_dr3NperG}
   \end{figure}

Based on DR3, not only 1619 high-priority ($Plx/e\_Plx$$>$5) but also 4228
low-priority (3$<$$Plx/e\_Plx$$<$5) HVS candidates with $PM$$>$20\,mas/yr,
$Plx$$>$0.1\,mas, and $vtan\_g$$>$500\,km/s were selected (without further
requirements on their available photometry). The proper motion distribution 
of all candidates (Fig.~\ref{Fig_dr3pm}) looks more compact than that of the
1520 high-priority HVS candidates (with measured $BP$ and $RP$) previously
selected in DR2 (Fig.~\ref{Fig_dr2pm}). There are less extreme HPM stars
among the DR3 HVS candidates and no obviously false HPM stars (cf. 
red symbols in Fig.~\ref{Fig_dr2pm} and HPM spikes 
seen in Fig.~\ref{Fig_fpm}). Except for only one star lacking a proper 
motion in DR2 (but including 5 high-priority and 49 low-priority candidates 
with negative DR2 parallaxes), the differences of the proper motions 
(DR3-DR2) of the DR3 nearby HVS candidates are 
much smaller than their total values. 
For high-priority candidates both
proper motion components agree within about $\pm$3\,mas/yr. The agreement
is still within about $\pm$5\,mas/yr for all low-priority candidates, except
for one object with a difference in $pmDE$ of about -10\,mas/yr and
a total DR3 proper motion of about 45\,mas/yr (with the largest negative
parallax in DR2).
Compared to the DR2 candidates (Fig.~\ref{Fig_dr2NperG}), the DR3 candidates
are expected to have more reliable astrometry according to their shift 
towards higher numbers of visibility periods $Nper$ (Fig.~\ref{Fig_dr3NperG}).
The sub-sample of low-priority DR3 HVS candidates is dominated by 
faint objects (see green histogram in Fig.~\ref{Fig_dr3NperG}).
The sky distribution of all DR3 HVS candidates is not shown here.
It appears similar to that
of the DR2 HVS candidates but without any
concentration of candidates near the GC. 
An overdensity in the
1st ($GLON$$<$90$\degr$) and 4th ($GLON$$>$270$\degr$) Galactic quadrants
is observed for both DR2 and DR3 HVS candidates. This concerns mainly
distant giants among the HVS candidates (see below).

\begin{figure*}
  \centering
  \includegraphics[angle=270,width=15cm]{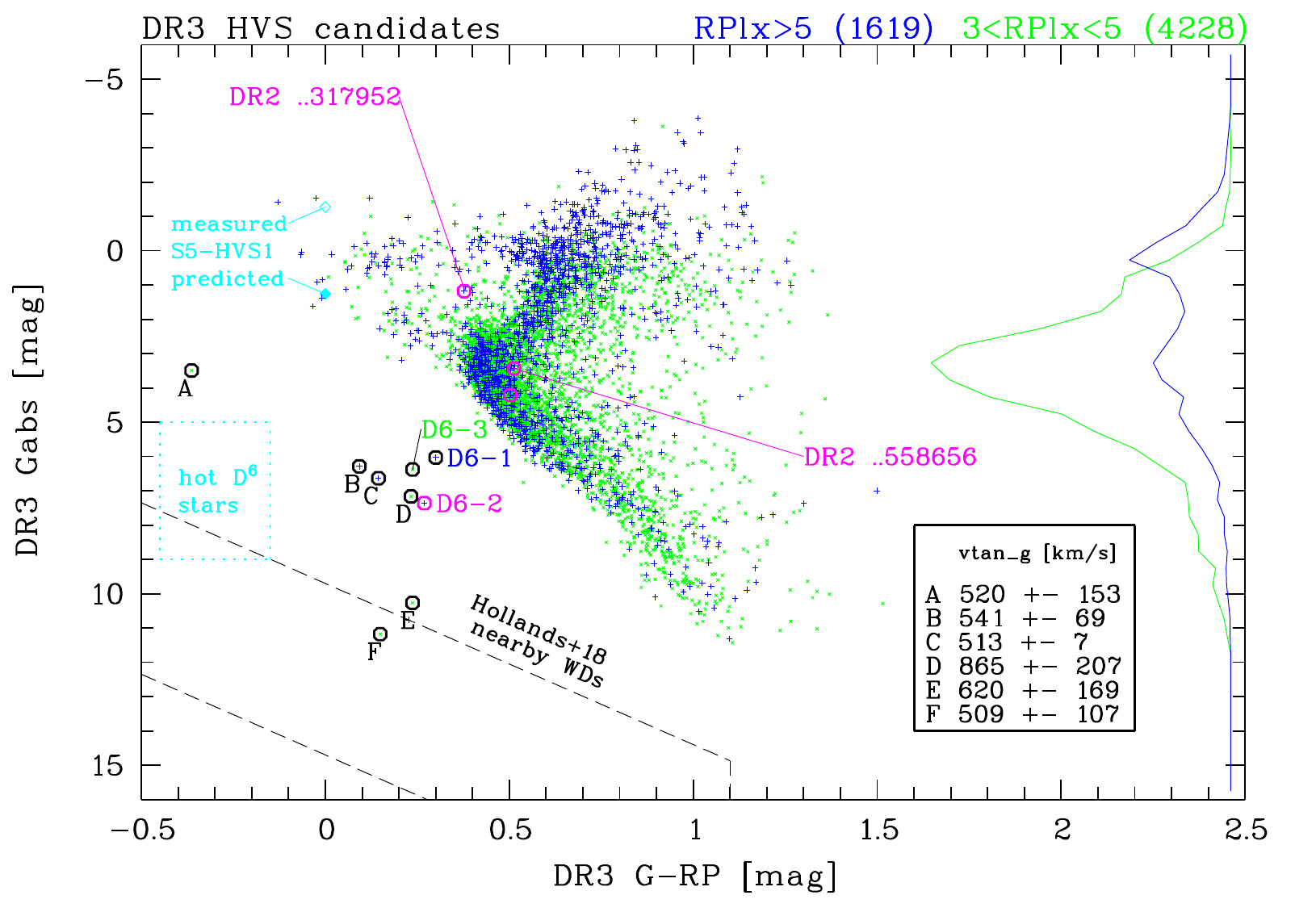}
      \caption{{\it Gaia} DR3 colour-magnitude diagram of high-priority
           (blue plus signs) and low-priority (green crosses)
           DR3 HVS candidates. The blue and green histograms on the right
           illustrate their absolute magnitude distributions, respectively.
           The region occupied by nearby WDs
\citepads{2018MNRAS.480.3942H} 
           is bounded by dashed lines, that of ''hot D$^6$ stars''
\citepads{2023OJAp....6E..28E} 
           by cyan dotted lines. The cyan lozenges show
           \object{S5-HVS1}
\citepads{2020MNRAS.491.2465K}, 
           not included in this study,
           with its predicted and measured absolute magnitudes (see text).
	   Magenta symbols and labelled objects
	   as in Figs.~\ref{Fig_dr2pm}--\ref{Fig_dr2vtangPM}. Two more D$^6$ stars from
\citetads{2018ApJ...865...15S}, 
	   \object{D6-1} and \object {D6-3}, and six other relatively blue
	   stars (A-F) are also labelled. The $vtan\_g$ of the latter six stars
	   are listed in the insert (see also text).
              }
      \label{Fig_dr3cmd}
\end{figure*}

   \begin{figure}
   \resizebox{\hsize}{!}{\includegraphics[angle=270]{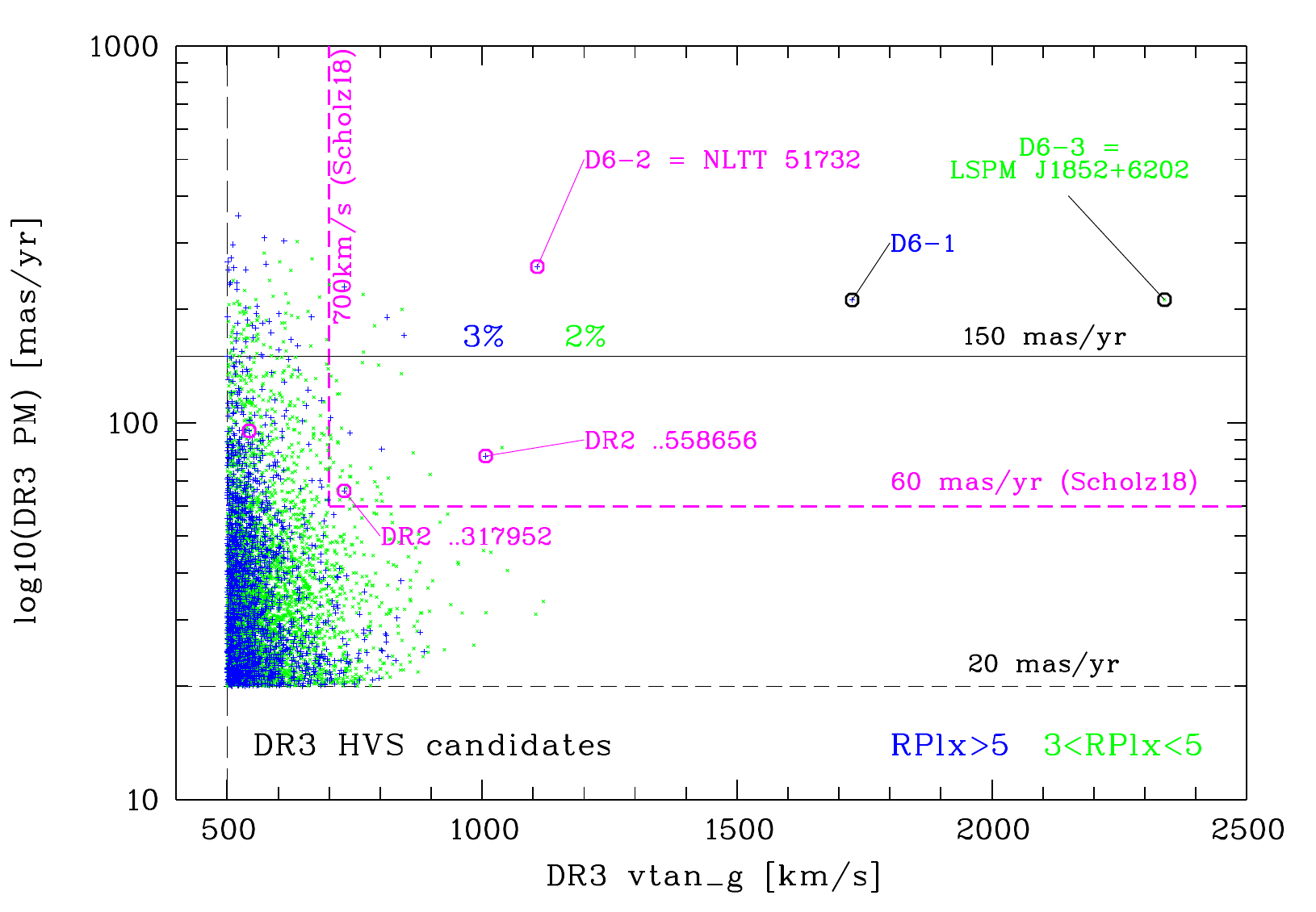}}
      \caption{Galactocentric tangential velocities and total proper motions
	   of high-priority (blue plus signs) and low-priority (green crosses)
	   DR3 HVS candidates. For tangential velocity errors, see
	   Fig.~\ref{Fig_DR3evtanG}. Overplotted magenta open circles and 
	   magenta labelled objects 
	   as in Figs.~\ref{Fig_dr2pm}--\ref{Fig_dr2vtangPM}. 
	   magenta dashed lines show the search limits of
\citetads{2018RNAAS...2..211S}. 
	   Also labelled are two other D$^6$ objects
\citepads{2018ApJ...865...15S}. 
              }
      \label{Fig_dr3vtangPM}
   \end{figure}

Only few of the high-priority (3\%) and low-priority (2\%) DR3 nearby HVS
candidates have proper motions as high as $PM$$>$150\,mas/yr
(Fig.~\ref{Fig_dr3vtangPM}). Among them are now also the other two of the
D$^6$ objects of
\citetads{2018ApJ...865...15S}, 
\object{D6-1} (= \object{Gaia DR3 5805243926609660032}) with high priority 
and \object{D6-3} (= \object{Gaia DR3 2156908318076164224}) with low priority,
which appear in Fig.~\ref{Fig_dr3vtangPM} with the most extreme Galactocentric
tangential velocities among all DR3 candidates. The third D$^6$ object,
\object{D6-2} (= \object{Gaia DR3 1798008584396457088}), 
was already included with high priority in the DR2
selection (Sect.~\ref{subS_DR2}) and listed by
\citetads{2018RNAAS...2..211S}. 
among six extreme HVS candidates (cf. overplotted magenta open circles in
Figs.~\ref{Fig_dr2vtangPM} and \ref{Fig_dr3vtangPM}). Two of them
have lower $vtan\_g$$<$500\,km/s in DR3 and are therefore not plotted
in Fig.\ref{Fig_dr3vtangPM}. Only three of these
six have still extreme $vtan\_g$$>$700\,km/s in DR3 and are labelled.
The velocity of $vtan\_g$$\approx$1000\,km/s of
one of these three, \object{Gaia DR2 3841458366321558656},
has not changed much between DR2 and DR3. However, for \object{D6-2}
$vtan\_g$ reduced from $\approx$1250\,km/s to $\approx$1100\,km/s,
and the DR2-based value of $\approx$1600\,km/s
for \object{Gaia DR2 6097052289696317952} became with $\approx$750\,km/s
much lower in DR3.
All large changes in the resulting $vtan\_g$ from DR2 to DR3
were caused by large differences in the measured parallaxes of the stars
but not in their proper motions. The HPM status of two of the D$^6$ objects
was known long before {\it Gaia}. One, \object{D6-2}, was already listed
as \object{NLTT 51732} in the catalogue of
\citetads{1995yCat.1098....0L}, 
whereas \object{D6-3} was originally named \object{LSPM J1852+6202} by
\citetads{2005AJ....129.1483L}. 

The CMD of the DR3 HVS candidates is shown in Fig.~\ref{Fig_dr3cmd}.
Only 5 out of 1619 high-priority and 13 out of 4228 low-priority DR3 HVS 
candidates lack $BP,RP$ photometry and are therefore not plotted as symbols 
in Fig.~\ref{Fig_dr3cmd}. 
As can be expected from the different parallax uncertainties and $G$
magnitude distributions (see histograms in Fig.~\ref{Fig_dr3NperG}), 
high-priority candidates dominate at the bright end and represent a sharper
picture of the MS and giants regions than the low-priority candidates.
As for DR2, with simply counting all high-priority DR3 candidates with 
absolute magnitudes brighter or fainter than the turnoff point
at $G_{abs}$$\approx$2.5\,mag, one gets about equal numbers of giants (841)
and dwarfs (778). However, the low-priority sample (see green histogram
on the right of Fig.~\ref{Fig_dr3cmd}) contains more dwarfs just below
the turnoff point. For both high- and low-priority samples, 95\% of these 
selected giants are located in the 1st and 4th Galactic quadrant and 
about 50\% at low Galactic latitudes ($|GLAT|$$<$20$\degr$). 
These fractions are considerably smaller for dwarfs, in particular 
the high-priority ones with 75\% and
28\%, respectively. These differences can be explained by
generally high numbers of distant giants in the Galactic 
bulge and inner disk regions (the mean parallax
and standard deviation of the selected high-priority giants is
0.17$\pm$0.05\,mas) compared to a more uniform sky distribution of
dwarfs in the solar 
neighbourhood 
(0.46$\pm$0.33\,mas for the
selected high-priority dwarfs). Both mean parallaxes and standard
deviations become smaller for low-priority candidates.

In comparison to the CMD of the DR2 HVS candidates (Fig.~\ref{Fig_dr2cmd}),
there are only few new blue DR3 HVS candidates left of the MS, whereas
from the two bluest DR2 objects and two luminous DR2 WDs discussed in 
Sect.~\ref{subS_DR2} only one (labelled C in Fig.~\ref{Fig_dr3cmd},
see below) is according to DR3 still counted as HVS
candidate. The already mentioned in DR2 high-priority HVS candidate
\object{D6-2} is now joined by \object{D6-1} (high priority in DR3)
and \object{D6-3} (low priority in DR3) from
\citetads{2018ApJ...865...15S}. 
When comparing the DR2 and DR3 CMDs, note that one of the
three objects labelled in magenta, \object{Gaia DR2 6097052289696317952},
was shifted from the giants region towards the MS, because of its
much larger parallax measured in DR3.

The six objects labelled with capital letters A-F in Fig.~\ref{Fig_dr3cmd},
three of which are located close to the D$^6$ objects and two falling in or 
at the border of the WD colour box\footnote{differences between DR2 and DR3
photometry are neglected here} defined for WDs in the Solar neighbourhood
\citepads{2018MNRAS.480.3942H}, 
deserve special attention. Their $vtan\_g$ and tangential velocity errors
are listed in the figure. Most of their $vtan\_g$ are not much higher than
the lower limit of 500\,km/s used in this study and much lower
than the extreme velocities of the D$^6$ objects. Only object D, appearing
next to the D$^6$ objects in the CMD, belongs to the small subsample of
the nearest extreme DR3 HVS candidates with $vtan\_g$$>$700\,km/s studied 
in more detail in Sect.~\ref{Sect_nearextrHVS} but with a relatively large 
velocity error. All tangential velocity errors will be shown in
Fig.~\ref{Fig_DR3evtanG}. The bright and very blue object A is 
the hot subdwarf \object{PG 1454+358} of spectral type sdO 
\citepads{1986ApJS...61..305G}  
also listed in the DR3-based catalogue of hot subluminous stars of
\citetads{2022A&A...662A..40C}. 
It was recently confirmed as sdO by
\citetads{2023OJAp....6E..28E}, 
who measured an RV of only about $-$170\,km/s.
Objects B and C are high-priority candidates and have relatively small
tangential velocity errors. Object C is \object{LP 40-365} 
\citepads{2017Sci...357..680V}, 
already discussed in Sects.~\ref{Sect_intro} and \ref{subS_DR2},
and object B is \object{Gaia DR2 5822236741381879040}, 
one of a few similar ''LP 40-365 stars'' found by
\citetads{2019MNRAS.489.1489R}. 
Both have also well-measured high RV of the order of 500\,km/s confirming 
their HVS status. Object D (= \object{Gaia DR3 3507697866498687232}) 
was included in the HVS search of
\citetads{2023OJAp....6E..28E} 
targeting 
DR3 sources with low-accurate parallaxes and spectroscopically
classified as another ''LP 40-365 star'' with an RV of only about $+$50\,km/s.
Objects E (= \object{Gaia DR3 5703888058542880896}) 
and F (= \object{Gaia DR3 3534629338669165440})
were already included in the DR2-based WD catalogue of
\citetads{2019MNRAS.482.4570G}. 
Object E was listed as HVS WD by
\citetads{2023MNRAS.518.6223I}. 
It was spectroscopically classified as DA WD by
\citetads{2023OJAp....6E..28E}. 

The region of ''hot D$^6$ stars'' discovered by 
\citetads{2023OJAp....6E..28E}. 
is also indicated
in Fig.~\ref{Fig_dr3cmd}, using their estimated absolute magnitudes.
Note that this region overlaps with the nearby WDs region. However, all
four ''hot D$^6$ stars'' did not even enter the sample of low-priority DR3
HVS candidates studied here, because of their extremely low $Plx/e\_Plx$$<$1.5
(including one negative parallax). If their true parallaxes will be
larger (following the observed trend in Sect.~\ref{subS_trends}), they may 
move into the WD region of the CMD with updated {\it Gaia} data. 
Also shown in Fig.~\ref{Fig_dr3cmd}
is the extreme HVS \object{S5-HVS1} (also not among our low-priority
candidates) with two different absolute magnitude estimates. The cyan 
filled lozenge corresponds to its predicted parallax of 0.11\,mas
\citepads{2020MNRAS.491.2465K} 
and the open lozenge to its still non-significant measured 
DR3 parallax of 0.035$\pm$0.040\,mas. This was already a much larger value 
than its negative DR2 parallax of $-$0.042$\pm$0.092, which was the only 
known parallax measurement available to
\citetads{2020MNRAS.491.2465K}. 
Their prediction was based on photometry, spectroscopy and the travel 
of \object{S5-HVS1} from the GC to its current position and may be proved 
with future {\it Gaia} data releases.

\subsection{Spurious HPMs in crowded regions}
\label{subS_falsepm}

According to the expected {\it Gaia} science
performance\footnote{https://www.cosmos.esa.int/web/gaia/science-performance},
not all stars down to the limiting magnitude
of $G$$\approx$20\,mag can be measured in crowded regions, when the
maximum transmission rate to the ground is reached. In addition, ''the
minimum separation to resolve a close, equal-brightness double star in the
on-board star-mapper detector is 0.23\,arcsec in the along-scan
and 0.70\,arcsec in the across-scan direction''
\citepads{2015A&A...576A..74D}. 
With the observing periods of 22 months in DR2 and 34 months in DR3,
and using the above along-scan resolution, mismatched objects detected
at the beginning and end of the observing period could mimic HPM stars with
125\,mas/yr and 80\,mas/yr, respectively. These false proper motions could
be even higher, if the epoch difference between mismatched detections was
smaller than the full observing period or if their separation was larger
than 0.23\,arcsec but not all detections could be transmitted to the ground.
For DR3, the angular resolution was investigated in more detail by
\citetads{2021A&A...649A...2L}. 
They found that it depends on many factors including the magnitudes of
close neighbours, their orientation on the sky, and the astrometric solution
type. In their Fig.~6, they illustrate three problematic ranges of separation.
Below 0.18\,arcsec always only one DR3 source is kept and flagged as
duplicated source. At separation 0.18-0.6\,arcsec most neighbours have
only two-parameter solutions (parallax and proper motion not measured).
Finally, with 0.6-2\,arcsec separation, many DR3 sources needed six-parameter
solutions, because they had no well-determined colour from the previous DR2,
and a pseudocolour had to be estimated as the sixth astrometric parameter.
In future {\it Gaia} data releases, the angular resolution will be reduced to
about 0.1\,arcsec. With a longer full observing period and higher numbers of
visibility periods $Nper$ for each detected object it will also be
easier to distinguish between mismatched detections of different faint
objects and real HPM objects in crowded fields.

   \begin{figure}
   \resizebox{\hsize}{!}{\includegraphics[angle=270]{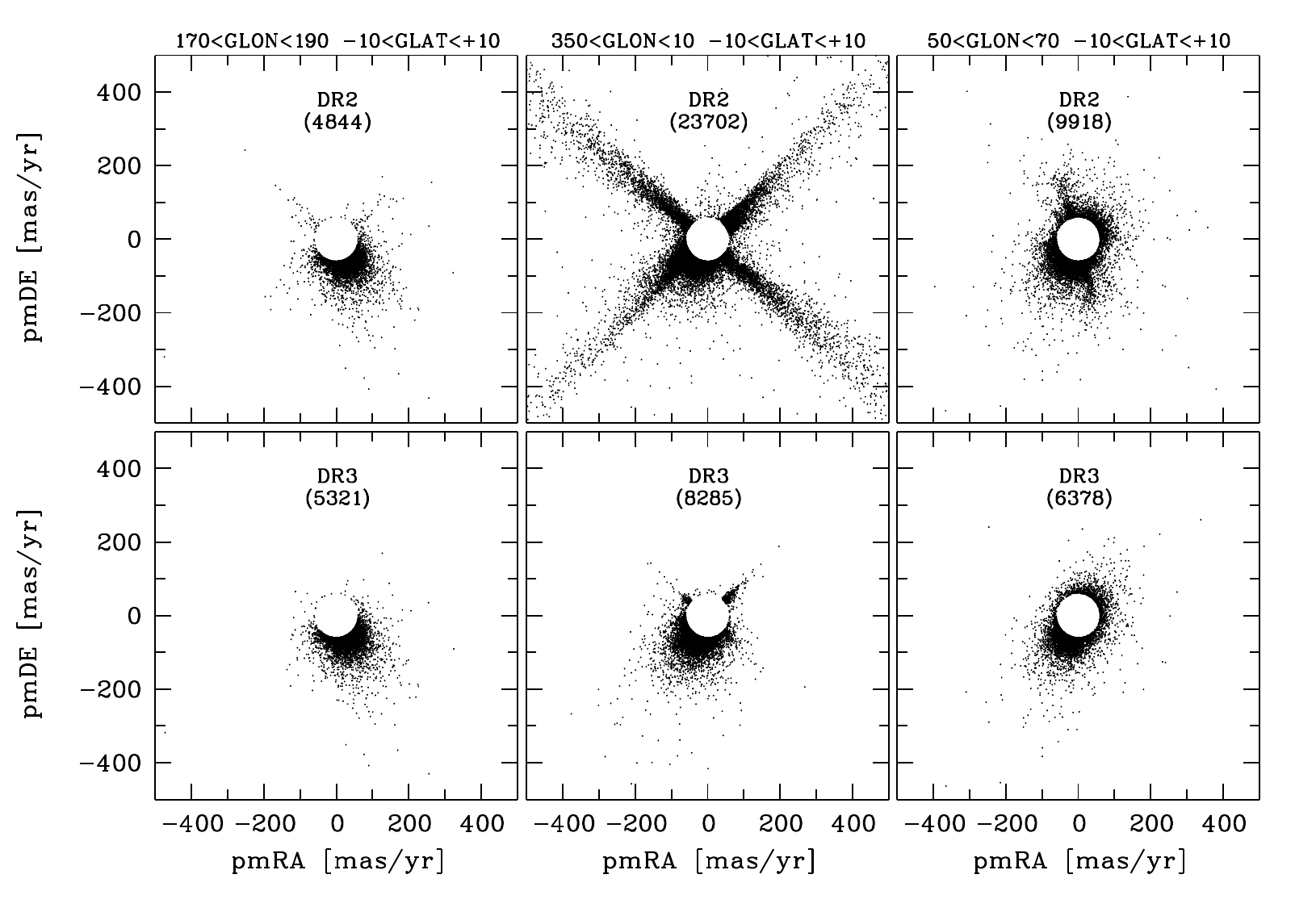}}
           \caption{Proper motions of all faint ($G$$>$18\,mag) HPM objects
           ($PM$$>$60\,mas/yr) in {\it Gaia} DR2 (upper row)
           and DR3 (lower row) in three selected 20$\degr$$\times$20$\degr$
           fields in the Galactic plane: at the Galactic anti-centre (left
           column), Galactic centre (middle), and at $GLON$$=$60$\degr$
           (right). The numbers of measured HPM stars in each field are
           given in round brackets, respectively. A few extreme HPM objects
           (in the upper middle panel more than 700) fall outside the frame.
              }
      \label{Fig_fpm}
   \end{figure}

Because of the already mentioned 25 false HPM objects among DR2 HVS candidates
(Sect.~\ref{subS_DR2}), the sky distribution of faint ($G$$>$18\,mag)
HPM stars measured in DR2 and DR3 were investigated. No further selection
with respect to their parallax significance or other astrometric quality
parameters was made at this point, to find out how well the cross-matching
of {\it Gaia} detections worked in fields of high stellar density. With a
proper motion limit of $PM$$>$60\,mas/yr, the DR2 HPM stars are strongly
concentrated in some areas towards low Galactic latitude, in particular in
the GC region, whereas the DR3 HPM stars are more uniformly distributed
on the sky.
Therefore, the proper motion diagrams in selected relatively
large and equal-size sky areas along the Galactic plane were studied
(Fig.~\ref{Fig_fpm}). In selected smaller sky areas of
high stellar density, containing e.g. the LMC, the Small Magellanic Cloud
(SMC), and some Galactic globular clusters, no obviously spurious HPMs
were found.

The strongest effect was uncovered for DR2 proper motions in the GC region
(upper middle panel in Fig.~\ref{Fig_fpm}) in form of unexpected cross-like
proper motion spikes. These spikes in the DR2 HPM data are getting sharper
and change their angles, when
one selects smaller sub-areas just north/south and east/west from the GC
(not shown here). A change in the direction of and angle between the
spikes can also be observed in the upper right panel, corresponding to a
sky area centered on ($GLON$$=$60$\degr$, $GLAT$$=$0$\degr$).
In the Galactic anti-centre area, only two weak spikes
can be seen in the DR2 data (upper left panel). Most of the spiky
structure in the DR2 proper motion diagrams is no longer visible in the
corresponding DR3 diagrams. Except for these obviously false proper motions,
one can also see that small numbers of other HPM stars in the outer parts
of the DR2 panels do not appear in the corresponding DR3 panels. The latter
can be expected from random pairing of unrelated object detections
(mismatching) in crowded fields, whereas the spikes in the proper motion
distributions may be the result of enhanced mismatching along certain
scan directions.

Out of the three selected fields, the Galactic anti-centre area is
probably less affected by crowding. If one considers all $\approx$5300
DR3 objects in this field without any obvious spiky proper motion structure
(lower left panel of Fig.~\ref{Fig_fpm}) real HPM stars, one can estimate
the number of false HPM objects in the other fields by assuming for
simplicity a uniform number density of real HPM stars on the sky. In the
Galactic plane at $GLON$$=$60$\degr$ (lower right panel), where crowding
effects are stronger, slightly more ($\approx$6400 $\leftrightarrow$ 120\%)
HPM objects were measured in DR3,
although no proper motion spikes are seen. The DR3 number of HPM objects
continues to rise up to $\approx$8300 $\leftrightarrow$ 156\%
in the GC region (lower middle panel),
where many HPMs concentrate along spikes and additional false HPM objects
may have been measured due to the strongest crowding. The corresponding
DR2 numbers rise to $\approx$9900 $\leftrightarrow$ 186\%
at $GLON$$=$60$\degr$ (upper right panel)
and even $\approx$23700 $\leftrightarrow$ 445\%
(upper middle panel), mainly because of many more
false HPM objects along the spikes. Interestingly, in the direction of the
Galactic anti-centre, the number of HPM objects in DR2 (upper left panel)
is with $\approx$4800 $\leftrightarrow$ 91\%
slightly lower than in DR3, although some spiky DR2
proper motions are visible. This indicates that DR3 contains less false
HPMs but may also be more complete with respect to real faint HPMs.

How does the picture shown in Fig.~\ref{Fig_fpm} change with brighter
HPM objects? This was investigated by selecting corresponding HPM objects
with $PM$$>$60\,mas/yr in the same sky areas but in the three
magnitude bins 17$<$$G$$<$18, 16$<$$G$$<$17, and 15$<$$G$$<$16 [mag].
Traces of the spike-like structure in the proper motion distribution,
containing dozens of false HPM objects, can then be found in the
DR2 data around the GC (cf. upper middle panel) down to the
magnitude interval 16$<$$G$$<$17 [mag]. A few individual false HPM objects
measured only in DR2 but not DR3 and located along the previously mentioned
spikes appear down to 15$<$$G$$<$16 [mag] in the GC and also in the Galactic
anti-centre field. The numbers of measured HPM objects become more and more
equal in DR2 and DR3 in each of the three fields, whereas the differences
between the fields are reduced with brighter magnitudes down to
about 120\% at $GLON$$=$60$\degr$ and
135\% in the GC region, in comparison to the Galactic anti-centre field.
These differences show that the assumption of a completely uniform
distribution of HPM stars on the sky is of course not true.
The non-uniform sky distributions of HVSs and slightly 
slower runaway stars were discussed by
\citetads{2014ApJ...793..122K} 
and
\citetads{2021A&A...646L...8N}. 

If the main astrometric quality criterion ($Plx/e\_Plx$$>$5) applied to HVS
candidates (Sect.~\ref{subS_DR2}) is used for the faint HPM objects shown in
Fig.~\ref{Fig_fpm}, strong proper motion spikes are still visible in the
GC region for both DR2 and DR3 data, comparable to what is seen in the two
middle panels. However, a different criterion can be used to exclude most of
the false HPM objects, namely an increased minimum number of visibility
periods $Nper$ (cf. Figs.~\ref{Fig_dr2NperG} and \ref{Fig_dr3NperG}). The
allowed numbers for objects with measured parallax and proper motion in DR2
and DR3 are $Nper$$\ge$6 and $Nper$$\ge$10, respectively. This is clearly
not yet sufficient in crowded fields, where faint objects appear in such high
numbers and with such small separations that a mismatching of downloaded
detections can not be avoided. For the faint HPM objects in the less
problematic DR3 data, the false proper motions along the spikes still
seen in the lower middle panel of Fig.~\ref{Fig_fpm} disappear only with
relatively high numbers of visibility periods ($Nper$$\ge$16).

   \begin{figure}
   \resizebox{\hsize}{!}{\includegraphics[angle=270]{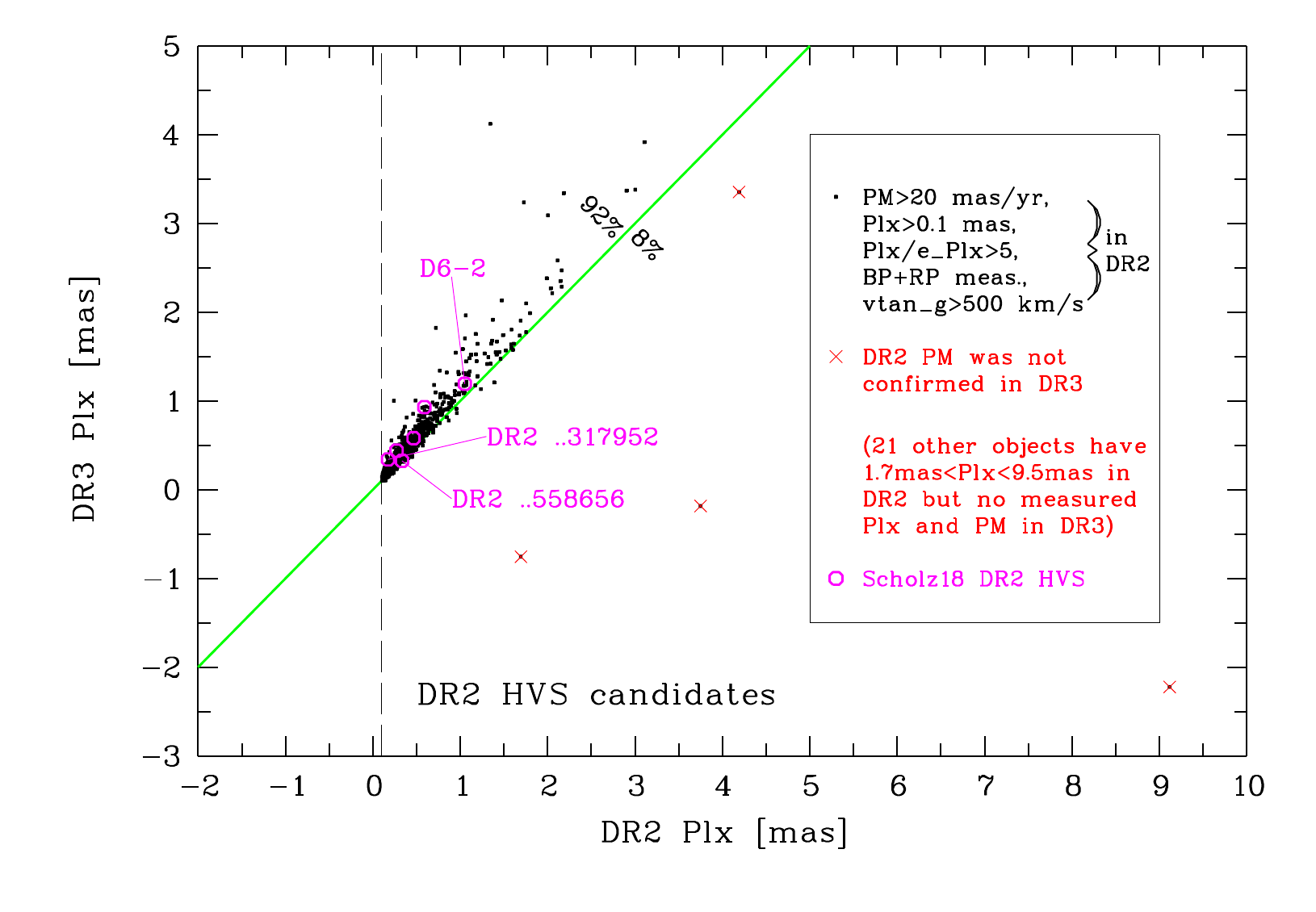}}
      \caption{Comparison of measured {\it Gaia} DR2 and DR3 parallaxes of
           high-priority HVS candidates selected from DR2. From 1520
           candidates 21 are not shown because of lacking DR3 parallaxes.
           Overplotted coloured symbols and labelled objects 
	   as in Figs.~\ref{Fig_dr2pm}--\ref{Fig_dr2vtangPM}. 
	   The green line indicates equality.
              }
      \label{Fig_dr2dr3plx}
   \end{figure}

\subsection{Trends from {\it Gaia} DR2 to DR3}
\label{subS_trends}

\begin{figure}
   \resizebox{\hsize}{!}{\includegraphics[angle=270]{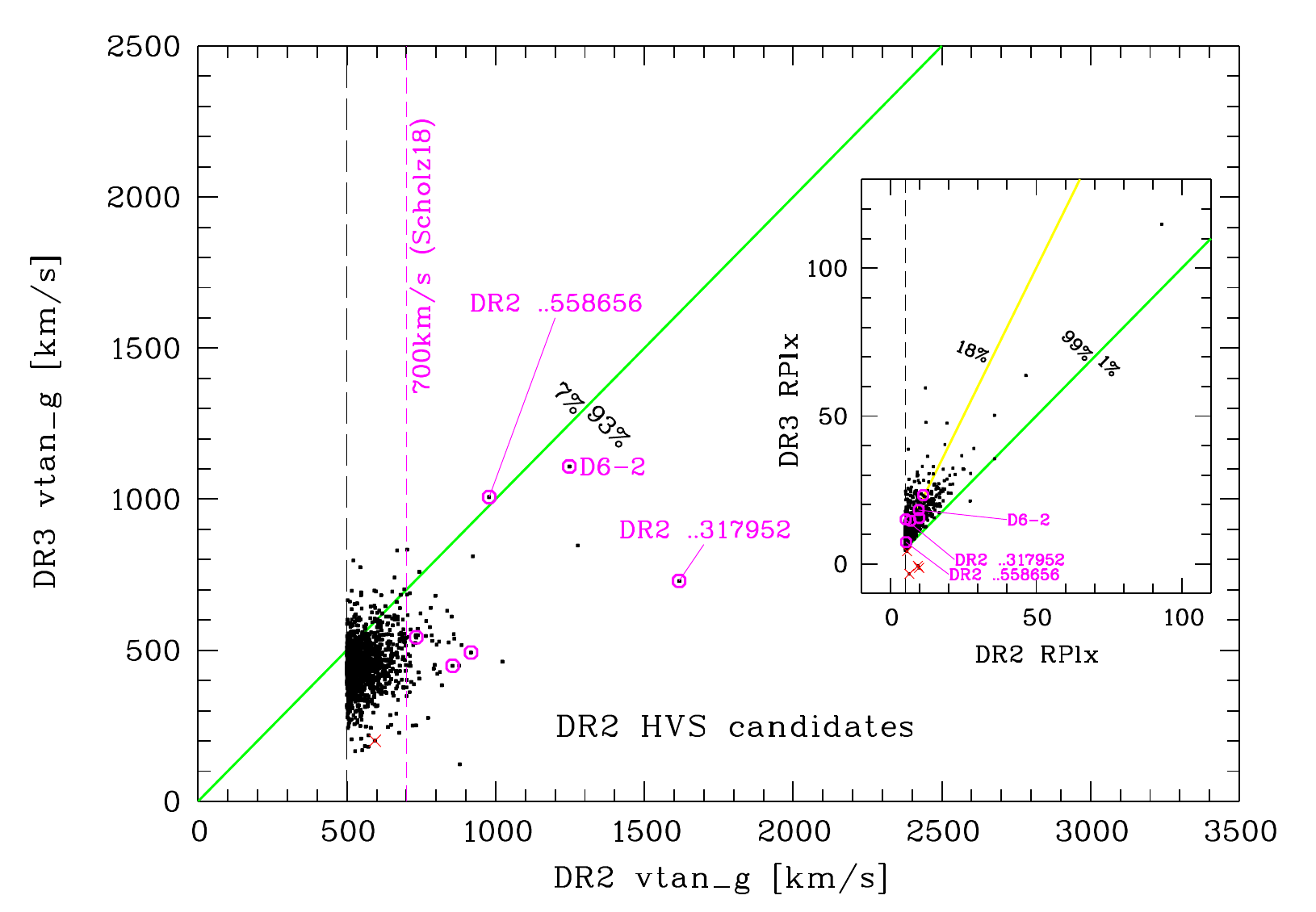}}
      \caption{Comparison of computed Galactocentric tangential velocities
           based on {\it Gaia} DR2 and DR3 data for
           high-priority HVS candidates selected from DR2.
           Overplotted coloured symbols and labelled objects 
	   as in Figs.~\ref{Fig_dr2pm}--\ref{Fig_dr2vtangPM}. 
	   The magenta dashed line marks the velocity limit set by
\citetads{2018RNAAS...2..211S}. 
           The green line indicates equality.
           The ratios $RPlx$$=$$Plx/e\_Plx$ are compared in the insert,
           where the green line again indicates equality, and the yellow
           line a 2:1 relation.
              }
      \label{Fig_dr2dr3vtang}
   \end{figure}

Individual objects were already mentioned in the previous subsections
for their changes towards larger DR3 parallax values with respect to
DR2 data. In Fig.~\ref{Fig_dr2dr3plx} the DR2 and DR3 parallax measurements
of all DR2-selected HVS candidates (Sect.~\ref{subS_DR2}) are compared. 
The DR3 parallaxes of four
objects with unconfirmed DR2 proper motions turned out to be much smaller
than in DR2, in three cases even negative DR3 parallaxes were measured.
For all other objects a general trend towards larger parallaxes in DR3
is observed. About 92\% of the DR2 HVS candidates lie above the green line
indicating equal parallaxes between DR2 and DR3. These changes lead also
to a general decrease in the computed $vtan\_g$ values 
(Fig.~\ref{Fig_dr2dr3vtang}). Again, for a large majority of 93\% of
the DR2-based HVS candidates the Galactocentric tangential velocities
became lower with DR3. The ratio $RPlx$$=$$Plx/e\_Plx$$>$5, used
as the only quality criterion in the DR2 HVS selection, increased for
99\% of all candidates, where an improvement by a factor 2 or more was
achieved for 18\% (see insert in Fig.~\ref{Fig_dr2dr3vtang}), as one
could expect from the extended {\it Gaia} observations and data
reductions resulting in DR3. Out of all 1520 high-priority DR2 HVS
candidates with $vtan\_g$$>$500\,km/s, only 348 (23\%) show up again 
among the 1619 high-priority DR3 HVS candidates. Concerning higher
velocities, there were 66 DR2 HVS candidates with $vtan\_g$$>$700\,km/s,
from which only seven (11\%) still appear among 64 high-priority DR3 HVS 
candidates with $vtan\_g$$>$700\,km/s. So, the numbers of HVS candidates
in DR2 and DR3 are similar, but the overlap of the samples reduces with
higher velocities.

   \begin{figure}
   \resizebox{\hsize}{!}{\includegraphics[angle=270]{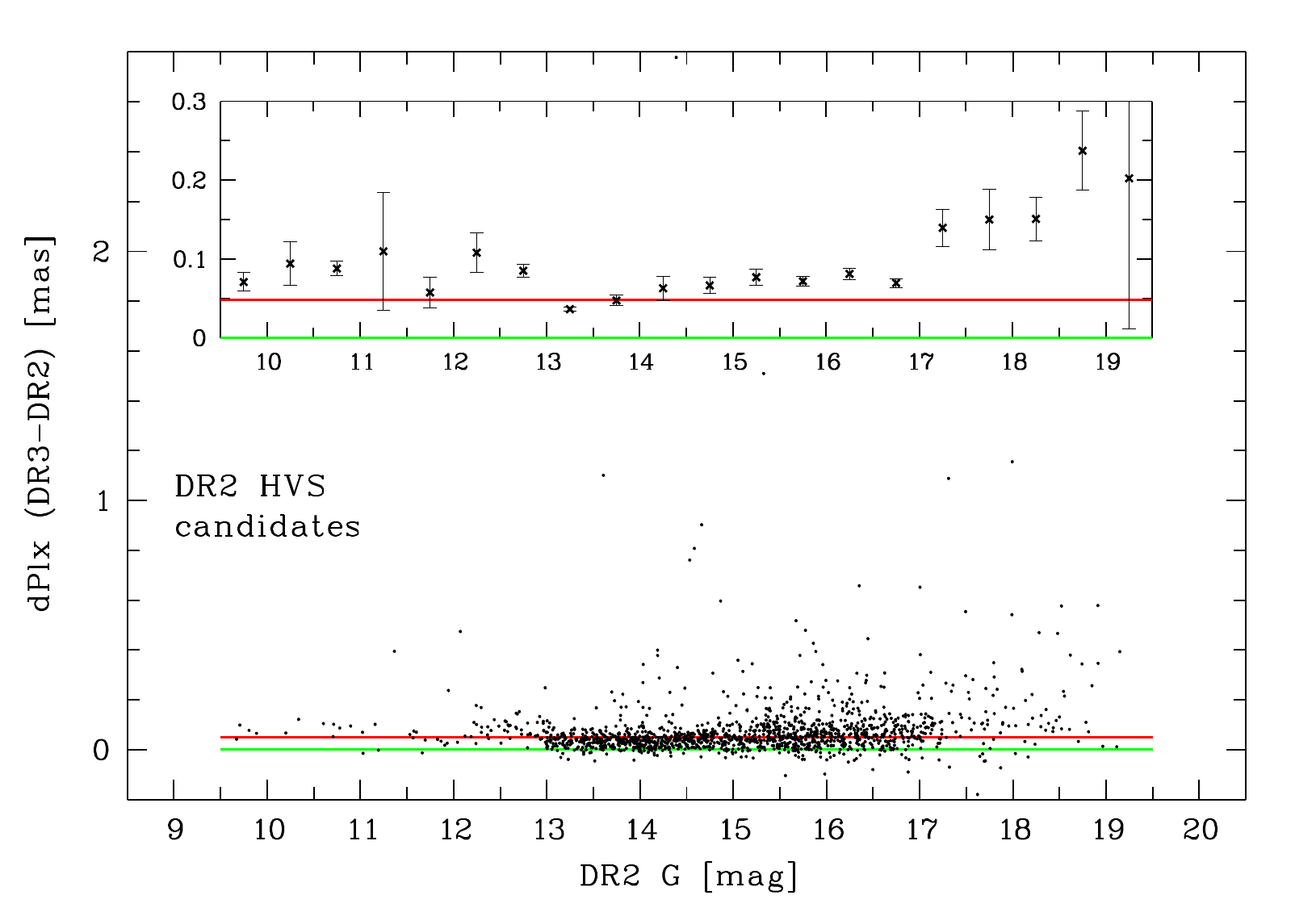}}
           \caption{Parallax differences $dPlx$ (DR3-DR2)
           vs. DR3 $G$ magnitudes for
           1495 high-priority DR2 HVS candidates with confirmed proper
           motions in DR3. The red line indicates
           their median parallax difference, the green line equal parallaxes.
           The insert shows mean parallax differences
           and standard deviations in 0.5\,mag bins.
              }
      \label{Fig_dPlxG}
   \end{figure}

Excluding stars with spurious DR2 HPMs, which were still used by
\citetads{2019ApJS..244....4D}, 
\citetads{2023AJ....166...12L}, 
by applying their own different quality criteria,
mentioned systematically larger distances of the remaining common stars
in the previous study of
\citetads{2019ApJS..244....4D}. 
\citetads{2023AJ....166...12L} 
argued that the official DR2 parallax zero-point correction of 
0.029\,mas, found from using quasars
\citepads{2018A&A...616A...2L} 
and taken into account by
\citetads{2019ApJS..244....4D}, 
was not sufficient. 
Since the HVSs from 
\citetads{2019ApJS..244....4D}. 
were relatively bright,
\citetads{2023AJ....166...12L} 
recommended using a larger correction of about 0.05\,mas found by
\citetads{2019MNRAS.487.3568S} 
and
\citetads{2019ApJ...878..136Z}, 
for bright stars. 
In Fig.~\ref{Fig_dPlxG}, the parallax differences (DR3-DR2) of
all DR2 HVS candidates, except for 21 objects without DR3 parallaxes and
4 objects with unconfirmed proper motions, are shown as a function of $G$ 
magnitude. The median parallax difference of 0.049\,mas agrees very well with 
the above suggested correction. However, individual parallax differences
are much larger (up to almost 3\,mas), and the mean values are larger than 
the overall median for both bright and faint stars. Only in the magnitude
interval 13$<$$G$$<$14 [mag] the mean parallax difference falls 
below 0.05\,mas.

   \begin{figure}
   \resizebox{\hsize}{!}{\includegraphics[angle=270]{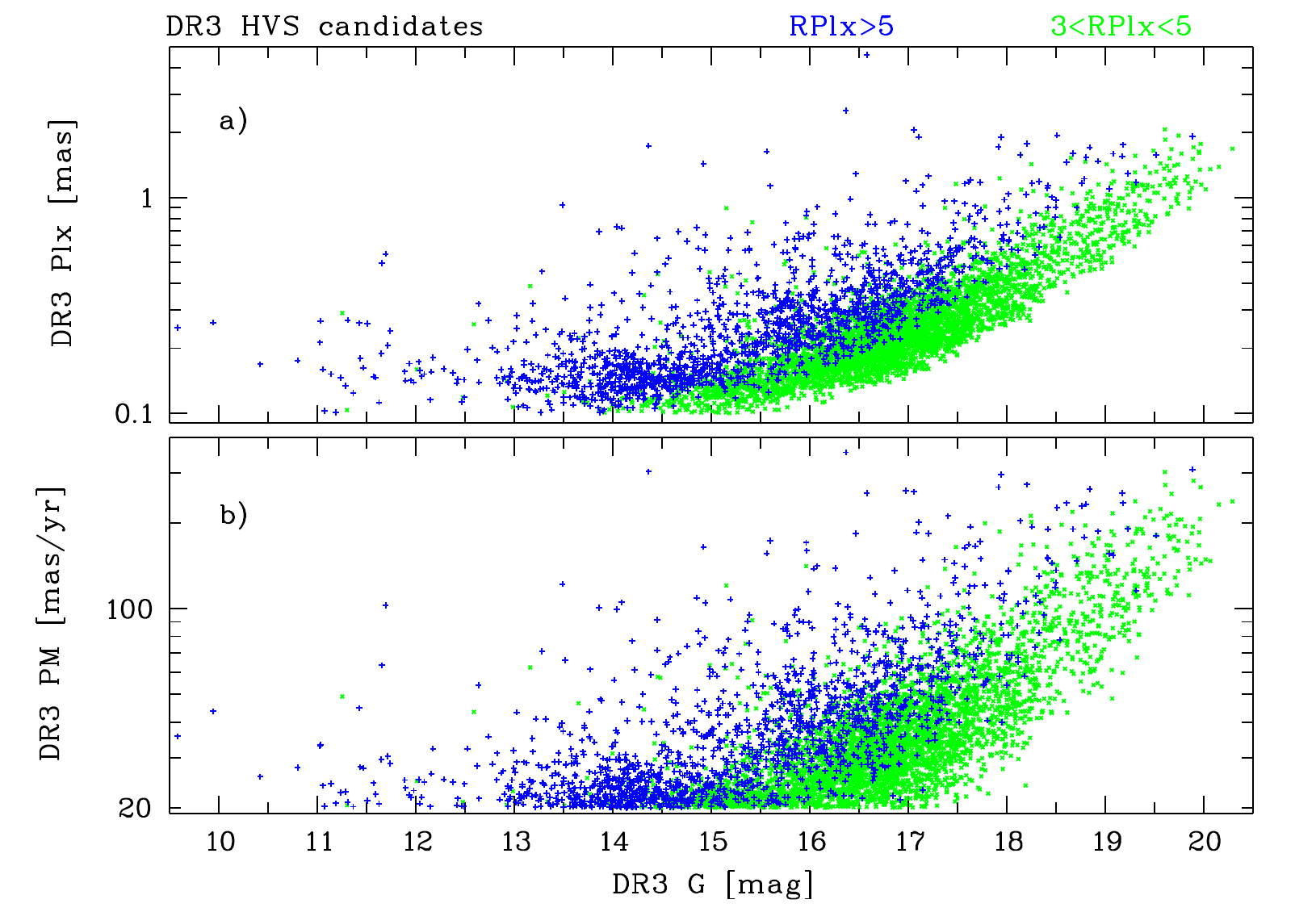}}
           \caption{Parallaxes (top) and total proper motions (bottom) of
           DR3 HVS candidates of high priority (blue plus signs)
           and low priority (green crosses) as a function of $G$ magnitude.
              }
      \label{Fig_DR3astmG}
   \end{figure}

   \begin{figure}
   \resizebox{\hsize}{!}{\includegraphics[angle=270]{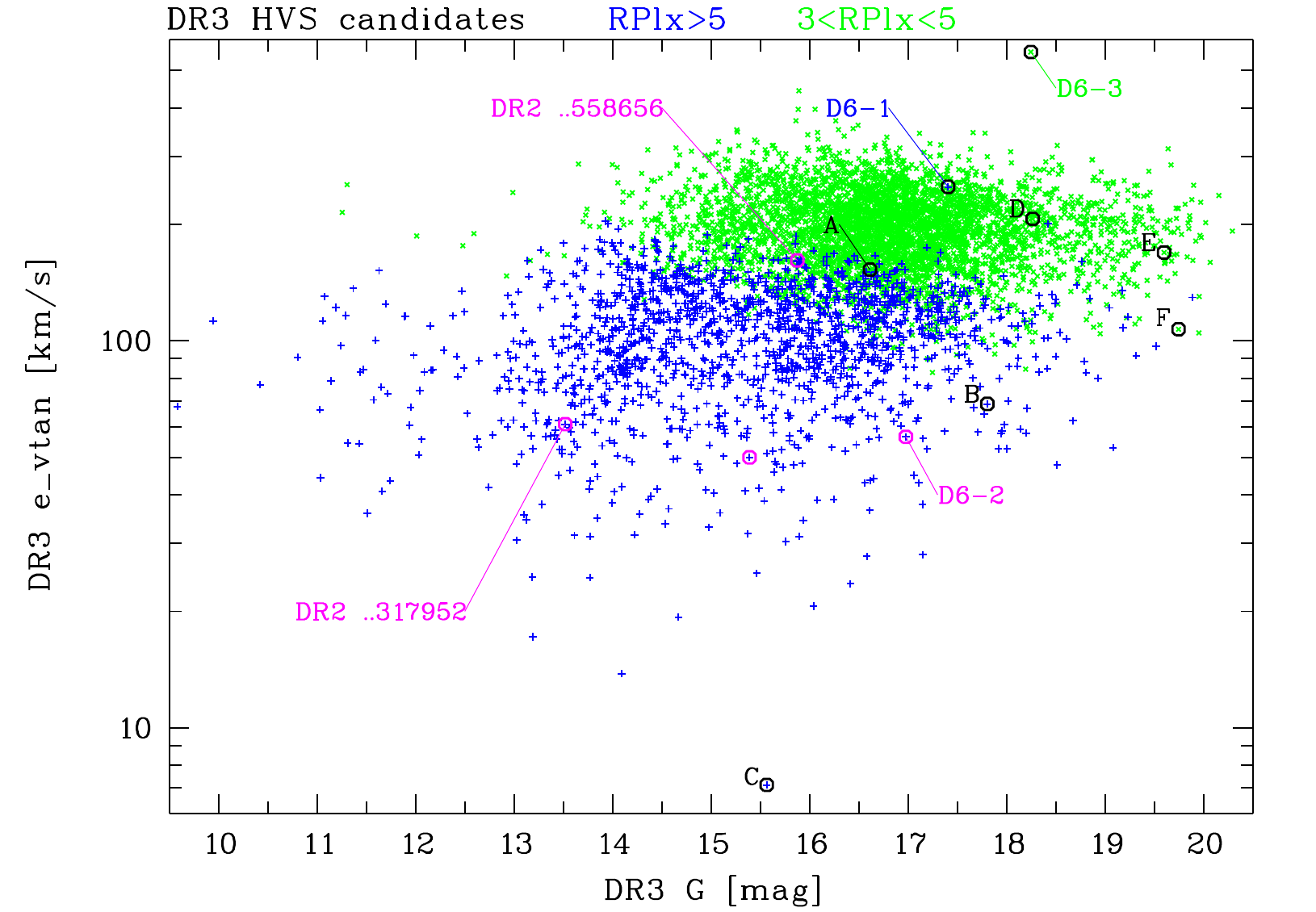}}
           \caption{Tangential velocity errors $e\_vtan$
           of high-priority (blue plus
           signs) and low-priority (green crosses)
           DR3 HVS candidates as a function of $G$ magnitude.
           For overplotted open circles and labelled objects compare
	   with Fig.~\ref{Fig_dr3cmd} 
	   and see text.
              }
      \label{Fig_DR3evtanG}
   \end{figure}

\subsection{DR3 HVS selection effects with magnitude}
\label{subS_DR3magsel}

The strongly magnitude-dependent parallax errors in DR3
\citetads[][see their Fig.~7]{2021A&A...649A...2L} 
and our HVS selection criteria, mainly the required
significance of parallaxes and minimum proper motion, lead to
some selection effects with $G$ magnitude. Figure~\ref{Fig_DR3astmG}
demonstrates that the parallaxes (panel a) of the HVS candidates
begin to rise with faint magnitudes, at $G$$\gtrsim$14.5\,mag for
high-priority and at $G$$\gtrsim$16.5\,mag for low-priority candidates.
There are also very few candidates at the lower parallax limit of 0.1\,mas.
Only small numbers of additional more distant HVS candidates were excluded
by the condition $Plx$$>$0.1\,mas, namely 5 (compared to 1619 selected)
high-priority and 49 (compared to 4228 selected) low-priority candidates.
Already after applying only two selection criteria, a proper motion limit
of $PM$$>$25\,mas/yr 
slightly
above the limit used in the HVS selection
and $RPlx$$>$3 as used for low-priority HVS candidates, to the whole DR3, 
there are no objects with parallaxes below 0.1\,mas left. Only about 1500 
out of 14.2 million selected objects have small parallaxes in the  
range of 0.1-0.2\,mas. With $RPlx$$>$5 the corresponding numbers reduce 
to about 400 out of 12.2 million. In both cases, the parallaxes of all
these HPM objects with $PM$$>$25\,mas/yr do also increase with fainter 
magnitudes in a similar way as seen in Fig.~\ref{Fig_DR3astmG} (panel a).
Panel b) of Fig.~\ref{Fig_DR3astmG} shows that the proper motions of our
HVS candidates also become higher with very faint magnitudes, 
at $G$$\gtrsim$17\,mag for
high-priority and at $G$$\gtrsim$18\,mag for low-priority ones. The faintest
objects of each category have $PM$$>$100\,mas/yr. Here, the reason is 
the HVS selection by high Galactocentric tangential velocity and the trend 
towards nearer objects with fainter magnitudes (panel a).

Figure~\ref{Fig_DR3evtanG} shows that the tangential velocity errors of
the DR3 HVS candidates do not change much with $G$ magnitude. The errors of
4228 low-priority candidates lie in the range of 80-560\,km/s (with a mean of
198\,km/s). Among high-priority candidates, the ten most significant
parallaxes reach ratios of about 20$<$$RPlx$$<$65. Therefore, the minimum
error of 7\,km/s, computed for the labelled object C (= \object{LP 40-365})
already discussed in Sects.~\ref{Sect_intro}, \ref{subS_DR2},
and \ref{subS_DR3}, is 36 times smaller than the maximum error of 250\,km/s.
The mean error of all 1619 high-priority candidates is with 104\,km/s
about two times smaller than that of low-priority candidates. It is 
remarkable (and logical) 
that the maximum errors in both groups belong to the objects
with the most extreme $vtan\_g$ (see Fig.~\ref{Fig_dr3vtangPM}), 
\object{D6-1} and \object{D6-3} (labelled). From three other extreme
HVS candidates previously selected in DR2 by
\citetads{2018RNAAS...2..211S} 
and still appearing with high priority 
in DR3, \object{Gaia DR2 3841458366321558656} exhibits a relatively 
large error of the tangential velocity, whereas the errors of \object{D6-2}
and \object{Gaia DR2 6097052289696317952} are smaller. These and some other
candidates, including object D that appeared next to the D$^6$ objects
in the CMD (Fig.~\ref{Fig_dr3cmd}) but was found by
\citetads{2023OJAp....6E..28E} 
to be spectroscopically similar to object C (\object{LP 40-365}),
are also labelled in Fig.~\ref{Fig_DR3evtanG}.

   \begin{figure}
   \resizebox{\hsize}{!}{\includegraphics[angle=270]{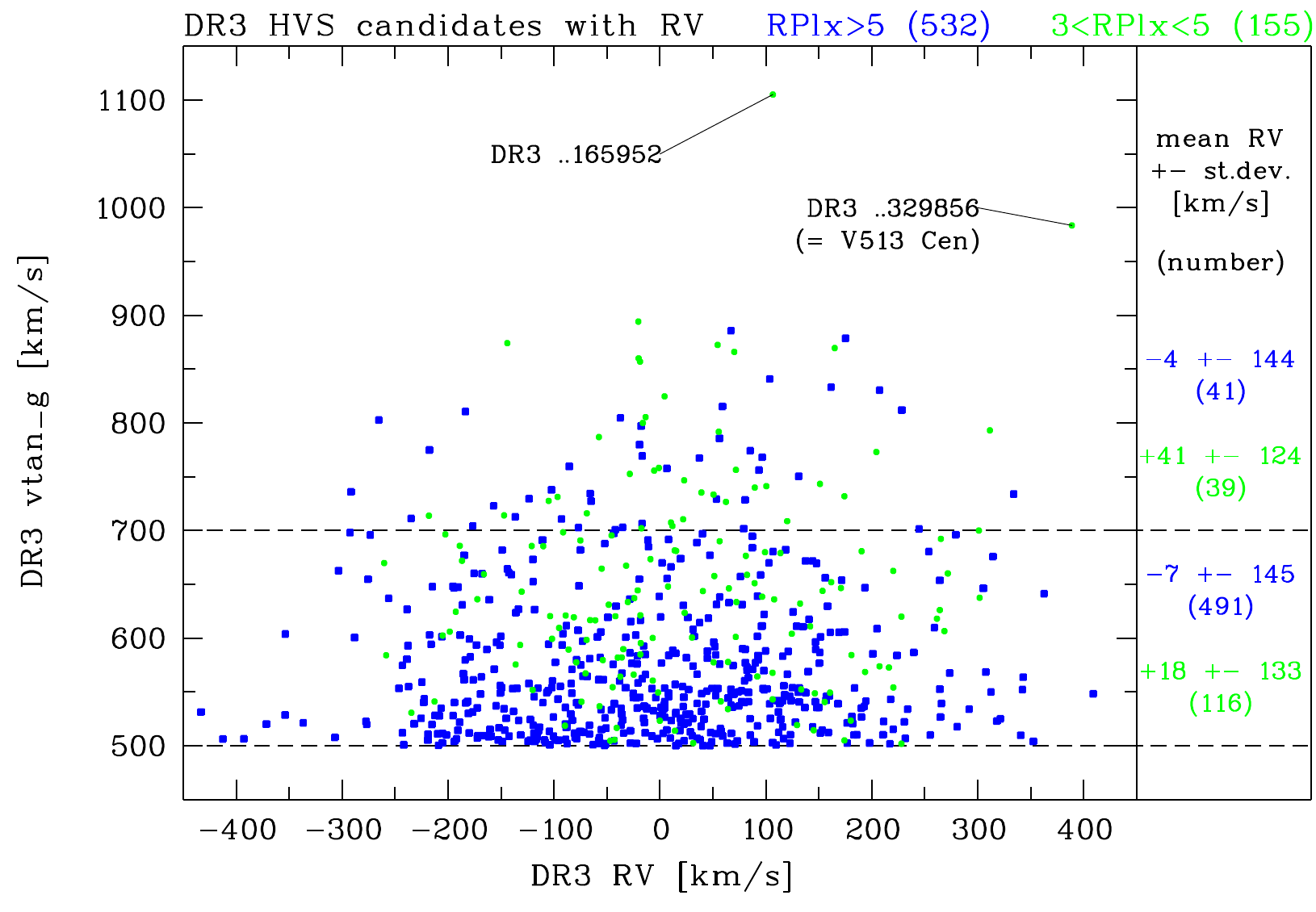}}
           \caption{Computed DR3 $vtan\_g$ vs. measured DR3 RVs of
           high-priority (blue filled squares) and low-priority (green dots)
           DR3 HVS candidates.
           Two objects with the highest $vtan\_g$ are labelled (see text).
           Right of the frame, the mean RVs and standard deviations in
           the velocity intervals
           500$<$$vtan\_g$$<$700 [km/s] and $vtan\_g$$>$700\,km/s are listed.
              }
      \label{Fig_RVvtang}
   \end{figure}

\subsection{Comparison with DR3 radial velocities}
\label{subS_compRV}  

With {\it Gaia} DR3, RV measurements of many of the brighter ($G$$<$15.5\,mag) 
HVS candidates became available. Out of 1619 high-priority DR3 HVS candidates
532 (33\%) have measured RVs, whereas among the mostly faint
4228 low-priority candidates, which show a peak in their magnitude 
distribution at $G$$\approx$17\,mag (see Fig.~\ref{Fig_dr3NperG}), 
only 155 (4\%) have RVs in DR3. Comparing these RVs with the $vtan\_g$
values (Fig.~\ref{Fig_RVvtang})
one can see that all RVs are within $\pm$500\,km/s, i.e. lower
than the minimum $vtan\_g$ used in the selection of the HVS
candidates. There is also no trend towards a higher RV dispersion with more 
extreme $vtan\_g$ values, as one could expect for a sample of real HVSs.
The RV standard deviation of high-priority candidates remains the same
($\pm$145\,km/s) for objects with $vtan\_g$ below and above 
700\,km/s. For low-priority candidates it is even smaller and slightly
decreases with higher $vtan\_g$. 

\begin{figure}
   \resizebox{\hsize}{!}{\includegraphics[angle=0]{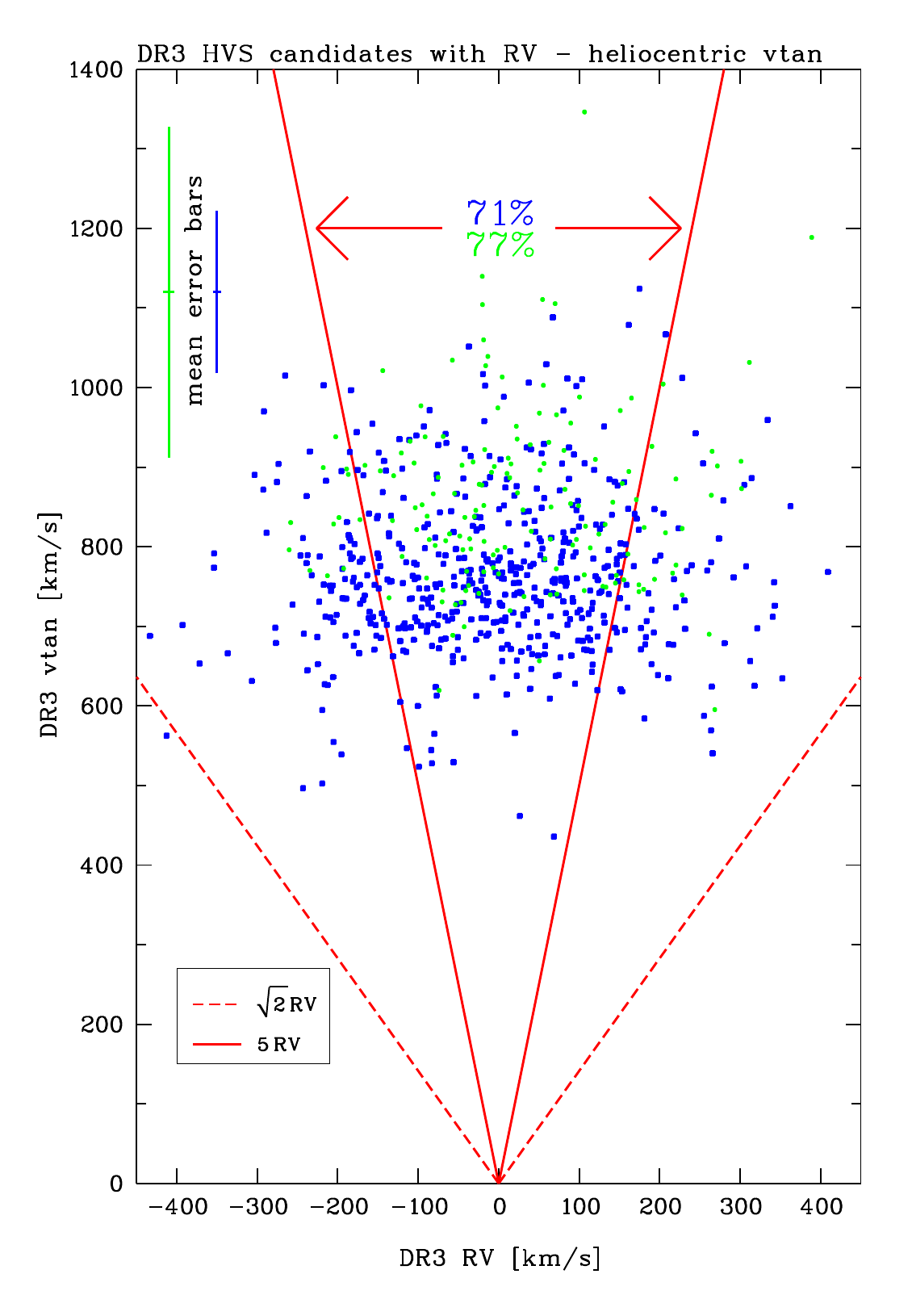}}
           \caption{Heliocentric tangential velocity $vtan$ vs. RV of
	   high-priority (blue filled squares) and low-priority (green dots)
           DR3 HVS candidates. Their mean error bars are shown on 
	   the top left. The red dashed line indicates a $vtan$
	   $\sqrt{2}$ times higher than the RV, expected for an
	   isotropic stellar distribution, the red solid line a 5 times 
	   higher $vtan$
\citepads[cf. Fig.~1 in][]{2014ApJ...780....7P}. 
              }
      \label{Fig_RVvtanh}
   \end{figure}

In any comparison of {\it Gaia} tangential velocities and RVs for objects
with only moderately significant parallaxes like our HVS candidates, one
should keep in mind the very different velocity errors. The mean error
bars of high-priority and low-priority HVS candidates with available RVs in
DR3 are shown in Fig.~\ref{Fig_RVvtanh}. The mean RV errors only slightly
change from about $\pm$5\,km/s to
$\pm$7\,km/s, respectively. On the other hand, the tangential velocity
errors are dominated by relatively large parallax errors, whereas the
proper motion errors of these HPM stars can be neglected. Therefore,
in this comparison with RVs,
the already very large mean tangential velocity errors of the
DR3 high-priority HVS candidates of about 102\,km/s are about twice
as large (208\,km/s) for the low-priority candidates.
Very similar mean tangential velocity errors were found for the complete
HVS samples (see Sect.~\ref{subS_DR3magsel}).  

The two most extreme ($vtan\_g$$>$900\,km/s) candidates with available 
DR3 RVs are of low priority only (labelled in Fig.~\ref{Fig_RVvtang}). Their 
tangential velocity errors are large (270-285\,km/s). Both are at parallactic 
distances of 9-10\,kpc. One, \object{Gaia DR3 6116771824584329856}, has
a relatively high RV of $+$389.0$\pm$7.7\,km/s measured in DR3 and is listed
in SIMBAD as the known RR Lyrae variable \object{V513 Cen}. However, its 
RV is not yet mentioned in SIMBAD. The other,
\object{Gaia DR3 6752707497293165952}, has an RV of $+$106.7$\pm$11.8\,km/s
and no SIMBAD entry. For \object{V513 Cen}, the All-Sky Automated Survey 
for Supernovae
\citepads[ASAS-SN;][]{2018MNRAS.477.3145J} 
catalogue of variable stars provides a period of about 
0.5 days\footnote{https://asas-sn.osu.edu/variables/107503}.
Its variability was also mentioned and classified in DR2, according to which
it belongs to the class of fundamental-mode RR Lyrae stars, but not in DR3. 

The heliocentric tangential velocity $vtan$ can be expected
to be only $\sqrt{2}$ times higher than the RV, if one assumes for
simplicity an isotropic stellar distribution. When
\citetads{2014ApJ...780....7P} 
compared $vtan$ and RV of their HVS candidates, they mentioned
much larger transverse-to-radial velocity ratios and concluded that
their HVS sample was strongly affected by large proper-motion errors
(see their Fig.~1). This conclusion was later confirmed by
\citetads[][see their Fig.~2]{2015A&A...576L..14Z}, 
who measured smaller proper motions for most of the HVS candidates of
\citetads{2014ApJ...780....7P}. 
In Fig.~\ref{Fig_RVvtanh} one can see that the majority of our DR3-selected
HVS candidates with available DR3 RVs, 376 of 532 high-priority candidates 
(71\%) and 119 of 155 low-priority candidates (77\%),
have $vtan$ values more than 5 times higher than their absolute RVs.
However, in this case the sample is likely not affected by proper motion
errors but underestimated parallaxes. The {\it Gaia} proper motion 
measurements of the DR3-selected HVS candidates will probably 
not change much, whereas the trend towards
larger parallaxes noted from DR2 to DR3 (Sect.~\ref{subS_trends}) may 
continue with the next {\it Gaia} data release.

\begin{figure*}
   \centering
   \includegraphics[angle=270,width=15cm]{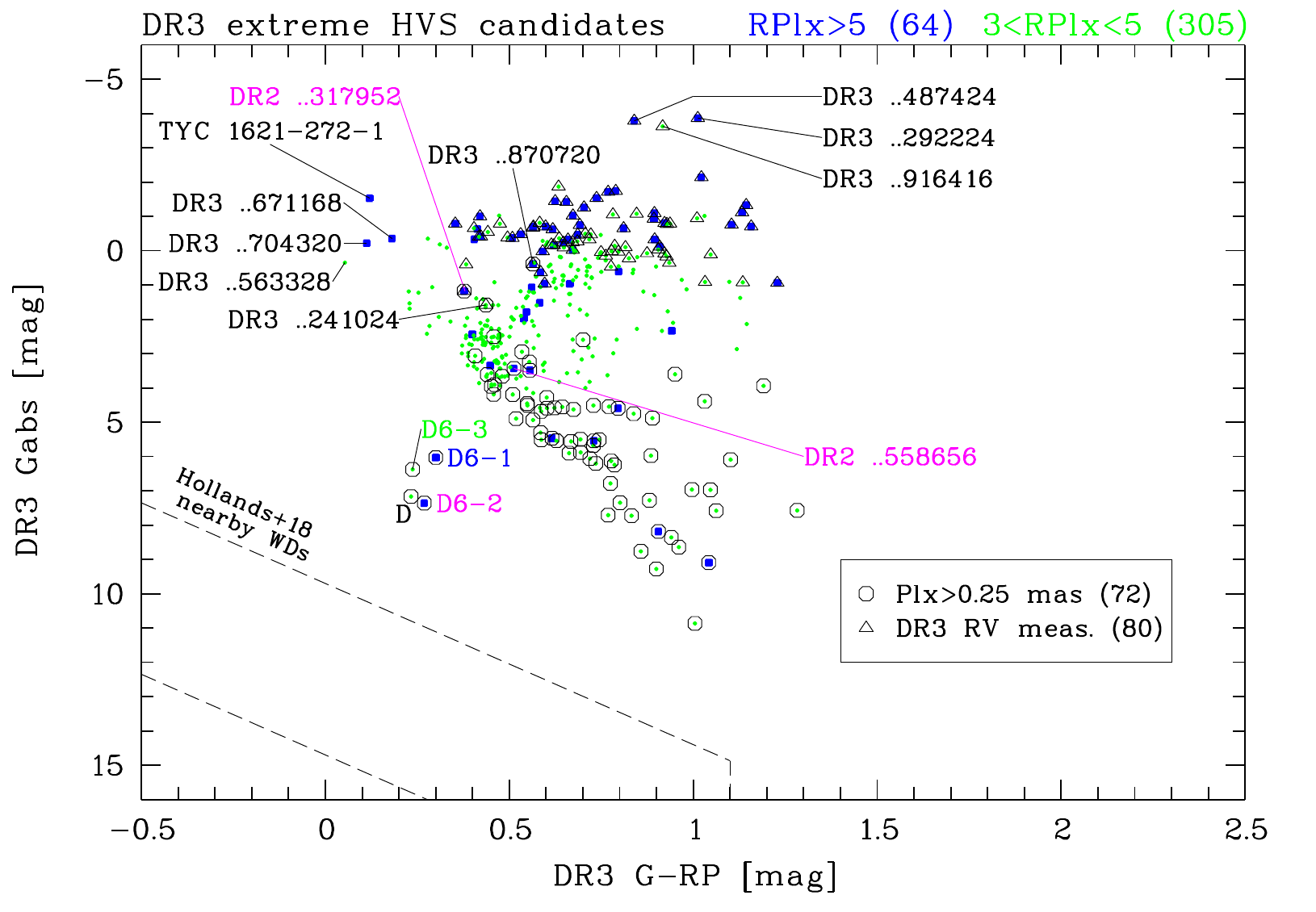}
           \caption{{\it Gaia} DR3 colour-magnitude diagram of
           extreme ($vtan\_g$$>$700\,km/s) HVS candidates of high priority
           (blue filled squares) and low priority (green dots). The 80
           objects with available DR3 RVs are overplotted by open triangles,
           the 72 nearest candidates ($Plx$$>$0.25\,mas) by open hexagons.
           For labelled objects and the WD region (dashed lines)
           compare with Figs.\ref{Fig_dr2cmd} and \ref{Fig_dr3cmd}
           and see text.
              }
      \label{Fig_CMDextr}
   \end{figure*}

Among the HVS candidates, only small numbers have extremely high
velocities $vtan\_g$$>$700\,km/s, 64 out of 1619 (4\%) high-priority and 
305 out of 4228 (7\%) low-priority candidates. Two of the low-priority
extreme HVS candidates lack $BP,RP$ photometry and are therefore not 
shown in the corresponding CMD (Fig.~\ref{Fig_CMDextr}).
Interestingly, for candidates with available DR3 
RVs (Fig.~\ref{Fig_RVvtang}) including only relatively bright 
stars ($G$$<$15.5\,mag), almost equal numbers of high- and low-priority 
objects, 41 out of 532 (8\%) and 39 out of 155 (25\%) appear as such 
extreme HVS candidates, respectively.
However, these 39 low-priority candidates are on average 0.7\,mag
fainter and 1.15 times more distant than the 41 high-priority candidates.
Both groups mainly consist of distant giants, as can be seen in the
{\it Gaia} DR3 CMD of extreme ($vtan\_g$$>$700\,km/s) HVS candidates
(Fig.~\ref{Fig_CMDextr}). Except for the bluest, the majority of 
high-priority candidates at the bright end of this CMD have DR3 RVs measured. 

The brightest ($G$$\approx$12.2\,mag) of the blue stars lacking DR3 RV 
measurements is \object{TYC 1621-272-1} (labelled in Fig.~\ref{Fig_CMDextr}). 
With its highly-significant DR3 parallax of 0.181$\pm$0.018\,mas it has
a relatively well-measured heliocentric 
tangential velocity of 842$\pm$84\,km/s but no references in 
SIMBAD\footnote{http://simbad.cds.unistra.fr/} and,
according to VizieR\footnote{http://vizier.cds.unistra.fr/},
also no non-{\it Gaia} RV measurements until now. Also labelled
are the blue high-priority HVS candidates
\object{Gaia DR3 6862302246501671168} and
\object{Gaia DR3 4445879669255704320}. Both are included in
the catalogue of blue horizontal-branch stars by
\citetads{2021A&A...654A.107C}, 
where their high heliocentric tangential velocities ($>$900\,km/s)
are listed, although without mentioning their large errors of the order
of $\pm$175\,km/s. The bluest low-priority candidate, 
\object{Gaia DR3 6885820662780563328}, is labelled as well 
in Fig.~\ref{Fig_CMDextr}.

The brightest giants ($G_{abs}$$<$$-$3\,mag) among the extreme DR3 HVS 
candidates, also labelled in Fig.~\ref{Fig_CMDextr}, have both parallaxes 
and proper motions just above the lower limits (0.1\,mas and 20\,mas/yr)
chosen for this study and $vtan\_g$$\lessapprox$760\,km/s, but tangential
velocity errors in the range 120-260\,km/s. Their RVs are much smaller and
precisely measured in DR3. For the two high-priority HVS candidates
of them, {\it Gaia} measured only moderately high RVs 
of $-$85.6$\pm$0.3\,km/s  (\object{Gaia DR3 1375165725506487424}) 
and $-$123.4$\pm$0.3\,km/s (\object{Gaia DR3 1591615309672292224}), for the
low-priority object \object{Gaia DR3 6666207818021916416} a small RV 
of $+$22.9$\pm$0.7\,km/s. The latter two have according to SIMBAD
external RV measurements confirming the {\it Gaia} results, whereas
the first of these three bright stars (all three have $G$$\approx$11\,mag)
had only a similar RV measured before in DR2 and no SIMBAD references.
According to DR3, \object{Gaia DR3 1591615309672292224} is a long-period 
variable candidate. It was also found to be a high-velocity star by
\citetads{2022AJ....164..187Q}. 
A spectroscopic binary classification is given in DR3 
for \object{Gaia DR3 6666207818021916416}, which is also included in DR6
of the Radial Velocity Experiment
\citepads[RAVE;][]{2020AJ....160...83S} 
and the Catalina Surveys Southern periodic variable star catalogue
\citepads{2017MNRAS.469.3688D}. 

\section{Verification of nearest extreme HVS candidates}
\label{Sect_nearextrHVS}

So far, in Sect.~\ref{Sect_vtan}, HVS candidates with parallactic
distances up to 10\,kpc and moderately high $vtan\_g$$>$500\,km/s were
considered using only one astrometric quality criterion
of $Plx/e\_Plx$$>$5 (high priority) or 3$<$$Plx/e\_Plx$$<$5 (low priority,
in DR3). In a more detailed validation of the HVS status, the 72 most 
extreme ($vtan\_g$$>$700\,km/s) DR3 HVS candidates with 
parallactic distances $<$4\,kpc were analysed. 
Most promising among those 
72 are only 11 candidates of high priority, but the majority of low-priority 
candidates were included for completeness. Their basic photometric and 
astrometric data, including astrometric quality parameters 
(see Sect.~\ref{subS_DR3quali}) and Table~\ref{Tab_quali}) 
from DR3, and their computed absolute magnitudes and tangential velocities 
are listed in Table~\ref{Tab_72dr3}. They have 32$<$$PM$$<$260 [mas/yr].
Most of the 72
objects are dwarfs according to their location in 
the CMD (Fig.~\ref{Fig_CMDextr}), for which the RV is not available 
(... in the tables).
Only three do have a measured RV in DR3. Among 
those three are \object{Gaia DR2 6097052289696317952} (labelled and already
mentioned earlier) with a relatively small RV of $+$53.5$\pm$4.5\,km/s and
another high-priority candidate (\object{Gaia DR3 1180569514761870720},
labelled), also with a small RV of $-$76.9$\pm$1.4\,km/s. The third
is a low-priority HVS candidate (\object{Gaia DR3 716216621590241024},
labelled) with a moderately high RV of $+$311.5$\pm$9.2\,km/s and
about three-times higher but uncertain heliocentric tangential velocity of
1032$\pm$263\,km/s (cf. Fig.~\ref{Fig_RVvtanh}). The DR3 RVs
of these three candidates do not strongly support their HVS status.

A possible problem leading to unreliable parallax measurements 
are close companions, which can occur everywhere on the sky, 
or overlapping background/foreground stars in crowded fields,
in particular in the Galactic plane and GC regions
(see Sect.~\ref{subS_falsepm}). After checking DR3 flags and
image parameters hinting at possibly disturbed stellar images,
each individual target was 
queried 
with a radial search 
for its closest next neighbours in DR3 (Sect.~\ref{subS_cnn}).
In addition, the most important astrometric quality parameters 
(Sect.~\ref{subS_DR3quali}) of objects with
similar magnitudes in the wider vicinity of each 
target (Sect.~\ref{subS_clocal}) and on the whole sky
(Sect.~\ref{subS_cglobal})
were investigated in a local and global comparison, respectively.

\subsection{Close next neighbours}
\label{subS_cnn}  
\subsubsection{DR3 flags}
\label{subS_DR3cnn}

The DR3 catalogue provides a number of flags and parameters, which can be
used to classify potential close binaries or overlapping stars. The flag
$Dup$ (= \textsf{duplicated\_source}) 
is set to 1, 
if multiple sources separated 
by less than 0.18\,arcsec were found. In such cases only one source was kept
\citepads{2021A&A...649A...2L}. 
This flag as well as the 
parameter $Solved$ (= \textsf{astrometric\_params\_solved}), indicating the
type of the astrometric solution (two-, five-, or six-parameter solution), 
were already briefly discussed in Sect.~\ref{subS_falsepm}. Six-parameter
solutions had to derive a 
pseudocolour 
in addition to the usual five
astrometric parameters, because of lacking colour information in DR2.
According to 
\citetads{2021A&A...649A...2L}, 
''the six-parameter solution is normally only used for sources that
are problematic in some respect, for example in very crowded
areas''.
At faint magnitudes ($G$$\gtrapprox$19\,mag), the six-parameter solutions 
begin to be more frequent than five-parameter solutions 
\citepads[][their Fig.~5]{2021A&A...649A...2L}. 

Other useful information obtained in the 
process of the DR3 image parameter determination (IPD) include the 
parameters $IPDfmp$ (= \textsf{ipd\_frac\_multi\_peak})
and $IPDfow$ (= \textsf{ipd\_frac\_odd\_win}). The first reports on the
fraction of observations of a given object where IPD detected more than 
one peak and may be a hint to resolved binaries that in 
some scan directions produce more than one peak in the window. As noted by
\citetads{2021A&A...649A...2L}, 
''towards the faint
magnitudes the \textsf{ipd\_frac\_multi\_peak} is always decreasing,
because the diminishing signal-to-noise ratio (S/N) makes the
detection of secondary peaks increasingly difficult''.
The second gives the fraction of field-of-view transits with 'odd'
windows and is sensitive to an overlap with another (usually brighter) 
source in the window 
\citepads{2021A&A...649A...2L}. 
Finally, there is a non-single star flag (= \textsf{non\_single\_star})
for objects with additional information in the various non-single star 
tables of DR3.

\begin{figure*}
   \centering
   \includegraphics[angle=270,width=15cm]{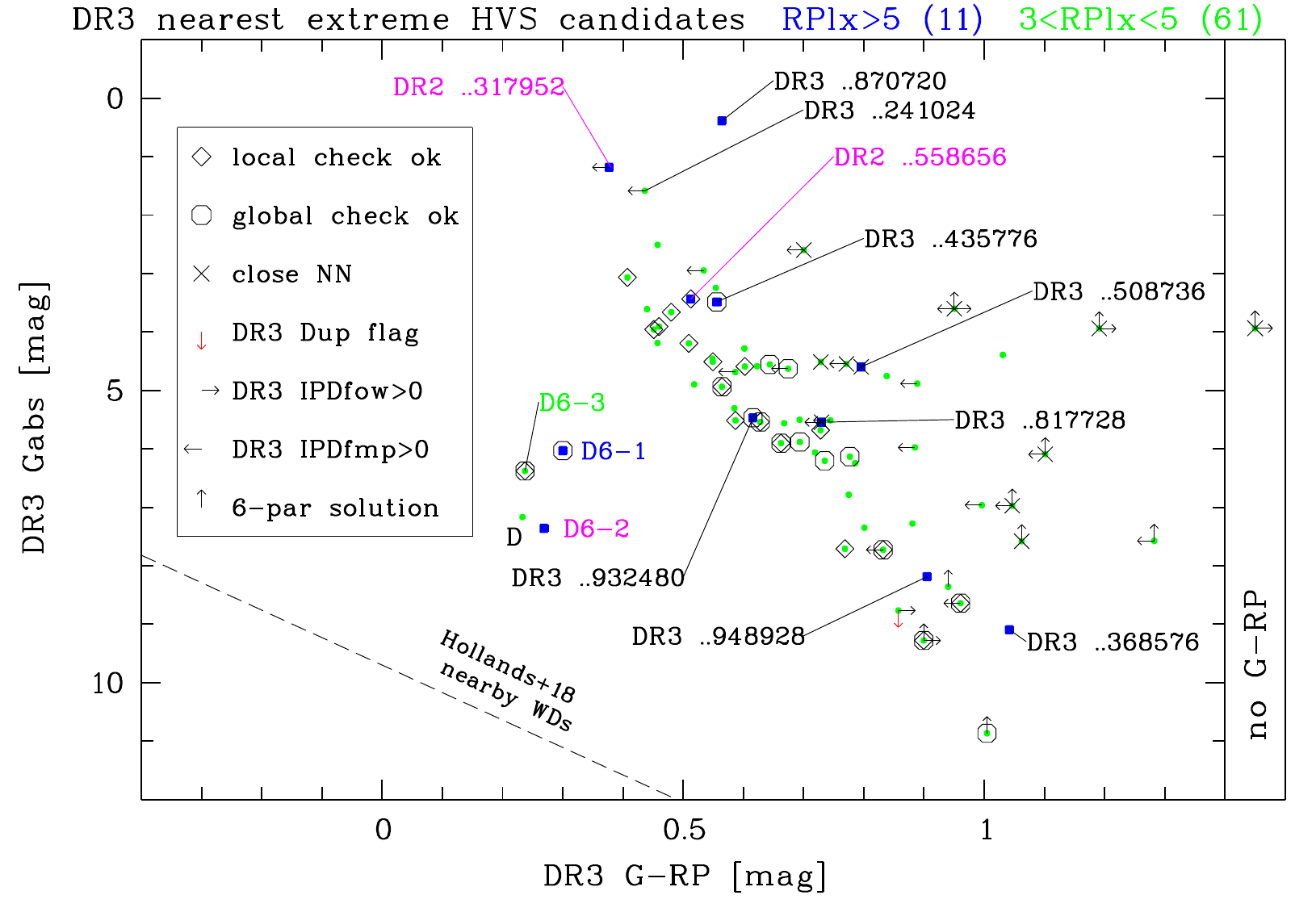}
           \caption{Zoomed CMD of the 72 nearest extreme DR3 HVS candidates
           of high priority (blue filled squares) and low priority
           (green dots). Only objects with overplotted open symbols passed
           the local (lozenges) and global (hexagons) checks of astrometric
           parameters (see Sects.~\ref{subS_clocal} and \ref{subS_cglobal}).
           One object lacking $G$$-$$RP$ colour is
           shown on the right. Overplotted crosses mark objects with a close
           next neighbour (NN) in DR3, arrows with different directions
           indicate conspicuous DR3 flags and parameters
           (see Sect.~\ref{subS_cnn}).
           In addition to objects already labelled in Fig.\ref{Fig_CMDextr},
           all other high-priority candidates are also labelled (see text).
           The upper boarder of the nearby WD region is indicated by the
           dashed line.
              }
      \label{Fig_cmd72nnGG}
   \end{figure*}

For none of the 72 nearest extreme HVS candidates a non-single star
flag was found in DR3. The occurrence of the four other flags and parameters 
mentioned above is indicated in the CMD (Fig.~\ref{Fig_cmd72nnGG}) by 
overplotted arrows in four different directions. A relatively large
number of candidates (18) have $IPDfmp$$>$0 (arrows left), 
followed by 10 candidates with six-parameter solutions (arrows up), 
a relatively small number (5) with $IPDfow$$>$0 (arrows right), 
and only one with $Dup$$>$0 (red arrow down). Altogether, 25 candidates
are marked by arrows, some with more than one of these four criteria. 
These 25 objects are located at the bright and faint end of the CMD or to 
the red side from the MS, but not in the region above the nearby WDs
occupied by the most extreme (with respect to their $vtan\_g$) D$^6$ objects
\citepads{2018ApJ...865...15S}. 

   \begin{figure}
   \resizebox{\hsize}{!}{\includegraphics[angle=270]{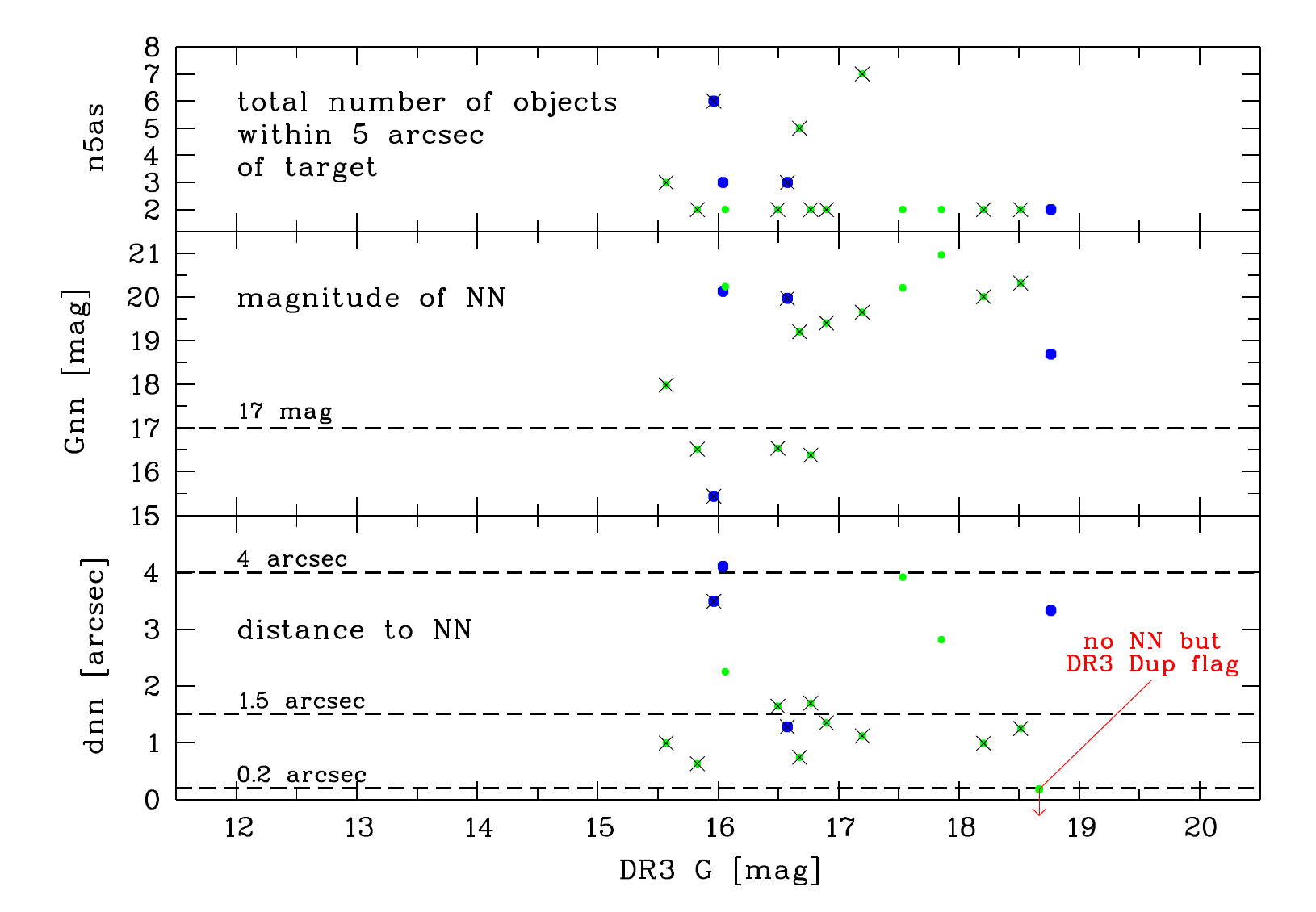}}
           \caption{Results of next neighbour (NN) search around
	   72 nearest extreme HVS candidates of high priority (blue filled 
	   squares) and low priority (green dots). Neighbours separated
	   by less than 5\,arcsec were found for only 16 objects.
	   Plotted over their $G$ magnitudes are the total number of
	   objects within 5\,arcsec (top), the magnitude of the NN
	   (middle), and the distance to the NN (bottom).
           Overplotted crosses mark the 11 objects with a close NN 
	   expected to have affected the astrometry.
	   One candidate without NN but with a DR3 Dup 
	   flag is shown for comparison (red arrow down).
              }
      \label{Fig_nn16}
   \end{figure}

\subsubsection{Next neighbour search in DR3}
\label{subS_owncnn}

To get even more information on possibly affected astrometric measurements,
an own close next neighbour (NN) search within DR3 was carried out around each 
of the 72 nearest extreme HVS candidates using VizieR. All candidates with at 
least one other DR3 object with a separation of less than 1.5\,arcsec from
the target (8 candidates) or with at least one relatively bright 
($G$$<$17\,mag) neighbour separated by less than 4\,arcsec
(3 candidates) were considered as having a close NN that potentially
had an influence on the astrometric measurements. Around 5 other candidates, 
only fainter neighbours were found at separations of 2-5\,arcsec, which
were not considered to be problematic. 

The $G$ magnitudes of the 16 nearby 
extreme HVS candidates and the number of detected objects (including the
target) within a search radius of 5\,arcsec $n5as$, the magnitude of
the nearest neighbour $Gnn$, and the distance to their nearest neighbour
$dnn$ are shown in Fig.~\ref{Fig_nn16} and listed in Table~\ref{Tab_72nnflg}.  
Among the 11 candidates defined 
to have a close NN and marked by overplotted crosses in 
Figs.~\ref{Fig_cmd72nnGG} and \ref{Fig_nn16}, 5 have two or more (up to 
six) DR3 neighbours within 5\,arcsec. All 11 appear as red objects in the 
CMD (Fig.~\ref{Fig_cmd72nnGG}), and 9 of 11 do also have suspicious DR3 
parameters or flags, as marked by the overplotted arrows in different 
directions. Thus the close NN search partly confirmed and slightly extended 
the list of 25 previously found problematic HVS candidates based on their 
DR3 flags and parameters.

\subsection{DR3 astrometric quality parameters}
\label{subS_DR3quali}

The {\it Gaia} DR3 provides a very large number of parameters and flags,
which can be used for a quality assessment of each individual source. 
The {\it Gaia} Catalogue of Nearby Stars
\citepads[GCNS;][]{2021A&A...649A...6G} served as a good example for
a careful DR3 quality check. More
than 40 catalogue columns were involved in the classification of good
astrometric solutions, when the GCNS
was constructed. From about 500000 stars with measured $G$ magnitudes
and $G$$-$$RP$ colours, and with parallaxes $Plx$$>$10\,mas, only about 296000
(59\%) entered after their random forest classification the 
reliable 100\,pc sample. The rejected about 204000 (41\%) objects are 
strongly concentrated in a wide region towards the GC ($|GLAT|$$<$15$\degr$,
$GLON$$<$90$\degr$ or $GLON$$>$270$\degr$) and two small regions towards the
LMC and SMC but do also show patterns corresponding to the scanning
law of the satellite with presumably low numbers of visibility periods $Nper$
\citepads[][see their Fig.~1]{2021A&A...649A...6G}. 
All 15 astrometric parameters found as the most important ones by
\citetads[][their Table A.1]{2021A&A...649A...6G}, 
except for the proper motion components $pmRA$ (= \textsf{pmra}) 
and $pmDE$ (= \textsf{pmdec}), were also used in our evaluation of
the 72 nearest extreme HVS candidates. Table~\ref{Tab_quali} 
summarises the used parameters and flags with fixed limits applied
throughout the paper ($RPlx$) and in the analysis of possibly disturbed
images caused by an unresolved binary or close NN (Sect.~\ref{subS_cnn}) 
in the upper part. In the lower part of Table~\ref{Tab_quali}
those parameters are listed, for which the target's values 
were compared to the 0.75 quantile $q75$
and (in case of $Nper$) median (0.5 quantile) values $med$
of large samples of objects with similar magnitudes selected for 
local (Sect.~\ref{subS_clocal}) and global (Sect.~\ref{subS_cglobal})
comparison.

\begin{table}
\caption{DR3 quality flags and parameters used for HVS candidates,
	in particular for the 72 nearest extreme candidates studied
	in Sect.~\ref{Sect_nearextrHVS}}  
\label{Tab_quali}      
\fontsize{7.0pt}{0.90\baselineskip}\selectfont
\centering                                      
\begin{tabular}{@{}l@{\hspace{2mm}}l@{\hspace{2mm}}c@{\hspace{2mm}}l@{}}     
\hline\hline                        
Quantity/  & DR3 column  &  Critical values & Remark \\    
criterion &             & (? among 72) \\
\hline                                   
	    & \textbf{throughout the paper:} \\
$RPlx$      & \textsf{parallax\_over\_error}  & $>$5 (11) & high priority \\
	    &                                 & 3-5 (61) & low priority \\
\hline                                   
	    & \textbf{Sect.~\ref{subS_cnn}:} \\
close NNs   & ...                         & (11) & own NN search in DR3\\
$IPDfmp$    & \textsf{ipd\_frac\_multi\_peak} & $>$0 (18) & \\
$IPDfow$    & \textsf{ipd\_frac\_odd\_win} & $>$0 (5) & \\
$Solved$    & \textsf{astrometric\_params\_solved} & $=$95 (10) & six-parameter solution \\
$Dup$       & \textsf{duplicated\_source} & $>$0 (1) & \\
\hline                                   
	    & \textbf{Sects.~\ref{subS_clocal}; \ref{subS_cglobal}:} \\
$Nper$      & \textsf{visibility\_periods\_used} & $<$$med$ (22; 41) \\
$e\_Plx$    & \textsf{parallax\_error} & $>$$q75$ (15; 26) & in [mas] \\
$e\_pmRA$   & \textsf{pmra\_error} & $>$$q75$ (15; 25) & in [mas/yr] \\
$e\_pmDE$   & \textsf{pmdec\_error} & $>$$q75$ (15; 26) & in [mas/yr] \\
$gofAL$     & \textsf{astrometric\_gof\_al} & $>$$q75$ (36; 36) & can be $<$0 \\
$epsi$      & \textsf{astrometric\_excess\_noise} & $>$$q75$ (33; 28) & $\geqq$0; in [mas] \\
$sepsi$     & \textsf{astrometric\_excess\_noise\_sig} & $>$$q75$ (35; 29) & $\geqq$0 \\
$amax$      & \textsf{astrometric\_sigma5d\_max} & $>$$q75$ (16; 23) & in [mas] \\
$IPDgofha$  & \textsf{ipd\_gof\_harmonic\_amplitude} & $>$$q75$ (16; 18) \\
$RUWE$      & \textsf{ruwe} & $>$$q75$ (37; 37) \\
\hline                                   
\end{tabular}
\end{table}

One can consider the ratio of parallax to parallax error $RPlx$
(= \textsf{parallax\_over\_error}) as the principal quality parameter 
of HVS candidates, because it has a direct influence on the significance
of the computed high tangential velocity. The important role of one of 
the other parameters included
in Table~\ref{Tab_quali}, $Nper$ (= \textsf{visibility\_periods\_used}),
was already discussed 
in Sects.~\ref{subS_DR2}, \ref{subS_DR3} and
\ref{subS_falsepm}. 
Its improvements from DR2 to DR3 were demonstrated 
in Figs.~\ref{Fig_dr2NperG} and \ref{Fig_dr3NperG}. The standard uncertainties
in the measured parallax and proper motion components of DR3,
$e\_Plx$ (= \textsf{parallax\_error}), $e\_pmRA$ (= \textsf{pmra\_error}), 
and $e\_pmDE$ (= \textsf{pmdec\_error}), are known to increase with fainter
magnitudes. As shown in 
\citetads[][their Tables 4 and 5]{2021A&A...649A...2L}, 
respectively for five- and six-parameter astrometric solutions,
the corresponding median errors in a given magnitude interval are very
similar to each other, with those of $e\_pmDE$ (in mas/yr) being always,
except for the faintest magnitude bin at $G$$=$21\,mag,
slightly smaller than of $e\_pmRA$ (in mas/yr) and $e\_Plx$ (in mas).
Six-parameter solutions, which are marked with a value of 95 for the
parameter $Solved$ (= \textsf{astrometric\_params\_solved}), generally
exhibit larger uncertainties.

According to 
\citetads{2021A&A...649A...2L}, 
the most relevant goodness-of-fit statistics describing how well the
observations correspond to a single-star model come on one hand from
the IPD. These are $IPDgofha$ (= \textsf{ipd\_gof\_harmonic\_amplitude}),
which ''could become large for sources that have
elongated images, such as partially resolved binaries'',
and the already mentioned (in Sect.~\ref{subS_cnn})
parameters $IPDfmp$ and $IPDfow$. In comparison to five-parameter
solutions, the latter are much larger for six-parameter solutions
\citepads[][their Tables 4 and 5]{2021A&A...649A...2L}, 
On the other hand,
the quality of the astrometric solution should be mainly described by the 
re-normalised 
unit weight error $RUWE$ (= \textsf{ruwe}) and
the excess source noise $epsi$ (= \textsf{astrometric\_excess\_noise})
and its significance $sepsi$ (= \textsf{astrometric\_excess\_noise\_sig}).
The excess source noise $epsi$ (in units of mas)
is usually considered to be significant, if $sepsi$$>$2.

Also given in units of mas is $amax$ (= \textsf{astrometric\_sigma5d\_max}),
which was already described in DR2 as the ''five-dimensional equivalent
to the semi-major axis of the position error ellipse [that is]
useful for filtering out cases where one of the five parameters, 
or some linear combination of several parameters, is particularly bad''
\citepads{2018A&A...616A...2L}. 
A parameter that was not given special attention by
\citetads{2021A&A...649A...2L}, 
is the goodness-of-fit statistic of model with respect to along-scan
observations $gofAL$ (= \textsf{astrometric\_gof\_al}). However, this
parameter was 
listed among the 15 relevant features for the construction of the GCNS
\citepads{2021A&A...649A...6G}. 
An upper limit of $gofAL$$<$3 was formerly used in the selection of HVS 
candidates in DR2 by
\citetads{2018ApJ...868...25B}, 
\citetads{2018RNAAS...2..211S}, 
and
\citetads{2019MNRAS.490..157M}. 

\subsection{Local comparison of astrometric parameters}
\label{subS_clocal}

The stellar density and probability of overlapping objects may change
over small sky areas, and a local comparison of astrometric parameters
may be useful.
To find out how typical in a given sky region the quality of the {\it Gaia} 
DR3 measurements was for each of the 72 nearest extreme HVS candidates,
all its wider NNs were extracted using VizieR with a maximum search
radius of 3600\,arcsec and a maximum number of 9999. Among these NNs only
those with roughly similar ($\pm$0.5\,mag) $G$ magnitudes to that of the 
target and with measured parallaxes and proper motions (i.e. five- or 
six-parameter solutions) were selected for the comparison of ten DR3 
quality parameters listed in the lower part of Table~\ref{Tab_quali}. 
The effective search radius was smaller (down to 216\,arcsec) in crowded 
regions, and the number of comparison objects within the 
effective search radius varied between about 50 for the brightest 
and 2800 for the faintest targets. 

All comparison objects, including the
target, were sorted by a given parameter. In case of the integer parameter 
$Nper$, where small values indicate problematic measurements, the median
$med$ corresponding to the 0.5 quantile was considered as an allowed 
lower limit for the target. The more negative the local difference $l\Delta$
($Nper$ value of the target minus the median $Nper$ of local comparison 
objects) is, the less reliable astrometric measurements can be expected.
These differences $l\Delta$ are listed in Table~\ref{Tab_72nnflg},
where $l\Delta$$<$0 can be considered critical.
The reason for locally small $Nper$ values may be a combination of image
crowding, telemetry limits and object-matching problems.
In case of the nine other parameters, where large values report on problems,
the 0.75 quantile $q75$, i.e. the limit below which
75\% of all comparison objects were falling, was considered as critical
for the target. Since some of the parameters ($epsi$, $sepsi$)
can have zero values, the corresponding $q75$ could be zero, too.
However, this was not the case for the comparison samples of 72 targets.
Therefore, for nine parameters the 
local ratio $lR$ (given in Table~\ref{Tab_72nnflg}) of the
value of the target to the 0.75 quantile of local comparison objects
describes quantitatively, how unusual the target's value is in its
sky region. All $lR$$>$1 indicate critical parameter values.

\subsection{Global comparison of astrometric parameters}
\label{subS_cglobal}

For a global comparison of the same ten DR3 quality parameters 
(lower part of Table~\ref{Tab_quali}) as used in the local 
comparison (Sect.~\ref{subS_clocal}), from all about 2 million
DR3 HPM stars with $PM$$>$60\,mas/yr only those with 
exactly the same magnitude ($\pm$0.005\,mag) as that of the target
were selected. In this case, the numbers of comparison objects varied 
between about 600 for the brightest and 3300 for the faintest targets.
As HPM objects represent a special class of objects with highly
significant proper motions, their properties may be different in
comparison to the hundreds of millions of slow-moving objects dominating 
the typical parameter quality as a function of magnitude described by
\citetads[][their Tables 4 and 5]{2021A&A...649A...2L}. 
Again, all our comparison objects including the target were sorted by
a given parameter. Negative global differences $g\Delta$ ($Nper$ 
value of the target minus median $Nper$ of all global comparison 
objects) given in Table~\ref{Tab_72nnflg} mainly indicated a still 
insufficient coverage of that sky region according to the scanning 
law of {\it Gaia}. For the other nine parameters, the 
global ratio $gR$ (Table~\ref{Tab_72nnflg}) of the value of the 
target to the 0.75 quantile of global comparison objects was computed.
Again, $g\Delta$$<$0 and $gR$$>$1 are critical values.

   \begin{figure}
   \resizebox{\hsize}{!}{\includegraphics[angle=270]{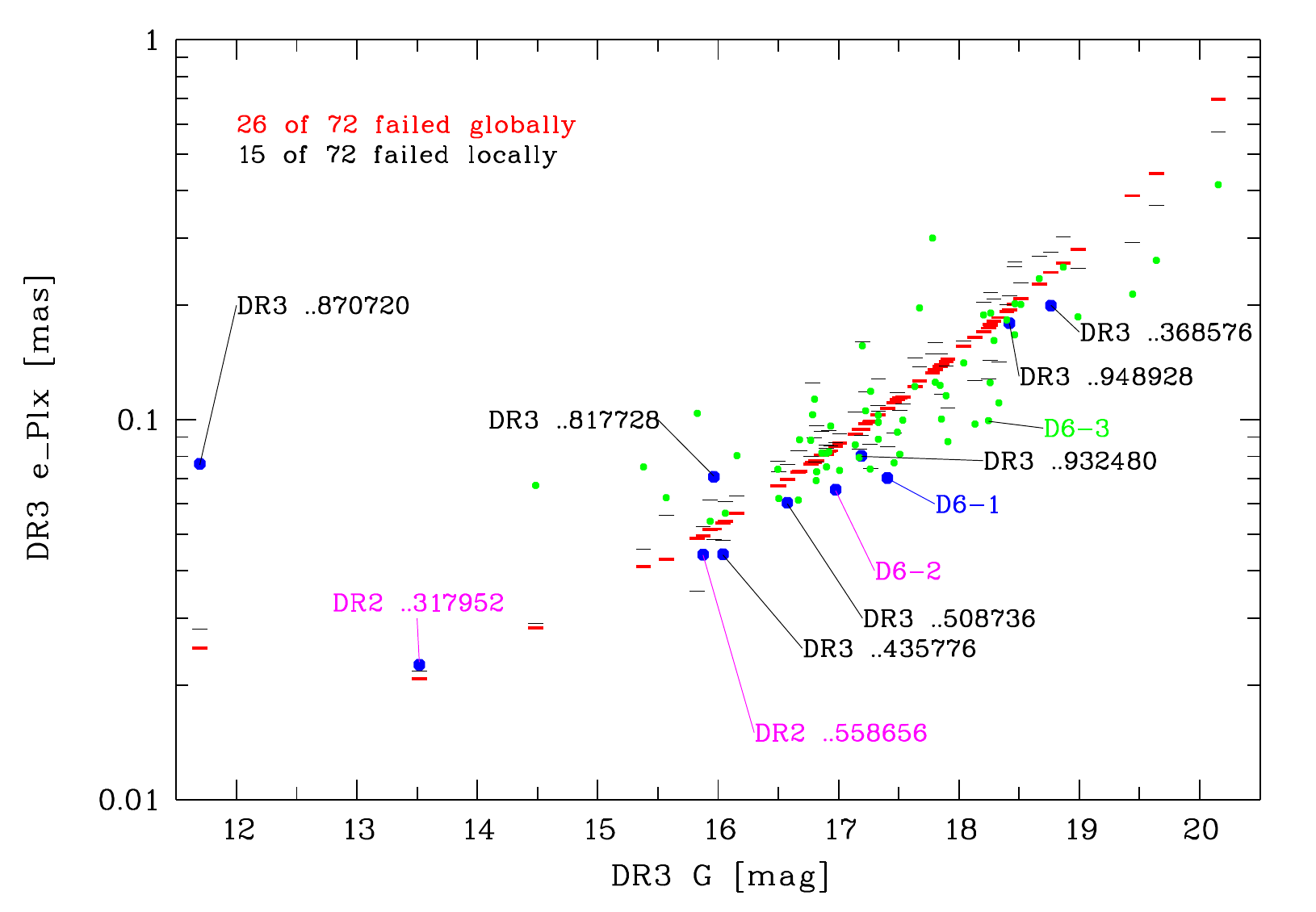}}
           \caption{DR3 $e\_Plx$ of
	   72 nearest extreme HVS candidates (high priority = blue 
	   filled hexagons; low priority = green dots) as a function
	   of $G$ magnitude. Thin black and thick red horizontal bars show
	   $q75$ values of local and global comparison
	   objects, respectively.
              }
      \label{Fig_ePlxGnnGG}
   \end{figure}

\subsection{No good candidates fulfilling all criteria}
\label{subS_doubts}

Taking into account $RPlx$ as the most important quality
parameter of HVS candidates joined by the exposure of
close NN (marked in Fig.~\ref{Fig_cmd72nnGG} with crosses), available
IPD and other flags (marked with arrows), and the analysis of ten
astrometric parameters in the local and global comparison,
no convincing candidates are left. In particular, none of the 11 high-priority
and only 7 out of 61 low-priority candidates (all 7 are faint, with
$G$$>$17.5\,mag, and 6 of 7 have $RPlx$$<$4) passed both the local and global
check of all ten parameters (Fig.~\ref{Fig_cmd72nnGG}). The three faintest
of these 7 low-priority candidates (with $G_{abs}$$>$7.5\,mag 
and $G$$\gtrapprox$19\,mag) have flags (overplotted different arrows
in Fig.~\ref{Fig_cmd72nnGG}) indicating crowding and IPD problems and
very small $RPlx$$\approx$3.

Original DR3 parameters (rounded) of all 72 HVS candidates are
listed in Table~\ref{Tab_72dr3}.
In the lower part of Table~\ref{Tab_quali} the relatively large numbers of 
objects among the 72 candidates with astrometric parameters exceeding the 
allowed $q75$ upper limits (or falling below the median of $Nper$) in the 
local and global comparison are summarised. Only for $epsi$ and $sepsi$, 
the local comparison revealed higher numbers of such outliers than the global 
one. In case of $gofAL$ and $RUWE$, the numbers of problematic candidates 
are equally high, 
whereas for the other six parameters, the global comparison turned out
to be more problematic, with up to almost twice as many failed candidates 
than in the local comparison in case of $Nper$. 
The individual results, including
the differences $l\Delta$ and $g\Delta$ for $Nper$ 
and ratios $lR$ and $gR$ for nine other astrometric parameters, 
are listed in Table~\ref{Tab_72nnflg}.

\subsubsection{$e\_Plx$, $e\_pmRA$, and $e\_pmDE$}

The DR3 parallax errors $e\_Plx$ of the 72 candidates are shown 
in Fig.~\ref{Fig_ePlxGnnGG} together with the corresponding $q75$
values found in the local and global comparison. The proper motion errors
$e\_pmRA$ and $e\_pmDE$ and their local and global comparison look
very similar and are not presented here. All high-priority candidates,
as well as the low-priority candidate \object{D6-3}, are
labelled in this and the following figures. The brightest six
candidates ($G$$<$15.85\,mag), including the two high priority 
candidates \object{Gaia DR3 1180569514761870720} 
and \object{Gaia DR2 6097052289696317952}, the latter of which appeared 
in DR2 as a much more extreme HVS
\citepads{2018RNAAS...2..211S}, 
exhibit
unusually large errors in both local and global comparisons. As outlined
at the beginning of Sect.~\ref{Sect_nearextrHVS}, the HVS status of these 
two candidates was also not supported by their DR3 RVs.
At the faint
end ($G$$>$18.9\,mag), there are four low-priority objects with errors 
well below the allowed local and global comparison limits.
At intermediate magnitudes, the errors of the targets 
(in particular of low-priority ones) are scattered
around these limits. Except for two  bright objects
already mentioned above, only one more high-priority candidate
(\object{Gaia DR3 1820931585123817728}) shows an unusually large $e\_Plx$.
These three high-priority HVS candidates appear
clearly questionable with respect to other parameters, too 
(see next figures and Table~\ref{Tab_72nnflg}).

   \begin{figure}
   \resizebox{\hsize}{!}{\includegraphics[angle=270]{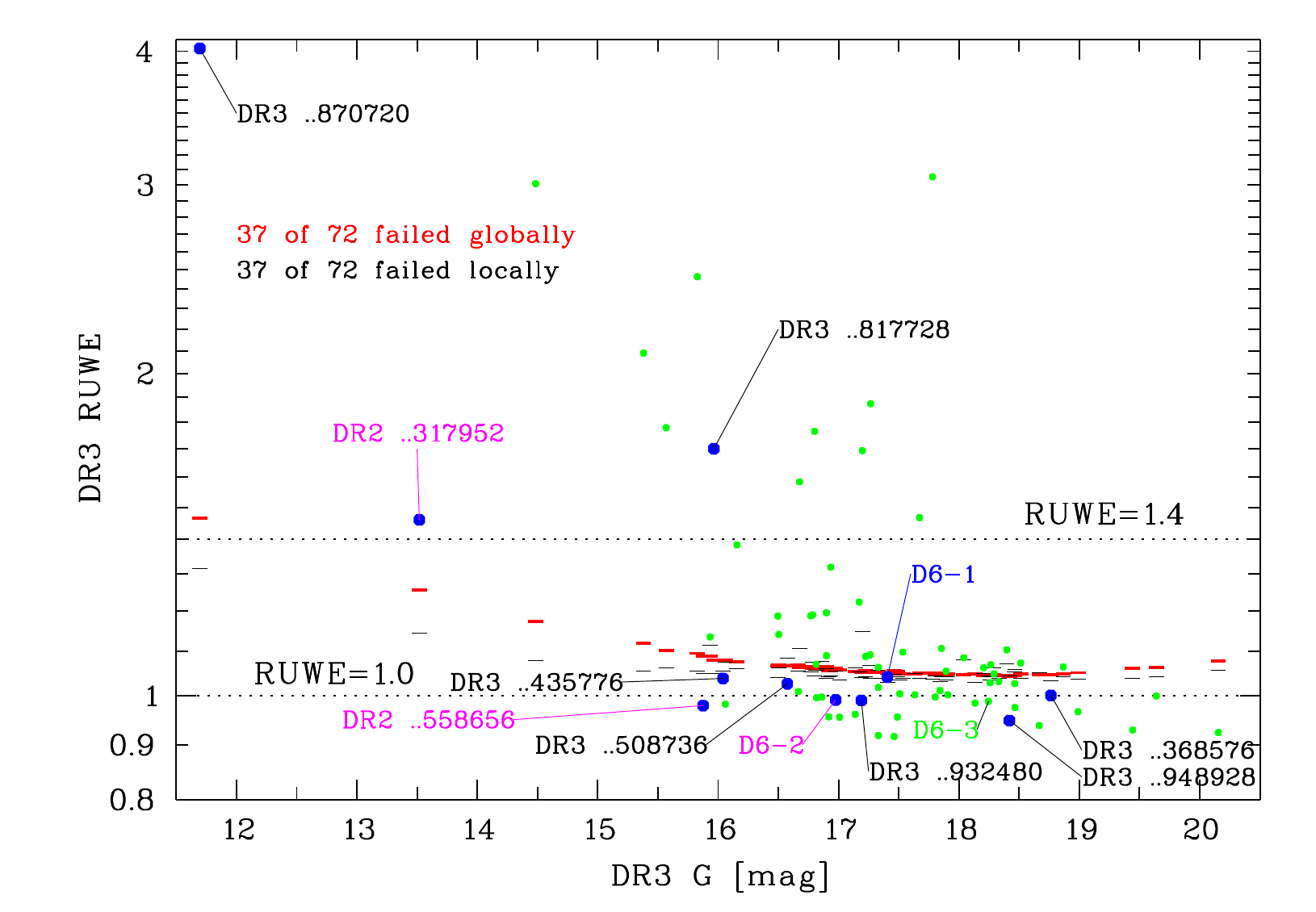}}
           \caption{Same as Fig.~\ref{Fig_ePlxGnnGG} for $RUWE$.
              }
      \label{Fig_RUWEGnnGG}
   \end{figure}

   \begin{figure}
   \resizebox{\hsize}{!}{\includegraphics[angle=270]{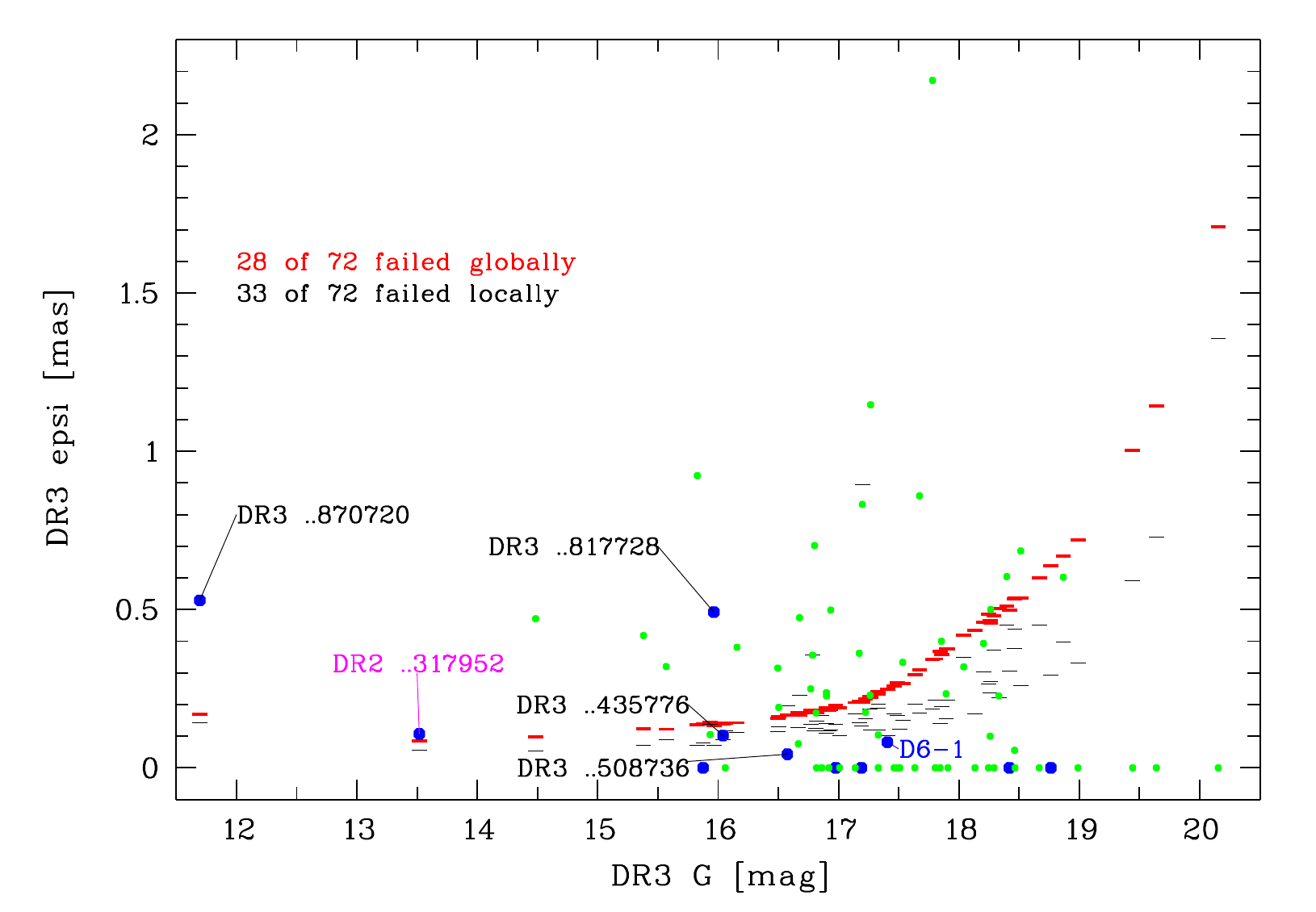}}
           \caption{Same as Fig.~\ref{Fig_ePlxGnnGG} for $epsi$.
	   Objects with $epsi$$=$0\,mas are not labelled.
              }
      \label{Fig_epsiGnnGG}
   \end{figure}

   \begin{figure}
   \resizebox{\hsize}{!}{\includegraphics[angle=270]{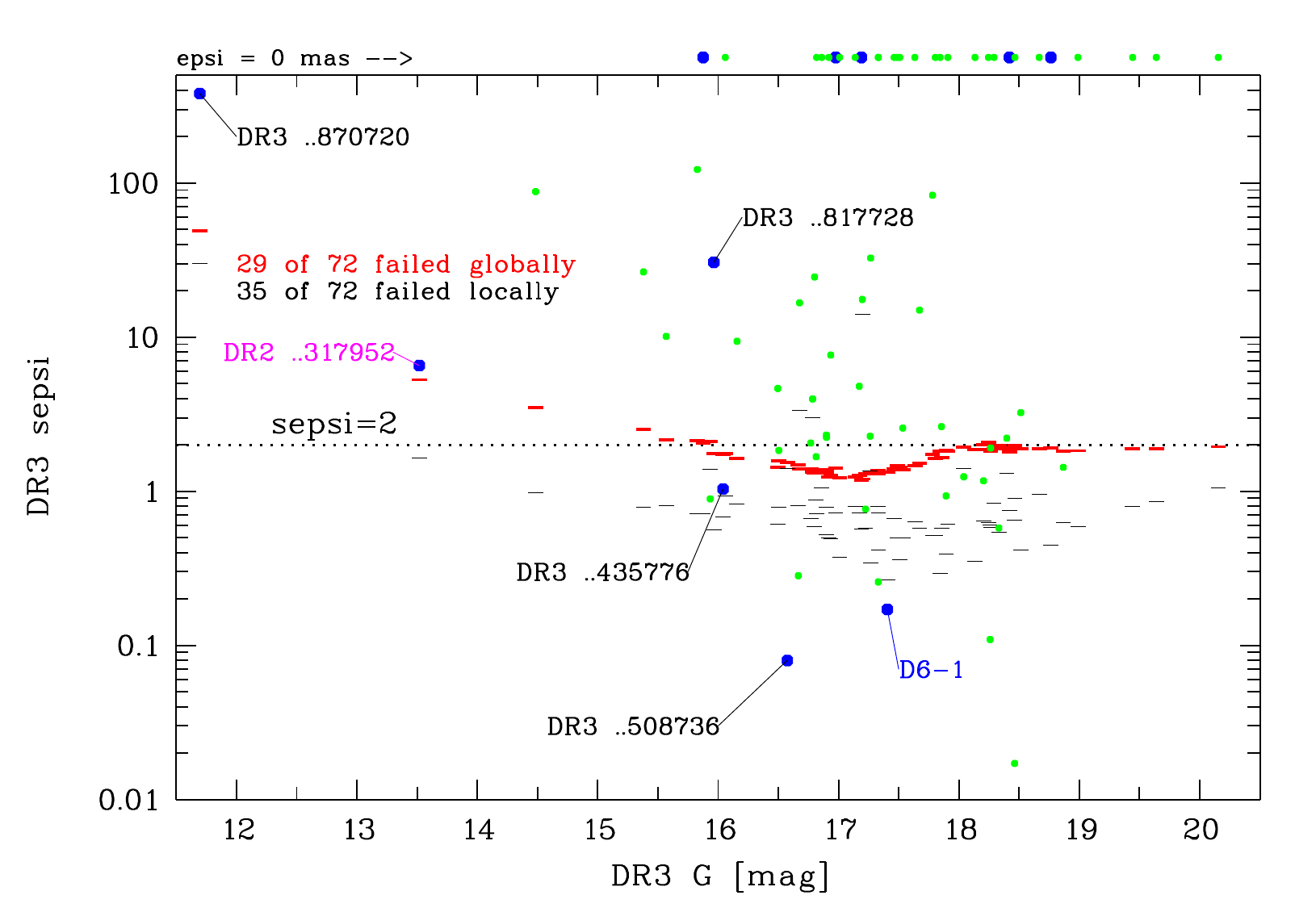}}
           \caption{Same as Fig.~\ref{Fig_ePlxGnnGG} for $sepsi$.
	   Objects with $epsi$$=$0\,mas are not labelled.
              }
      \label{Fig_sepsiGnnGG}
   \end{figure}

   \begin{figure}
   \resizebox{\hsize}{!}{\includegraphics[angle=270]{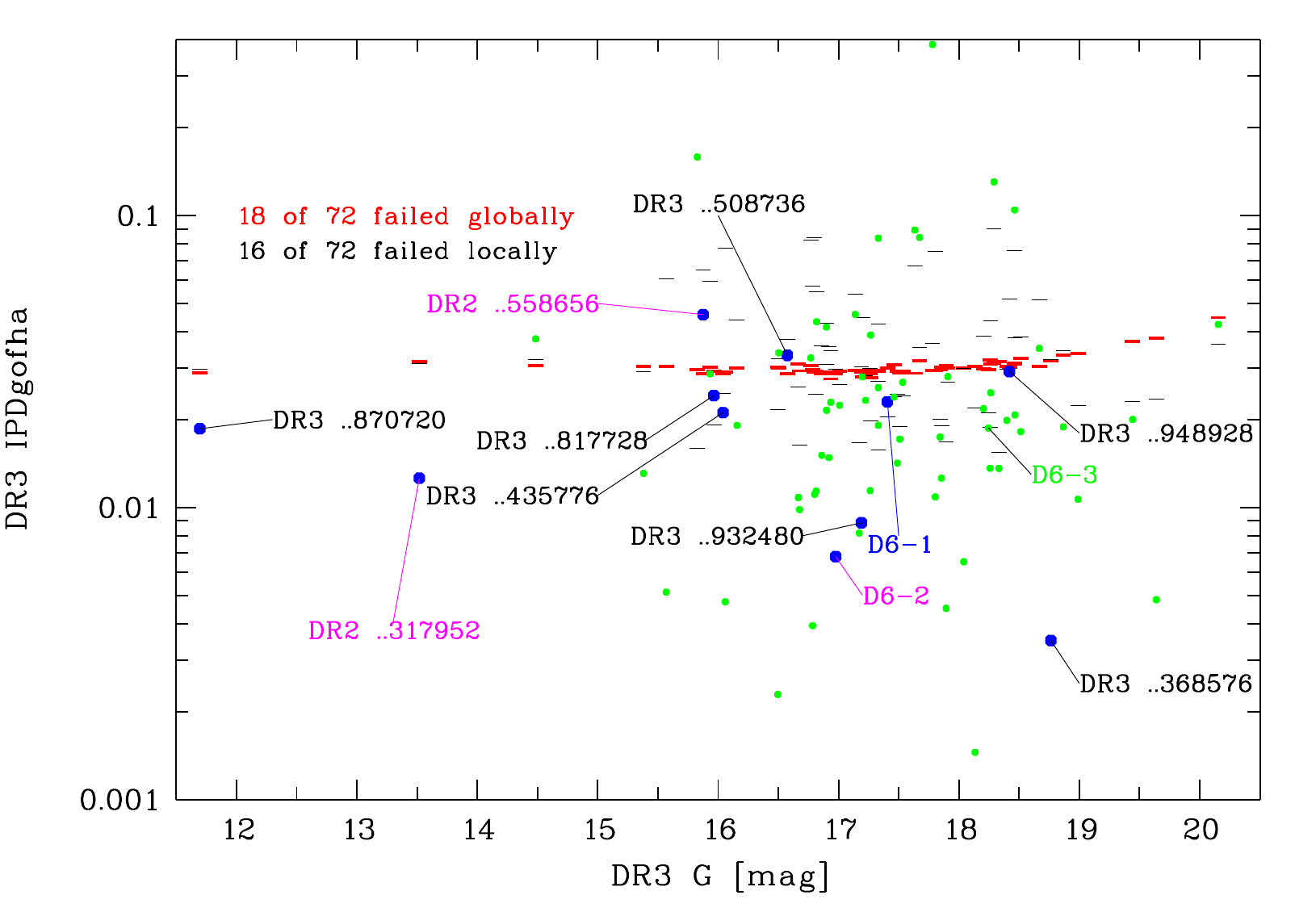}}
           \caption{Same as Fig.~\ref{Fig_ePlxGnnGG} for $IPDgofha$.
              }
      \label{Fig_IPDgofhaGnnGG}
   \end{figure}

\subsubsection{$RUWE$}

For one of the most important astrometric quality parameters, $RUWE$
\citepads{2021A&A...649A...2L}, 
Fig.~\ref{Fig_RUWEGnnGG} demonstrates how close the 0.75 quantiles of our
comparison objects come to the ideal value of 1.0 with fainter magnitudes.
This can be observed in the local comparison with all objects as well as
in the global comparison with HPM stars at $G$$\gtrapprox$16\,mag.
Already with $G$$>$13\,mag the DR3 0.75 quantiles of $RUWE$ are 
much smaller than 1.4, the critical value recommended for DR2 
by Lennart Lindegren in the {\it Gaia} technical note 
GAIA-C3-TN-LU-LL-124\footnote{http://www.rssd.esa.int/doc\_fetch.php?id=3757412}
in 2018. 

The median $RUWE$ values of all five-parameter solutions, including
predominantly slow-moving and 
non-significant 
proper motion objects, in DR3
\citepads[][their Table 4]{2021A&A...649A...2L}, 
are in the narrow range of 1.01-1.02 for $G$$=$14-21\,mag and rise only 
to $\approx$1.04 for brighter objects ($G$$=$9-12\,mag). In our global
comparison, the $q75$ values of $RUWE$ reach a minimum of 1.045 
at $G$$\approx$18\,mag, rise up to about 1.08 both at the faint end
at $G$$\approx$20\,mag and towards brighter objects at $G$$\approx$16\,mag, 
and rapidly increase to about 1.4 at $G$$\approx$12\,mag. These values
are better comparable (but not equal) to the DR3 median values 
of all six-parameter solutions
\citepads[][their Table 5]{2021A&A...649A...2L}. 
Except for three bright stars already mentioned in the preceding 
paragraph, which have very large $RUWE$$>$1.4, almost all other high-priority 
candidates pass the $RUWE$ quality check both locally and globally. 
Again, the low-priority candidates show a large spread around the allowed
limits.

\subsubsection{$epsi$ and $sepsi$}

The results of the local and global comparison of the excess source 
noise $epsi$ and its significance $sepsi$, described by
\citetads{2021A&A...649A...2L} 
as the next most relevant parameters from the astrometric 
fit, are shown in Figs.~\ref{Fig_epsiGnnGG} and \ref{Fig_sepsiGnnGG}. For 29 
of 72 objects DR3 provides zero values of these two parameters. Whereas 
the 0.75 quantiles of $epsi$ are below 0.2\,mas for the brighter
($G$$<$17\,mag) global comparison HPM objects, they rise up to about 1.5\,mas
at the faint end ($G$$\approx$20\,mag). The local comparison, which included
objects of small (non-significant) proper motions provided systematically
smaller 0.75 quantiles of $epsi$ around most of the targets. Similarly,
the local 0.75 quantiles of $sepsi$ are smaller than the global ones. The
global 0.75 quantiles of $sepsi$ exceed the level of 2.0, where the
excess source noise starts to be statistically significant according to
\citetads{2021A&A...649A...2L}, 
at brighter magnitudes only ($G$$<$15.5\,mag). A dip 
slightly below that level can be seen at $G$$\approx$17\,mag in
Fig.~\ref{Fig_sepsiGnnGG}.

A cut at $sepsi$$=$2.0, applied e.g. by
\citetads{2018ApJ...868...25B}, 
provided similar
results to the used 0.75 quantiles of $sepsi$ in the global comparison.
Among the 43 candidates with non-zero excess source noise, there are
six high-priority ones, including three with large significant values.
These three were already mentioned as problematic with respect to
other parameters. Among 37 low-priority candidates with $epsi$$>$0\,mas 
the majority do also show $sepsi$$>$2.0, with slightly growing 
numbers exceeding the 0.75 quantiles of $sepsi$ in the global and 
local comparison, respectively. If an $epsi$$>$0\,mas was measured, 
it was typically much larger than the parallax error $e\_Plx$
(Fig.~\ref{Fig_ePlxGnnGG}). About two times larger values than
$e\_Plx$ were found for $epsi$, even when it was formally 
non-significant ($sepsi$$<$2.0).

\subsubsection{$IPDgofha$}

Figure~\ref{Fig_IPDgofhaGnnGG} shows that the parameter $IPDgofha$,
generated in the IPD before the astrometric solution, is almost
constant over a wide magnitude range with its $q75$ values at a 
level of 0.03 in the global comparison. Only at very faint magnitudes 
($G$$>$19\,mag), the $q75$ values are slightly increased. Smaller
and similar (and also almost constant) median values, of 0.02 for 
five-parameter and 0.03 for six-parameter solutions, were reported 
for the complete DR3 in
\citetads[][their Tables 4 and 5]{2021A&A...649A...2L}, 
respectively. The corresponding $q75$
from the local comparison appear widespread around the global level,
except for brighter objects ($G$$<$15.5\,mag), where they are almost
equal. Interestingly, the formerly mentioned three bright high-priority 
candidates with problematic astrometric parameters are inconspicuous
at least in the global comparison, whereas some other high-priority 
candidates seem to have problems with $IPDgofha$. However, with this
parameter, the number of failed candidates is relatively small,
in particular in the global comparison (Table~\ref{Tab_quali}).

\subsubsection{$gofAL$ and $amax$}

The distribution of $gofAL$ with $G$ magnitude (not shown) appears very 
similar to that of $RUWE$. Almost all $q75$ values fall clearly below the
critical level of $gofAL$$=$3, used e.g. by
\citetads{2018ApJ...868...25B} 
and
\citetads{2019MNRAS.490..157M}. 
The location of individual $gofAL$ above and 
below the comparison $q75$ values is nearly identical to that of 
the corresponding $RUWE$ (Fig.~\ref{Fig_RUWEGnnGG}). 

Also not shown
are the $amax$ data. In this case, the general trend with magnitude
looks similar to that of $e\_Plx$ (Fig.~\ref{Fig_ePlxGnnGG}). Individual
$amax$, in particular of some high-priority candidates, appear to be more
critical with respect to their $q75$ values than in case of $e\_Plx$.
The amplitude of the global comparison $q75$ of $amax$ (given in mas,
as for $e\_Plx$) is over the whole magnitude range about 1.65 times larger 
than that of the corresponding $e\_Plx$.

\subsubsection{$Nper$}

\begin{figure}
   \resizebox{\hsize}{!}{\includegraphics[angle=270]{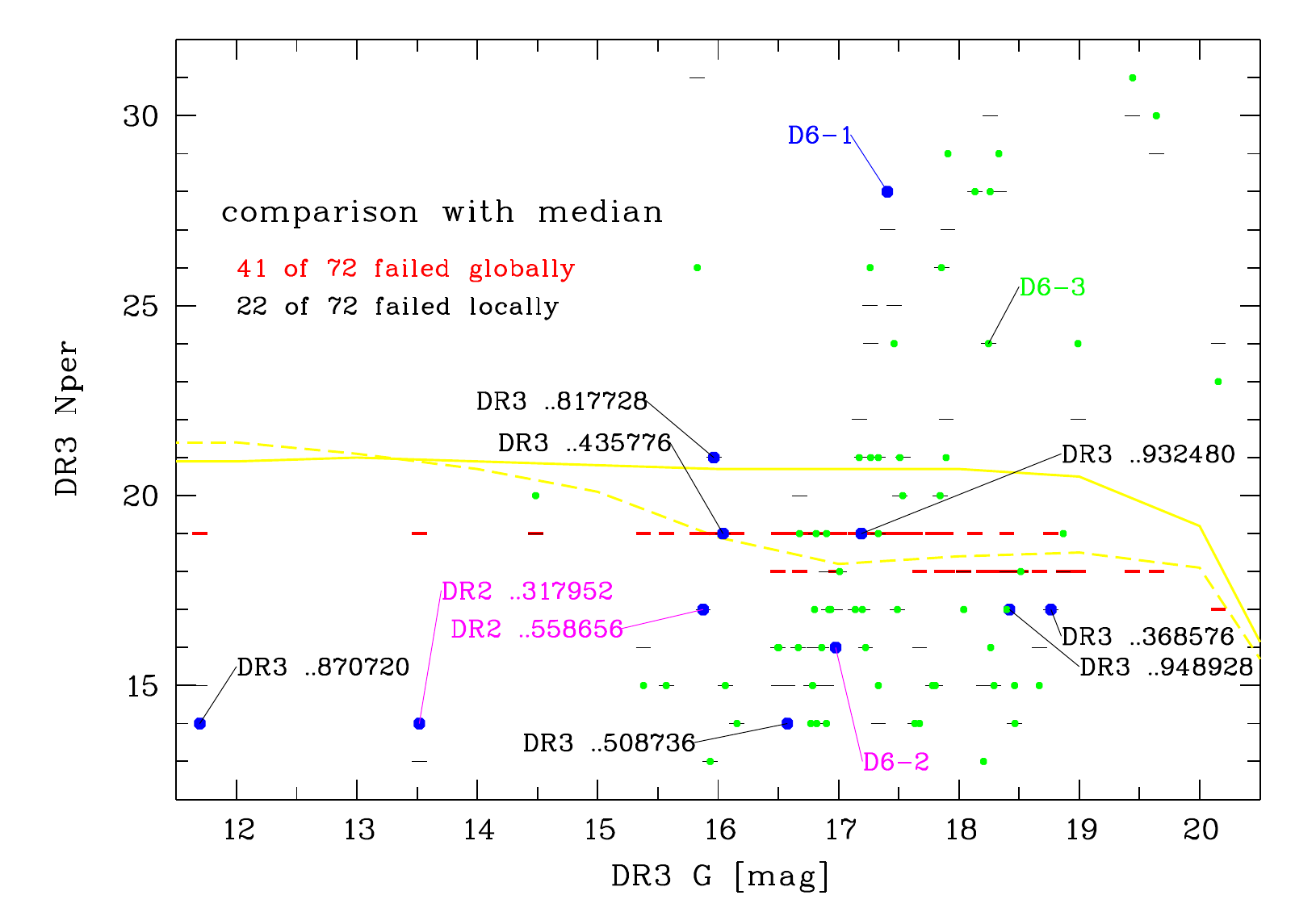}}
           \caption{DR3 $Nper$ of
           72 nearest extreme HVS candidates (high priority = blue
           filled hexagons; low priority = green dots) as a function
           of $G$ magnitude. Thin black and thick red horizontal bars show
           median values of our local and global comparison
           objects, respectively. Yellow solid and dashed lines show
	   mean $Nper$ of all DR3 five- and six-parameter solutions
\citepads[][their Tables 4 and 5]{2021A&A...649A...2L} 
	   for comparison.
              }
      \label{Fig_NperGnnGG}
   \end{figure}

Finally, in Fig.~\ref{Fig_NperGnnGG} the $Nper$ of 72 candidates are 
compared to median values of similarly bright objects in their vicinity 
and of equally bright HPM stars globally. The global median $Nper$ of
HPM stars are mostly equal to 19 (42 of 72 = 68\%) or 18 (22 of 72 = 31\%).
Only for the faintest target ($G$$=$20.15\,mag), it is 17. 
These integer values are systematically smaller than
the mean (non-integer) $Nper$ of all five-parameter solutions in DR3 and
partly comparable to those of all six-parameter solutions
\citepads[][their Tables 4 and 5]{2021A&A...649A...2L}. 
On the other hand, the median $Nper$
of local comparison stars vary a lot, from 13 to 31. 

The target $Nper$
also occupy the whole parameter space between these two extremes. The
majority of all candidates, including those of high priority, have $Nper$
equal to or above the median found in the local comparison but below the 
global comparison median values. With this parameter, the number of failed 
candidates (41 of 72 = 57\%) in the global comparison is higher than with
any of the other nine parameters considered alone either locally or
globally (Table~\ref{Tab_quali}). Compared to the mean $Nper$ of all 
five-parameter solutions even more (52 of 72 = 72\%) objects fall below
this limit. This means that most of the 72 HVS candidates are located in
sky regions, which have had relatively few {\it Gaia} observations so far 
(in DR3).
As outlined in Sect.~\ref{subS_falsepm}, there may be even spurious HPMs 
in crowded regions, which could lead to false HVS candidates. Such spurious 
HPMs could still be detected in DR3 with $Nper$$<$16. Note that out of the 72 
candidates 22 (31\%) have such extremely low $Nper$ values.

\subsection{Best candidates with relaxed criteria}
\label{subS_best}

There are candidates that failed only few of the tests summarised 
in Table~\ref{Tab_quali}, where they at the same time almost reached the 
allowed limits. For a new selection with relaxed criteria, one can, on
one hand, decrease the limit for $RPlx$, to include the best 
of the low-priority candidates. On the other hand, one can allow for 
smaller $Nper$ and larger values of the other nine astrometric parameters 
in the local and global comparison, to include the majority of the most
extreme ($vtan\_g$$>$1000\,km/s) high-priority
candidates. The following set of relaxed criteria
\begin{itemize}
	\item $RPlx > 4$ 
	\item $l\Delta \geqq -2$ and $g\Delta \geqq -2$ (for $Nper$)
	\item $lR < 1.6$ and $gR < 1.6$ (for nine other parameters)
\end{itemize}
is fulfilled by 11 out of 72 candidates (see Table~\ref{Tab_72nnflg}). 
Conveniently, these eleven (7 high-priority and 4 low-priority) candidates
are also 'clean' with respect to close NNs, IPD, 
and other flags (described in columns 3-9 of Table~\ref{Tab_72nnflg}).
Based on the data in Tables~\ref{Tab_72dr3} and \ref{Tab_72nnflg},
the reader may select an alternative set of best HVS candidates,
e.g. by using different weights for $RPlx$ or other parameters.
Our shortlist of eleven HVS candidates briefly described below is sorted 
by decreasing Galactocentric tangential velocity $vtan\_g$ 
(given in Table~\ref{Tab_72dr3}). Previously measured DR2 parallaxes
and relatively low Galactic latitudes $|GLAT|$$<$25$\degr$
(not included in Table~\ref{Tab_72dr3}) of four of the objects will
also be mentioned.
The seven high-priority and the first listed low-priority object are 
among the labelled objects in the zoomed CMD (Fig.~\ref{Fig_cmd72nnGG}) 
and in Figs.~\ref{Fig_ePlxGnnGG}--\ref{Fig_NperGnnGG}, describing 
the local and global comparison. The first four objects were already
mentioned before and shown in previous figures. Only the first five have
an entry in the SIMBAD data base. The running number
in Tables~\ref{Tab_72dr3} and \ref{Tab_72nnflg} is given first
for each of the eleven objects:

\paragraph{No.56 = \object{D6-3} = \object{Gaia DR3 2156908318076164224}}

Alias \object{LSPM J1852+6202}, this is one of four 
low-priority (3$<$$RPlx$$<$5) candidates that
passed all local and global quality checks (overplotted open 
lozenges and hexagons in Fig.~\ref{Fig_cmd72nnGG}), were not found to have
close NNs (no crosses) and not flagged 
concerning IPD parameters, duplicity or six-parameter solutions (no arrows). 
It has a relatively high $RPlx$$\approx$4.3 in DR3 (3.4
in DR2) and is therefore included in 
this shortlist. It was observed quite often ($Nper$$=$24). The uncertain
heliocentric and Galactocentric tangential velocities are equally high
($>$2300\,km/s, with an error of $>$500\,km/s).
With $G$$\approx$18.2\,mag it belongs to the faint candidates, for
which the binary detection in the IPD (using $IPDfmp$) becomes difficult
\citepads{2021A&A...649A...2L}. 
The object \object{D6-3} is one of two D$^6$ objects 
(the other was \object{D6-1}), for which a zero RV was measured by
\citetads{2018ApJ...865...15S}. 
They mentioned the possibility that the DR2 parallaxes used by them could be
systematically underestimated and the tangential velocities much smaller.
Then these two D$^6$ objects could change their location in the CMD 
(cf. Figs.~\ref{Fig_dr3cmd} and \ref{Fig_cmd72nnGG}) and move to the 
region of faint nearby WDs. However, the DR3 parallax of \object{D6-3} 
(0.423$\pm$0.099\,mas) did not change much in comparison to DR2 
(0.427$\pm$0.126\,mas), but just became more significant.
With $GLAT$$\approx$$+$24$\degr$ this object is located relatively close
to the Galactic plane.
According to its \textsf{phot\_variable\_flag} in DR3 \object{D6-3} seems 
to be variable and is even listed in DR3 as an eclipsing binary candidate.
However, 
\citetads{2023A&A...674A..25H} 
included it in their list of spurious signals.

\paragraph{No.40 = \object{D6-1} = \object{Gaia DR3 5805243926609660032}}

This $\approx$1\,mag brighter HPM star
\citepads[already listed in][]{2003AJ....125..984M} 
has even more observations ($Nper$$=$28) 
than \object{D6-3} and a more precise high $vtan\_g$ of over 1700\,km/s. 
The $RPlx$$\approx$7.6 of \object{D6-1} represents the second largest 
value among all eleven candidates described in this subsection. 
However, in the local comparison (with other objects in
this frequently observed sky field) this candidate shows 1\%
larger $RUWE$, $>$10\% larger $IPDgofha$, and 35\% larger $gofAL$ 
than the corresponding $q75$ of comparison objects. 
Both $RUWE$ and $IPDgofha$ may be indicative of binaries
\citepads{2021A&A...649A...2L}. 
It appears puzzling that, together with \object{D6-3}, 
\object{D6-1} has a measured RV consistent with a zero value
\citepads{2018ApJ...865...15S}. 
The DR3 parallax of \object{D6-1} (0.531$\pm$0.070\,mas)
is 13\% larger than measured in DR2 (0.471$\pm$0.102\,mas), 
in agreement with the trend found for 
DR2-selected high-priority HVS 
candidates (Sect.~\ref{subS_trends}). With 
$GLAT$$\approx$$-$18$\degr$, \object{D6-1} lies even closer
to the Galactic plane than \object{D6-3}.

\paragraph{No.28 = \object{D6-2} = \object{Gaia DR3 1798008584396457088}}

This is the highest-priority candidate ($RPlx$$\approx$18.2) in the shortlist
and also in the full list of 72 candidates, with the most precise extreme
$vtan\_g$ of $\approx$1100\,km/s. But it is not stainless.
With a relatively small $Nper$$=$16, it lies below the median of comparison
objects both locally and globally. So, contrary to the previous two
candidates in the shortlist, \object{D6-2} (= \object{NLTT 51732})
still suffers from too few
observations in the {\it Gaia} survey. With a parallactic distance
of $\approx$840\,pc it is the nearest candidate in the shortlist. It is 
the brightest of the three D$^6$ HVS candidates discovered by
\citetads{2018ApJ...865...15S} 
in DR2, and only for this one
they measured a large RV confirming its HVS status. Selected DR2 astrometric
quality parameters of \object{D6-2} were considered by
\citetads{2018RNAAS...2..211S}, 
as coming close to critical limits ($sepsi$$=$1.8, $gofAL$$=$2.7 $Nper$$=$9)
but, except for $Nper$, this is no longer valid in DR3. Nevertheless,
as in the case of \object{D6-1}, the DR3 parallax of \object{D6-2} 
(1.194$\pm$0.065\,mas) turned out to be 13\% larger than previously measured 
in DR2 (1.052$\pm$0.109\,mas), where it already appeared as a clear
high-priority candidate (Sect.~\ref{subS_DR2}).
The Galactic latitude ($GLAT$$\approx$$-$20$\degr$) is similar to that
of \object{D6-1}. The infrared excess found for \object{D6-2}, but not
for \object{D6-1} and \object{D6-3}, was attributed to circumstellar
material by
\citetads{2022MNRAS.512.6122C}, 
who ruled out a red companion.

%
\begin{table*}
\caption{Basic DR3 data on 72 nearest extreme HVS candidates with high ($RPlx$$>$5) and low (3$<$$RPlx$$<$5) priority, sorted by $G$ magnitude}
\label{Tab_72dr3}
\centering
\fontsize{6.4pt}{0.80\baselineskip}\selectfont
\begin{tabular}{@{}l@{\hspace{2mm}}l@{\hspace{2mm}}c@{\hspace{2mm}}c@{\hspace{1mm}}r@{\hspace{1mm}}r@{\hspace{2mm}}r@{\hspace{2mm}}r@{\hspace{2mm}}r@{\hspace{2mm}}r@{\hspace{2mm}}r@{\hspace{2mm}}c@{\hspace{2mm}}c@{\hspace{2mm}}c@{\hspace{2mm}}c@{\hspace{2mm}}c@{\hspace{2mm}}c@{\hspace{2mm}}c@{\hspace{2mm}}c@{}}     
\hline\hline
	No.& Gaia DR3		& $G$   & $Nper$ & $Plx$ & $RPlx$ & $pmRA$  	& $pmDE$  	& $gofAL$ & $epsi$ & $sepsi$ & $amax$ & $IPDgofha$ & $RUWE$ & $G_{abs}$ &$G$$-$$RP$ & RV   & $vtan$ & $vtan\_g$  \\
	   &         		& [mag] &        &[mas]  &        & [mas/yr]	& [mas/yr]	&         & [mas]  &         & [mas]  &            &        & [mag]  & [mag]     &[km/s]& [km/s] & [km/s] \\ 
\hline
       01 & 1180569514761870720 & 11.695 & 14 & 0.548 &  7.16 & -100.47$\pm$0.06 &  -22.23$\pm$0.07 & 47.73 & 0.529 & 380.11 & 0.105 & 0.0186 & 4.02 &  0.39 & 0.56 &  -76.9$\pm$1.4 &  891$\pm$124 &  703\\
       02 & 6097052289696317952 & 13.519 & 14 & 0.341 & 15.06 &  -61.14$\pm$0.02 &  -24.65$\pm$0.03 &  8.15 & 0.107 &   6.54 & 0.039 & 0.0126 & 1.46 &  1.18 & 0.38 &  +53.5$\pm$4.5 &  917$\pm$ 61 &  729\\
       03 & 716216621590241024  & 14.485 & 20 & 0.263 &  3.92 &  -10.80$\pm$0.06 &  -56.20$\pm$0.05 & 41.79 & 0.471 &  87.93 & 0.098 & 0.0378 & 3.01 &  1.58 & 0.44 & +311.5$\pm$9.2 & 1032$\pm$263 &  793\\
       04 & 6181489766183501952 & 15.380 & 15 & 0.267 &  3.55 &  -36.95$\pm$0.07 &  -37.83$\pm$0.08 & 20.00 & 0.418 &  26.48 & 0.135 & 0.0131 & 2.09 &  2.51 & 0.46 & ...            &  940$\pm$265 &  750\\
       05 & 4081142250396115584 & 15.569 & 15 & 0.255 &  4.09 &  -33.69$\pm$0.07 &  -37.62$\pm$0.05 & 13.74 & 0.320 &  10.10 & 0.093 & 0.0051 & 1.78 &  2.60 & 0.70 & ...            &  940$\pm$230 &  730\\
       06 & 1492081744690617344 & 15.827 & 26 & 0.359 &  3.46 &  -59.81$\pm$0.07 &  +18.78$\pm$0.11 & 34.97 & 0.923 & 122.35 & 0.154 & 0.1587 & 2.46 &  3.60 & 0.95 & ...            &  828$\pm$239 &  741\\
       07 & 3841458366321558656 & 15.875 & 17 & 0.325 &  7.38 &   +7.07$\pm$0.05 &  -81.29$\pm$0.04 & -0.51 & 0.000 &   0.00 & 0.078 & 0.0458 & 0.98 &  3.44 & 0.51 & ...            & 1190$\pm$161 & 1007\\
       08 & 2710296407281612544 & 15.934 & 13 & 0.253 &  4.68 &  +17.03$\pm$0.05 &  +27.88$\pm$0.05 &  2.56 & 0.105 &   0.89 & 0.079 & 0.0287 & 1.14 &  2.95 & 0.53 & ...            &  613$\pm$131 &  730\\
       09 & 1820931585123817728 & 15.965 & 21 & 0.824 & 11.65 &  -82.45$\pm$0.06 & -149.46$\pm$0.06 & 19.83 & 0.492 &  30.54 & 0.096 & 0.0242 & 1.70 &  5.54 & 0.73 & ...            &  982$\pm$ 84 &  847\\
       10 & 4294774301679435776 & 16.041 & 19 & 0.308 &  6.98 &  -31.34$\pm$0.04 &  -47.54$\pm$0.04 &  0.98 & 0.102 &   1.03 & 0.065 & 0.0211 & 1.04 &  3.49 & 0.56 & ...            &  875$\pm$125 &  709\\
       11 & 2606168116350328448 & 16.060 & 15 & 0.252 &  4.44 &  +38.91$\pm$0.06 &  -29.48$\pm$0.05 & -0.28 & 0.000 &   0.00 & 0.087 & 0.0048 & 0.98 &  3.07 & 0.41 & ...            &  918$\pm$207 &  728\\
       12 & 6293563806740171776 & 16.157 & 14 & 0.262 &  3.25 &  +19.19$\pm$0.08 &  -50.92$\pm$0.06 &  7.40 & 0.381 &   9.40 & 0.117 & 0.0191 & 1.38 &  3.25 & 0.55 & ...            &  986$\pm$303 &  918\\
       13 & 6018549427233722112 & 16.496 & 16 & 0.308 &  4.17 &  -33.98$\pm$0.09 &  -52.35$\pm$0.07 &  3.95 & 0.315 &   4.64 & 0.129 & 0.0023 & 1.19 &  3.94 & 1.19 & ...            &  960$\pm$230 &  721\\
       14 & 1171729441274021376 & 16.504 & 16 & 0.264 &  4.26 &  -48.08$\pm$0.07 &   -3.62$\pm$0.06 &  2.90 & 0.191 &   1.84 & 0.097 & 0.0339 & 1.14 &  3.61 & 0.44 & ...            &  866$\pm$204 &  707\\
       15 & 4265540383431508736 & 16.576 & 14 & 0.402 &  6.66 &  -37.25$\pm$0.07 &  -76.46$\pm$0.07 &  0.54 & 0.043 &   0.08 & 0.103 & 0.0332 & 1.03 &  4.60 & 0.80 & ...            & 1002$\pm$151 &  803\\
       16 & 297028668197924352  & 16.666 & 16 & 0.251 &  4.09 &  +42.39$\pm$0.08 &  -33.96$\pm$0.05 &  0.24 & 0.076 &   0.28 & 0.106 & 0.0108 & 1.01 &  3.66 & 0.48 & ...            & 1027$\pm$251 &  824\\
       17 & 5932954268193447424 & 16.677 & 19 & 0.375 &  4.24 &  -12.13$\pm$0.09 &  -70.86$\pm$0.06 & 20.15 & 0.474 &  16.68 & 0.126 & 0.0098 & 1.58 &  4.55 & 0.77 & ...            &  909$\pm$214 &  742\\
       18 & 6868645088567296128 & 16.770 & 14 & 0.271 &  3.07 &  -30.83$\pm$0.10 &  -46.57$\pm$0.07 &  3.40 & 0.249 &   2.05 & 0.143 & 0.0325 & 1.19 &  3.94 & ...  & ...            &  976$\pm$318 &  764\\
       19 & 4255853136265401216 & 16.785 & 15 & 0.332 &  3.23 &  -37.98$\pm$0.09 &  -58.42$\pm$0.08 &  3.65 & 0.356 &   3.97 & 0.131 & 0.0039 & 1.19 &  4.39 & 1.03 & ...            &  994$\pm$308 &  790\\
       20 & 2742538589334386176 & 16.801 & 17 & 0.390 &  3.44 &  +64.88$\pm$0.14 &  -41.03$\pm$0.10 & 18.41 & 0.702 &  24.54 & 0.209 & 0.0111 & 1.77 &  4.75 & 0.84 & ...            &  934$\pm$271 &  742\\
       21 & 1518135183809719552 & 16.814 & 19 & 0.299 &  4.32 &  -49.64$\pm$0.04 &  -38.05$\pm$0.07 &  2.04 & 0.173 &   1.67 & 0.098 & 0.0114 & 1.07 &  4.19 & 0.46 & ...            &  993$\pm$230 &  765\\
       22 & 2332854092898442112 & 16.819 & 14 & 0.262 &  3.59 &   +0.91$\pm$0.08 &  -57.17$\pm$0.07 & -0.05 & 0.000 &   0.00 & 0.106 & 0.0432 & 1.00 &  3.91 & 0.46 & ...            & 1036$\pm$288 &  822\\
       23 & 2421003524941335168 & 16.861 & 16 & 0.293 &  3.59 &   +8.53$\pm$0.08 &  -58.39$\pm$0.05 & -0.06 & 0.000 &   0.00 & 0.107 & 0.0151 & 1.00 &  4.20 & 0.51 & ...            &  955$\pm$266 &  740\\
       24 & 6094201393481158656 & 16.899 & 14 & 0.333 &  4.09 &  -82.68$\pm$0.08 &  -23.19$\pm$0.08 &  4.70 & 0.238 &   2.32 & 0.126 & 0.0415 & 1.20 &  4.51 & 0.73 & ...            & 1221$\pm$298 & 1039\\
       25 & 3516618964544300928 & 16.900 & 19 & 0.325 &  4.32 &  -64.11$\pm$0.08 &   -8.65$\pm$0.06 &  2.42 & 0.227 &   2.24 & 0.123 & 0.0215 & 1.09 &  4.46 & 0.55 & ...            &  944$\pm$218 &  801\\
       26 & 6642586734844528512 & 16.920 & 17 & 0.256 &  3.11 &  +32.70$\pm$0.07 &  -30.45$\pm$0.06 & -0.90 & 0.000 &   0.00 & 0.106 & 0.0148 & 0.96 &  3.96 & 0.45 & ...            &  828$\pm$266 &  733\\
       27 & 6642338313936288256 & 16.935 & 17 & 0.472 &  4.91 &   +9.14$\pm$0.10 &  -92.13$\pm$0.08 &  6.21 & 0.498 &   7.64 & 0.140 & 0.0229 & 1.32 &  5.31 & 0.58 & ...            &  929$\pm$189 &  703\\
       28 & 1798008584396457088 & 16.977 & 16 & 1.194 & 18.24 &  +98.28$\pm$0.07 & +240.18$\pm$0.06 & -0.21 & 0.000 &   0.00 & 0.094 & 0.0068 & 0.99 &  7.36 & 0.27 & ...            & 1031$\pm$ 56 & 1108\\
       29 & 3488721051716722048 & 17.009 & 18 & 0.316 &  4.31 &  -61.88$\pm$0.08 &   -1.58$\pm$0.08 & -1.16 & 0.000 &   0.00 & 0.121 & 0.0224 & 0.95 &  4.51 & 0.55 & ...            &  927$\pm$215 &  804\\
       30 & 6724844047285780224 & 17.139 & 17 & 0.309 &  3.60 &  -39.41$\pm$0.08 &  -54.95$\pm$0.06 & -0.92 & 0.000 &   0.00 & 0.121 & 0.0458 & 0.96 &  4.59 & 0.60 & ...            & 1036$\pm$288 &  795\\
       31 & 5706246274764433408 & 17.173 & 21 & 0.264 &  3.32 &  +27.53$\pm$0.07 &  -37.13$\pm$0.07 &  6.19 & 0.362 &   4.80 & 0.104 & 0.0082 & 1.22 &  4.28 & 0.60 & ...            &  829$\pm$250 &  732\\
       32 & 1451652599056932480 & 17.190 & 19 & 0.453 &  5.64 &  -69.40$\pm$0.09 &  -63.20$\pm$0.05 & -0.26 & 0.000 &   0.00 & 0.121 & 0.0089 & 0.99 &  5.47 & 0.62 & ...            &  983$\pm$174 &  741\\
       33 & 4053722045323914880 & 17.198 & 17 & 0.601 &  3.84 &  -94.99$\pm$0.16 &  -71.84$\pm$0.11 & 14.02 & 0.832 &  17.55 & 0.221 & 0.0280 & 1.69 &  6.09 & 1.10 & ...            &  939$\pm$244 &  716\\
       34 & 4336938064311520256 & 17.225 & 16 & 0.340 &  3.23 &  -60.66$\pm$0.11 &  -33.48$\pm$0.08 &  2.01 & 0.175 &   0.77 & 0.155 & 0.0233 & 1.09 &  4.88 & 0.89 & ...            &  965$\pm$299 &  755\\
       35 & 1938121687585625728 & 17.263 & 26 & 0.291 &  3.94 &  +44.49$\pm$0.06 &  +19.23$\pm$0.06 &  3.54 & 0.229 &   2.28 & 0.098 & 0.0114 & 1.09 &  4.59 & 0.62 & ...            &  788$\pm$200 &  714\\
       36 & 4664584145613345280 & 17.267 & 21 & 0.552 &  4.65 &  +34.46$\pm$0.18 &  -82.00$\pm$0.16 & 18.55 & 1.147 &  32.57 & 0.260 & 0.0390 & 1.87 &  5.98 & 0.89 & ...            &  764$\pm$164 &  717\\
       37 & 1014344556600754688 & 17.330 & 19 & 0.326 &  3.18 &   +2.52$\pm$0.09 &  -66.15$\pm$0.07 &  1.46 & 0.104 &   0.26 & 0.143 & 0.0257 & 1.06 &  4.90 & 0.52 & ...            &  962$\pm$303 &  716\\
       38 & 3909682994105224448 & 17.331 & 15 & 0.296 &  3.01 &  -15.62$\pm$0.11 &  -58.16$\pm$0.09 & -1.57 & 0.000 &   0.00 & 0.158 & 0.0837 & 0.92 &  4.69 & 0.59 & ...            &  965$\pm$321 &  748\\
       39 & 4485697275968661504 & 17.331 & 21 & 0.288 &  3.25 &   -0.66$\pm$0.08 &  -52.92$\pm$0.06 &  0.57 & 0.000 &   0.00 & 0.124 & 0.0191 & 1.02 &  4.63 & 0.67 & ...            &  871$\pm$268 &  709\\
       40 & 5805243926609660032 & 17.406 & 28 & 0.531 &  7.56 &  -80.23$\pm$0.06 & -195.96$\pm$0.06 &  1.22 & 0.081 &   0.17 & 0.092 & 0.0230 & 1.04 &  6.03 & 0.30 & ...            & 1891$\pm$250 & 1726\\
       41 & 5843685464461889792 & 17.462 & 24 & 0.262 &  3.41 &  -46.52$\pm$0.08 &  -12.56$\pm$0.08 & -2.16 & 0.000 &   0.00 & 0.114 & 0.0239 & 0.92 &  4.56 & 0.64 & ...            &  871$\pm$256 &  726\\
       42 & 6620452702488805504 & 17.489 & 17 & 0.403 &  4.35 &  -10.77$\pm$0.08 &  -77.73$\pm$0.08 & -1.03 & 0.000 &   0.00 & 0.126 & 0.0142 & 0.96 &  5.51 & 0.59 & ...            &  923$\pm$212 &  709\\
       43 & 6633599563021390080 & 17.509 & 21 & 0.306 &  3.78 &  -17.54$\pm$0.07 &  -70.75$\pm$0.07 &  0.14 & 0.000 &   0.00 & 0.103 & 0.0171 & 1.00 &  4.94 & 0.56 & ...            & 1129$\pm$299 &  898\\
       44 & 6080324938269559552 & 17.534 & 20 & 0.394 &  3.96 &  -71.65$\pm$0.09 &   +5.33$\pm$0.08 &  2.82 & 0.333 &   2.57 & 0.129 & 0.0268 & 1.10 &  5.51 & 0.74 & ...            &  864$\pm$218 &  721\\
       45 & 2685667415539171968 & 17.634 & 14 & 0.375 &  3.06 &  -18.34$\pm$0.13 &  -73.53$\pm$0.09 &  0.07 & 0.000 &   0.00 & 0.188 & 0.0892 & 1.00 &  5.50 & 0.69 & ...            &  959$\pm$313 &  781\\
       46 & 2350668693824590080 & 17.673 & 14 & 0.722 &  3.67 & +128.29$\pm$0.19 &  -34.71$\pm$0.16 & 13.57 & 0.859 &  14.95 & 0.309 & 0.0841 & 1.47 &  6.96 & 1.00 & ...            &  873$\pm$238 &  704\\
       47 & 6623387436461113216 & 17.781 & 15 & 0.910 &  3.03 & +153.13$\pm$0.27 & -128.34$\pm$0.31 & 39.30 & 2.172 &  83.32 & 0.518 & 0.3853 & 3.05 &  7.57 & 1.28 & ...            & 1041$\pm$344 &  843\\
       48 & 4401964212769155840 & 17.804 & 15 & 0.376 &  3.00 &  -68.92$\pm$0.14 &  +18.09$\pm$0.13 & -0.03 & 0.000 &   0.00 & 0.215 & 0.0109 & 1.00 &  5.68 & 0.73 & ...            &  898$\pm$299 &  816\\
       49 & 6448449707644184320 & 17.844 & 20 & 0.409 &  3.32 &  -33.14$\pm$0.09 &  -73.46$\pm$0.10 &  0.33 & 0.000 &   0.00 & 0.158 & 0.0174 & 1.01 &  5.90 & 0.66 & ...            &  934$\pm$281 &  702\\
       50 & 5349492955284690048 & 17.855 & 26 & 0.348 &  3.47 &   -5.39$\pm$0.10 &  +57.37$\pm$0.09 &  3.11 & 0.400 &   2.62 & 0.146 & 0.0126 & 1.11 &  5.56 & 0.67 & ...            &  785$\pm$226 &  788\\
       51 & 6148289080574141056 & 17.894 & 21 & 0.430 &  3.72 &  -74.61$\pm$0.09 &  -14.92$\pm$0.08 &  1.69 & 0.234 &   0.93 & 0.148 & 0.0045 & 1.05 &  6.06 & 0.72 & ...            &  838$\pm$225 &  702\\
       52 & 4715594303954641920 & 17.908 & 29 & 0.394 &  4.50 &  +20.99$\pm$0.10 &  -71.55$\pm$0.09 &  0.09 & 0.000 &   0.00 & 0.143 & 0.0281 & 1.00 &  5.88 & 0.69 & ...            &  898$\pm$200 &  726\\
       53 & 766482238762265088  & 18.041 & 17 & 0.438 &  3.10 &  -20.62$\pm$0.11 &  -85.88$\pm$0.10 &  2.21 & 0.319 &   1.24 & 0.175 & 0.0065 & 1.08 &  6.25 & 0.79 & ...            &  956$\pm$308 &  718\\
       54 & 6364080744469693056 & 18.134 & 28 & 0.303 &  3.11 &  -49.61$\pm$0.09 &   +5.48$\pm$0.12 & -0.41 & 0.000 &   0.00 & 0.180 & 0.0015 & 0.98 &  5.54 & 0.63 & ...            &  781$\pm$251 &  722\\
       55 & 3896957079021169024 & 18.205 & 13 & 0.567 &  3.01 &  +48.21$\pm$0.23 &  -83.84$\pm$0.20 &  1.28 & 0.393 &   1.17 & 0.369 & 0.0218 & 1.06 &  6.97 & 1.05 & ...            &  808$\pm$269 &  724\\
       56 & 2156908318076164224 & 18.246 & 24 & 0.423 &  4.26 &   +9.41$\pm$0.14 & +211.79$\pm$0.15 & -0.27 & 0.000 &   0.00 & 0.225 & 0.0187 & 0.99 &  6.38 & 0.24 & ...            & 2375$\pm$558 & 2339\\
       57 & 6497257617211406976 & 18.260 & 28 & 0.388 &  3.10 &  +67.17$\pm$0.10 &   +0.82$\pm$0.14 &  0.96 & 0.100 &   0.11 & 0.213 & 0.0136 & 1.03 &  6.21 & 0.74 & ...            &  820$\pm$264 &  733\\
       58 & 3507697866498687232 & 18.264 & 16 & 0.603 &  3.16 &  +64.57$\pm$0.23 &  +52.35$\pm$0.14 &  1.64 & 0.500 &   1.91 & 0.329 & 0.0247 & 1.07 &  7.17 & 0.23 & ...            &  653$\pm$207 &  865\\
       59 & 2608631816965886720 & 18.291 & 15 & 0.500 &  3.10 &  -27.28$\pm$0.21 &  -86.47$\pm$0.19 &  0.99 & 0.000 &   0.00 & 0.312 & 0.1304 & 1.05 &  6.79 & 0.78 & ...            &  860$\pm$278 &  703\\
       60 & 5436981164224577920 & 18.332 & 29 & 0.363 &  3.28 &  -61.68$\pm$0.10 &   +7.45$\pm$0.10 &  1.05 & 0.227 &   0.58 & 0.149 & 0.0136 & 1.03 &  6.13 & 0.78 & ...            &  810$\pm$247 &  793\\
       61 & 5133267224811090048 & 18.397 & 17 & 0.597 &  3.27 & +105.22$\pm$0.21 &  -56.25$\pm$0.18 &  2.41 & 0.604 &   2.21 & 0.303 & 0.0199 & 1.10 &  7.28 & 0.88 & ...            &  947$\pm$290 &  720\\
       62 & 3730485894679948928 & 18.420 & 17 & 0.898 &  5.01 &  -14.28$\pm$0.22 & -189.84$\pm$0.16 & -1.58 & 0.000 &   0.00 & 0.319 & 0.0292 & 0.95 &  8.19 & 0.91 & ...            & 1005$\pm$201 &  814\\
       63 & 6303664161071398272 & 18.463 & 15 & 0.599 &  3.58 & -117.06$\pm$0.20 &  -66.92$\pm$0.17 &  0.54 & 0.055 &   0.02 & 0.295 & 0.1045 & 1.03 &  7.35 & 0.80 & ...            & 1068$\pm$298 &  842\\
       64 & 5165642241611476480 & 18.466 & 14 & 0.706 &  3.50 & +137.42$\pm$0.20 &  -44.33$\pm$0.20 & -0.45 & 0.000 &   0.00 & 0.294 & 0.0207 & 0.97 &  7.71 & 0.77 & ...            &  969$\pm$277 &  765\\
       65 & 4323925515994734976 & 18.514 & 18 & 0.651 &  3.24 & -110.90$\pm$0.25 &  -89.02$\pm$0.18 &  1.96 & 0.685 &   3.23 & 0.353 & 0.0182 & 1.07 &  7.58 & 1.06 & ...            & 1036$\pm$320 &  797\\
       66 & 2374938763739778944 & 18.667 & 15 & 1.047 &  4.46 & +172.33$\pm$0.25 & -135.21$\pm$0.19 & -1.34 & 0.000 &   0.00 & 0.345 & 0.0351 & 0.94 &  8.77 & 0.86 & ...            &  992$\pm$223 &  766\\
       67 & 6884930019709368576 & 18.764 & 17 & 1.164 &  5.83 &  -60.35$\pm$0.25 & -221.74$\pm$0.27 &  0.04 & 0.000 &   0.00 & 0.484 & 0.0035 & 1.00 &  9.09 & 1.04 & ...            &  936$\pm$160 &  729\\
       68 & 6792735110724346880 & 18.867 & 19 & 0.792 &  3.14 & -113.96$\pm$0.23 & -103.55$\pm$0.20 &  1.53 & 0.602 &   1.43 & 0.344 & 0.0189 & 1.06 &  8.36 & 0.94 & ...            &  922$\pm$293 &  767\\
       69 & 2930227320146255104 & 18.989 & 24 & 0.559 &  3.00 &  -50.66$\pm$0.17 &  +51.11$\pm$0.19 & -1.27 & 0.000 &   0.00 & 0.274 & 0.0107 & 0.97 &  7.73 & 0.83 & ...            &  610$\pm$203 &  740\\
       70 & 1371498575053704064 & 19.443 & 31 & 0.692 &  3.23 &  -51.70$\pm$0.22 & -118.53$\pm$0.24 & -2.21 & 0.000 &   0.00 & 0.348 & 0.0200 & 0.93 &  8.64 & 0.96 & ...            &  886$\pm$274 &  714\\
       71 & 6394155823463179264 & 19.640 & 30 & 0.847 &  3.23 &  +93.41$\pm$0.25 & -154.87$\pm$0.27 &  0.00 & 0.000 &   0.00 & 0.406 & 0.0048 & 1.00 &  9.28 & 0.90 & ...            & 1012$\pm$314 &  789\\
       72 & 4623587651977758208 & 20.154 & 23 & 1.389 &  3.34 &  +48.53$\pm$0.47 & +227.36$\pm$0.45 & -1.87 & 0.000 &   0.00 & 0.670 & 0.0424 & 0.92 & 10.87 & 1.00 & ...            &  794$\pm$237 &  700\\
\hline
\end{tabular}
\tablefoot{\fontsize{6.4pt}{0.90\baselineskip}\selectfont
The data in Tables~\ref{Tab_72dr3} and \ref{Tab_72nnflg} will be 
combined and made available in electronic form at the CDS.
}
\end{table*}

%
\begin{table*}
	\caption{DR3 next neighbours, IPD parameters, flags, and results from local$+$global parameter comparison of 72 HVS candidates}
\label{Tab_72nnflg}
\centering
\fontsize{6.4pt}{0.80\baselineskip}\selectfont
\begin{tabular}{@{}l@{\hspace{1mm}}r@{\hspace{1mm}}r@{\hspace{1mm}}r@{\hspace{1mm}}r@{\hspace{1mm}}r@{\hspace{1mm}}r@{\hspace{1mm}}r@{\hspace{1mm}}c@{\hspace{1mm}}c@{\hspace{1mm}}c@{\hspace{2mm}}c@{\hspace{2mm}}c@{\hspace{2mm}}c@{\hspace{2mm}}c@{\hspace{2mm}}c@{\hspace{2mm}}c@{\hspace{2mm}}c@{\hspace{2mm}}c@{\hspace{1mm}}r@{}}     
\hline\hline
	No.& $G$   & $Gnn$ & $dnn$   &$n5as$&$IPD$&$IPD$&$Sol$&$Dup$&$Nper$             & $e\_Plx$    & $e\_pmRA$   & $e\_pmDE$   & $gofAL$     & $epsi$        & $sepsi$      & $amax$      & $IPDgofha$  & $RUWE$     & $RPlx$ \\
	&       &       & [arc    &      &$fmp$&$fow$&$ved$&     &$l\Delta$ $g\Delta$& $lR$ $gR$   & $lR$ $gR$   & $lR$ $gR$   & $lR$ $gR$   & $lR$ $gR$     & $lR$ $gR$    & $lR$ $gR$   & $lR$ $gR$   & $lR$ $gR$  &         \\
	   & [mag] & [mag] &     sec]&      &     &     &     &     &                   & \\
\hline
       01 & 11.695 & ...    & ...  & 1&  0& 0&31&0& -1  -5 & 2.722 3.058 & 2.783 2.389 & 2.654 3.213 &  7.795  3.695 &  3.699  3.144 &  12.637   7.737 & 2.605 2.604 &  0.624  0.644 & 3.060 2.747 &  7.16 \\
       02 & 13.519 & ...    & ...  & 1&  1& 0&31&0& +1  -5 & 1.038 1.088 & 1.038 0.942 & 0.923 1.386 &  3.080  1.343 &  1.948  1.255 &   3.991   1.237 & 0.897 1.129 &  0.405  0.398 & 1.276 1.163 & 15.06 \\
       03 & 14.485 & ...    & ...  & 1&  1& 0&31&0& +1  +1 & 2.306 2.375 & 2.143 1.943 & 2.250 2.051 & 20.001  9.382 &  8.887  4.801 &  89.537  25.208 & 2.207 2.094 &  1.174  1.234 & 2.790 2.563 &  3.92 \\
       04 & 15.380 & ...    & ...  & 1&  0& 0&31&0& -1  -4 & 1.650 1.829 & 1.730 1.686 & 2.009 2.088 & 16.210  6.565 &  5.969  3.363 &  33.479  10.497 & 2.001 2.039 &  0.448  0.429 & 1.983 1.867 &  3.55 \\
       05 & 15.569 & 17.985 & 0.99 & 3& 38& 0&31&0&  0  -4 & 1.112 1.452 & 1.138 1.420 & 1.133 1.292 & 11.421  5.301 &  3.636  2.627 &  12.508   4.685 & 1.111 1.335 &  0.085  0.169 & 1.676 1.615 &  4.09 \\
       06 & 15.827 & 16.518 & 0.63 & 2& 63& 1&95&0& -5  +7 & 2.935 2.134 & 2.517 1.385 & 3.057 2.396 & 20.567 14.278 & 13.000  6.795 & 170.880  57.270 & 2.967 1.942 &  9.981  5.342 & 2.337 2.248 &  3.46 \\
       07 & 15.875 & ...    & ...  & 1&  0& 0&31&0&  0  -2 & 0.845 0.892 & 0.879 0.952 & 0.976 0.902 & -0.398 -0.219 &  0.000  0.000 &   0.000   0.000 & 0.898 0.971 &  0.701  1.590 & 0.933 0.898 &  7.38 \\
       08 & 15.934 & ...    & ...  & 1&  2& 0&31&0&  0  -6 & 0.878 1.050 & 0.746 0.840 & 0.781 1.060 &  1.248  1.097 &  0.772  0.735 &   0.647   0.424 & 0.737 0.936 &  0.482  0.948 & 1.019 1.043 &  4.68 \\
       09 & 15.965 & 15.439 & 3.49 & 6&  4& 0&31&0&  0  +2 & 1.464 1.372 & 1.429 1.080 & 1.422 1.353 & 13.145  9.511 &  7.029  3.693 &  54.677  17.502 & 1.415 1.142 &  1.259  0.824 & 1.622 1.576 & 11.65 \\
       10 & 16.041 & 20.134 & 4.10 & 3&  0& 0&31&0&  0   0 & 0.917 0.829 & 0.896 0.738 & 0.875 0.717 &  0.681  0.470 &  1.159  0.740 &   1.517   0.593 & 0.922 0.748 &  0.859  0.735 & 0.984 0.962 &  6.98 \\
       11 & 16.060 & 20.239 & 2.25 & 2&  0& 0&31&0&  0  -4 & 0.934 1.050 & 0.922 1.013 & 0.877 1.003 & -0.194 -0.137 &  0.000  0.000 &   0.000   0.000 & 0.885 0.987 &  0.062  0.163 & 0.914 0.910 &  4.44 \\
       12 & 16.157 & ...    & ...  & 1&  0& 0&31&0&  0  -5 & 1.274 1.422 & 1.135 1.343 & 1.073 1.146 &  5.825  3.803 &  3.430  2.676 &  11.369   5.756 & 1.087 1.263 &  0.435  0.637 & 1.306 1.286 &  3.25 \\
       13 & 16.496 & 16.537 & 1.64 & 2&  0&13&95&0&  0  -2 & 0.954 1.107 & 1.000 1.285 & 0.986 1.124 &  4.650  2.243 &  2.739  2.004 &   7.610   3.247 & 0.980 1.190 &  0.106  0.076 & 1.141 1.113 &  4.17 \\
       14 & 16.504 & ...    & ...  & 1&  0& 0&31&0& +1  -3 & 0.848 0.927 & 0.885 0.946 & 0.868 0.962 &  2.273  1.639 &  1.481  1.175 &   2.343   1.167 & 0.867 0.883 &  1.048  1.128 & 1.075 1.068 &  4.26 \\
       15 & 16.576 & 19.972 & 1.28 & 3&  0& 0&31&0& -1  -5 & 0.792 0.869 & 0.787 0.932 & 0.825 1.045 &  0.340  0.316 &  0.221  0.258 &   0.057   0.052 & 0.783 0.910 &  0.884  1.157 & 0.946 0.963 &  6.66 \\
       16 & 16.666 & ...    & ...  & 1&  0& 0&31&0&  0  -3 & 0.742 0.844 & 0.916 0.955 & 0.839 0.775 &  0.189  0.136 &  0.594  0.436 &   0.353   0.190 & 0.866 0.886 &  0.419  0.349 & 0.957 0.944 &  4.09 \\
       17 & 16.677 & 19.203 & 0.74 & 5&  6& 0&31&0& -1  +1 & 1.069 1.210 & 1.072 1.125 & 0.984 0.932 &  5.301 12.560 &  2.070  2.851 &   4.985  11.910 & 1.073 1.055 &  0.600  0.335 & 1.431 1.492 &  4.24 \\
       18 & 16.770 & 16.379 & 1.70 & 2&  0&52&95&0& -1  -5 & 1.100 1.153 & 1.099 1.192 & 1.010 1.052 &  2.981  2.074 &  1.808  1.381 &   3.077   1.476 & 1.042 1.144 &  0.395  1.063 & 1.128 1.116 &  3.07 \\
       19 & 16.785 & ...    & ...  & 1&  0& 0&31&0&  0  -4 & 0.823 1.336 & 0.852 1.096 & 0.859 1.120 &  2.480  2.378 &  1.000  2.015 &   1.329   2.984 & 0.835 1.036 &  0.069  0.132 & 1.107 1.122 &  3.23 \\
       20 & 16.801 & ...    & ...  & 1&  0& 0&31&0& +1  -2 & 1.263 1.467 & 1.245 1.640 & 1.202 1.434 & 15.980 11.491 &  6.052  4.056 &  41.728  18.674 & 1.256 1.678 &  0.132  0.381 & 1.693 1.666 &  3.44 \\
       21 & 16.814 & ...    & ...  & 1&  0& 0&31&0&  0   0 & 0.895 0.887 & 0.843 0.509 & 0.870 0.944 &  1.240  1.222 &  1.395  0.983 &   1.903   1.267 & 0.853 0.771 &  0.466  0.388 & 1.013 1.007 &  4.32 \\
       22 & 16.819 & ...    & ...  & 1&  0& 0&31&0&  0  -5 & 0.755 0.937 & 0.844 0.896 & 0.821 0.972 & -0.044 -0.030 &  0.000  0.000 &   0.000   0.000 & 0.774 0.844 &  0.791  1.485 & 0.943 0.939 &  3.59 \\
       23 & 16.861 & ...    & ...  & 1&  0& 0&31&0&  0  -3 & 0.876 1.014 & 0.784 0.877 & 0.786 0.748 & -0.029 -0.036 &  0.000  0.000 &   0.000   0.000 & 0.760 0.818 &  0.423  0.527 & 0.926 0.940 &  3.59 \\
       24 & 16.899 & 19.405 & 1.35 & 2&  0& 0&31&0& -1  -5 & 0.970 0.999 & 0.978 0.957 & 0.891 1.116 &  3.096  2.898 &  1.697  1.259 &   2.925   1.710 & 0.922 0.952 &  0.972  1.427 & 1.129 1.124 &  4.09 \\
       25 & 16.900 & ...    & ...  & 1&  0& 0&31&0& +1   0 & 0.875 0.925 & 0.866 0.957 & 0.875 0.850 &  2.209  1.498 &  2.064  1.213 &   4.310   1.710 & 0.869 0.936 &  0.699  0.750 & 1.049 1.026 &  4.32 \\
       26 & 16.920 & ...    & ...  & 1&  0& 0&31&0&  0  -2 & 0.877 0.996 & 0.888 0.822 & 0.845 0.768 & -0.985 -0.633 &  0.000  0.000 &   0.000   0.000 & 0.883 0.790 &  0.416  0.515 & 0.917 0.906 &  3.11 \\
       27 & 16.935 & ...    & ...  & 1&  0& 0&31&0&  0  -2 & 1.127 1.161 & 1.183 1.071 & 1.174 1.065 &  6.563  4.220 &  4.153  2.706 &  15.510   5.994 & 1.185 1.024 &  0.664  0.831 & 1.264 1.249 &  4.91 \\
       28 & 16.977 & ...    & ...  & 1&  0& 0&31&0& -1  -2 & 0.752 0.769 & 0.791 0.738 & 0.763 0.740 & -0.150 -0.139 &  0.000  0.000 &   0.000   0.000 & 0.757 0.675 &  0.229  0.236 & 0.940 0.935 & 18.24 \\
       29 & 17.009 & ...    & ...  & 1&  0& 0&31&0&  0  -1 & 0.802 0.847 & 0.810 0.866 & 0.857 0.998 & -1.277 -0.780 &  0.000  0.000 &   0.000   0.000 & 0.806 0.860 &  0.851  0.765 & 0.924 0.903 &  4.31 \\
       30 & 17.139 & ...    & ...  & 1&  0& 0&31&0&  0  -2 & 0.818 0.937 & 0.809 0.846 & 0.787 0.740 & -0.628 -0.668 &  0.000  0.000 &   0.000   0.000 & 0.793 0.807 &  0.853  1.556 & 0.905 0.912 &  3.60 \\
       31 & 17.173 & ...    & ...  & 1&  0& 0&31&0& -1  +2 & 0.951 0.842 & 0.893 0.658 & 0.935 0.832 &  4.061  4.462 &  2.549  1.716 &   6.662   3.766 & 0.885 0.673 &  0.489  0.277 & 1.161 1.162 &  3.32 \\
       32 & 17.190 & ...    & ...  & 1&  0& 0&31&0& -2   0 & 0.884 0.853 & 0.827 0.836 & 0.855 0.611 & -0.225 -0.201 &  0.000  0.000 &   0.000   0.000 & 0.832 0.782 &  0.291  0.305 & 0.952 0.942 &  5.64 \\
       33 & 17.198 & 19.649 & 1.12 & 7&  7& 0&95&0&  0  -2 & 0.974 1.657 & 0.935 1.524 & 0.905 1.306 &  4.312 10.191 &  0.930  4.013 &   1.243  14.614 & 0.918 1.434 &  0.627  1.001 & 1.476 1.610 &  3.84 \\
       34 & 17.225 & ...    & ...  & 1&  2& 0&31&0&  0  -3 & 0.893 1.082 & 0.879 1.042 & 0.835 0.903 &  2.085  1.402 &  1.129  0.801 &   1.324   0.591 & 0.885 0.985 &  0.823  0.791 & 1.044 1.031 &  3.23 \\
       35 & 17.263 & ...    & ...  & 1&  0& 0&31&0& +1  +7 & 0.859 0.752 & 0.864 0.533 & 0.926 0.698 &  1.393  2.575 &  1.225  1.020 &   1.684   1.757 & 0.906 0.606 &  0.381  0.398 & 1.023 1.037 &  3.94 \\
       36 & 17.267 & ...    & ...  & 1&  1& 0&31&0& -3  +2 & 1.598 1.199 & 1.856 1.679 & 1.608 1.813 & 19.893 13.703 &  9.639  5.133 &  95.048  24.742 & 1.682 1.608 &  1.969  1.399 & 1.805 1.782 &  4.65 \\
       37 & 17.330 & ...    & ...  & 1&  0& 0&31&0&  0   0 & 0.941 0.999 & 0.931 0.852 & 0.737 0.776 &  1.081  1.041 &  0.550  0.434 &   0.322   0.190 & 0.917 0.849 &  0.953  0.877 & 1.007 1.008 &  3.18 \\
       38 & 17.331 & ...    & ...  & 1&  1& 0&31&0& +1  -4 & 0.769 0.957 & 0.783 0.970 & 0.773 0.977 & -1.364 -1.149 &  0.000  0.000 &   0.000   0.000 & 0.729 0.938 &  1.968  2.837 & 0.868 0.872 &  3.01 \\
       39 & 17.331 & ...    & ...  & 1&  2& 0&31&0&  0  +2 & 0.844 0.865 & 0.839 0.701 & 0.842 0.679 &  0.477  0.421 &  0.000  0.000 &   0.000   0.000 & 0.877 0.734 &  1.218  0.649 & 0.980 0.968 &  3.25 \\
       40 & 17.406 & ...    & ...  & 1&  0& 0&31&0& +1  +9 & 0.827 0.655 & 0.857 0.510 & 0.800 0.650 &  1.353  0.929 &  0.802  0.326 &   0.642   0.128 & 0.812 0.523 &  1.128  0.766 & 1.011 0.992 &  7.56 \\
       41 & 17.462 & ...    & ...  & 1&  0& 0&31&0& -1  +5 & 0.836 0.697 & 0.879 0.666 & 0.788 0.775 & -1.659 -1.576 &  0.000  0.000 &   0.000   0.000 & 0.791 0.632 &  0.995  0.777 & 0.875 0.869 &  3.41 \\
       42 & 17.489 & ...    & ...  & 1&  0& 0&31&0&  0  -2 & 0.789 0.822 & 0.705 0.646 & 0.721 0.719 & -1.098 -0.757 &  0.000  0.000 &   0.000   0.000 & 0.697 0.676 &  0.579  0.485 & 0.918 0.907 &  4.35 \\
       43 & 17.509 & ...    & ...  & 1&  0& 0&31&0& +1  +2 & 0.764 0.712 & 0.802 0.586 & 0.838 0.662 &  0.136  0.110 &  0.000  0.000 &   0.000   0.000 & 0.787 0.558 &  0.907  0.594 & 0.970 0.958 &  3.78 \\
       44 & 17.534 & 20.210 & 3.91 & 2&  4& 0&31&0& -1  +1 & 0.908 0.868 & 0.860 0.692 & 0.870 0.725 &  2.526  2.241 &  2.235  1.248 &   5.143   1.865 & 0.842 0.684 &  1.109  0.918 & 1.057 1.048 &  3.96 \\
       45 & 17.634 & ...    & ...  & 1&  0& 0&31&0&  0  -5 & 0.843 0.999 & 0.829 0.999 & 0.742 0.792 &  0.065  0.056 &  0.000  0.000 &   0.000   0.000 & 0.787 0.947 &  1.329  3.097 & 0.955 0.957 &  3.06 \\
       46 & 17.673 & ...    & ...  & 1& 13& 0&31&0& -2  -4 & 1.432 1.560 & 1.206 1.361 & 1.146 1.351 & 11.358 10.348 &  4.937  2.786 &  25.851   9.851 & 1.251 1.492 &  2.375  2.641 & 1.412 1.396 &  3.67 \\
       47 & 17.781 & ...    & ...  & 1& 46& 0&95&0&  0  -4 & 2.020 2.267 & 2.037 1.875 & 2.012 2.528 & 42.056 29.947 & 11.676  6.322 & 160.658  48.281 & 2.025 2.369 & 10.585 13.048 & 2.944 2.907 &  3.03 \\
       48 & 17.804 & ...    & ...  & 1&  0& 0&31&0&  0  -4 & 0.785 0.926 & 0.844 0.960 & 0.881 1.010 & -0.029 -0.022 &  0.000  0.000 &   0.000   0.000 & 0.866 0.966 &  0.144  0.369 & 0.956 0.953 &  3.00 \\
       49 & 17.844 & ...    & ...  & 1&  0& 0&31&0&  0  +1 & 0.826 0.890 & 0.808 0.619 & 0.932 0.819 &  0.357  0.253 &  0.000  0.000 &   0.000   0.000 & 0.832 0.696 &  0.868  0.590 & 0.981 0.964 &  3.32 \\
       50 & 17.855 & 20.961 & 2.82 & 2&  0& 0&31&0&  0  +8 & 0.859 0.719 & 0.853 0.657 & 0.883 0.715 &  2.535  2.476 &  2.073  1.119 &   4.548   1.589 & 0.852 0.645 &  0.660  0.417 & 1.063 1.056 &  3.47 \\
       51 & 17.894 & ...    & ...  & 1&  0& 0&31&0& -1  +2 & 0.837 0.815 & 0.842 0.612 & 0.799 0.619 &  1.596  1.381 &  1.508  0.620 &   2.386   0.504 & 0.853 0.639 &  0.269  0.147 & 1.020 1.008 &  3.72 \\
       52 & 17.908 & ...    & ...  & 1&  0& 0&31&0& +2 +11 & 0.817 0.607 & 0.783 0.646 & 0.781 0.679 &  0.077  0.070 &  0.000  0.000 &   0.000   0.000 & 0.758 0.612 &  1.047  0.935 & 0.963 0.956 &  4.50 \\
       53 & 18.041 & ...    & ...  & 1&  0& 0&31&0& -1  -1 & 0.876 0.904 & 0.820 0.676 & 0.808 0.703 &  0.999  1.805 &  0.911  0.762 &   0.880   0.644 & 0.809 0.684 &  0.217  0.216 & 1.004 1.037 &  3.10 \\
       54 & 18.134 & ...    & ...  & 1&  0& 0&31&0&  0  +9 & 0.768 0.592 & 0.738 0.510 & 0.859 0.812 & -0.480 -0.340 &  0.000  0.000 &   0.000   0.000 & 0.868 0.666 &  0.066  0.048 & 0.956 0.939 &  3.11 \\
       55 & 18.205 & 20.006 & 0.99 & 2&  5& 0&95&0& -2  -5 & 0.928 1.107 & 0.966 1.204 & 1.171 1.293 &  1.260  1.096 &  1.293  0.853 &   1.826   0.584 & 1.073 1.304 &  0.563  0.729 & 1.013 1.017 &  3.01 \\
       56 & 18.246 & ...    & ...  & 1&  0& 0&31&0&  0  +6 & 0.777 0.571 & 0.800 0.715 & 0.920 0.924 & -0.248 -0.221 &  0.000  0.000 &   0.000   0.000 & 0.890 0.782 &  0.886  0.631 & 0.946 0.944 &  4.26 \\
       57 & 18.260 & ...    & ...  & 1&  0& 0&31&0& -2 +10 & 0.873 0.703 & 0.894 0.542 & 0.994 0.837 &  0.834  0.793 &  0.420  0.215 &   0.180   0.059 & 1.001 0.718 &  0.725  0.427 & 0.995 0.983 &  3.10 \\
       58 & 18.264 & ...    & ...  & 1&  0& 0&31&0& +1  -2 & 0.885 1.077 & 0.884 1.191 & 0.833 0.874 &  1.600  1.375 &  1.832  1.093 &   3.287   1.054 & 0.876 1.120 &  0.566  0.797 & 1.026 1.022 &  3.16 \\
       59 & 18.291 & ...    & ...  & 1&  0& 0&31&0&  0  -3 & 0.777 0.890 & 0.926 1.080 & 0.880 1.130 &  0.840  0.810 &  0.000  0.000 &   0.000   0.000 & 0.862 1.037 &  1.447  4.174 & 0.988 1.001 &  3.10 \\
       60 & 18.332 & ...    & ...  & 1&  0& 0&31&0& +1 +11 & 0.779 0.596 & 0.793 0.474 & 0.757 0.597 &  0.846  0.870 &  1.027  0.452 &   1.071   0.293 & 0.758 0.487 &  0.879  0.431 & 0.993 0.986 &  3.28 \\
       61 & 18.397 & ...    & ...  & 1&  0& 0&31&0& -1  -2 & 0.913 0.953 & 0.991 1.010 & 0.894 0.998 &  1.330  2.101 &  1.336  1.181 &   1.701   1.172 & 0.961 0.964 &  0.669  0.655 & 1.031 1.058 &  3.27 \\
       62 & 18.420 & ...    & ...  & 1&  0& 0&31&0& -1  -1 & 0.847 0.925 & 0.807 1.042 & 0.645 0.890 & -1.167 -1.406 &  0.000  0.000 &   0.000   0.000 & 0.716 0.988 &  0.563  0.969 & 0.908 0.909 &  5.01 \\
       63 & 18.463 & ...    & ...  & 1&  0& 0&31&0& +1  -3 & 0.661 0.833 & 0.720 0.912 & 0.739 0.926 &  0.536  0.448 &  0.147  0.104 &   0.026   0.009 & 0.697 0.886 &  1.375  3.363 & 0.976 0.982 &  3.58 \\
       64 & 18.466 & ...    & ...  & 1&  0& 0&31&0&  0  -4 & 0.774 1.004 & 0.848 0.907 & 0.795 1.065 & -0.395 -0.386 &  0.000  0.000 &   0.000   0.000 & 0.771 0.879 &  0.543  0.664 & 0.922 0.933 &  3.50 \\
       65 & 18.514 & 20.320 & 1.25 & 2&  0& 0&95&0&  0   0 & 0.879 0.965 & 0.890 1.108 & 0.890 0.926 &  1.857  1.563 &  2.635  1.275 &   7.758   1.724 & 0.878 1.031 &  0.475  0.562 & 1.034 1.024 &  3.24 \\
       66 & 18.667 & ...    & ...  & 1&  0& 2&31&1& -1  -3 & 0.873 1.032 & 0.829 0.997 & 0.878 0.920 & -1.062 -1.122 &  0.000  0.000 &   0.000   0.000 & 0.794 0.915 &  0.681  1.155 & 0.892 0.897 &  4.46 \\
       67 & 18.764 & 18.694 & 3.33 & 2&  0& 0&31&0&  0  -2 & 0.723 0.818 & 0.773 0.932 & 1.027 1.200 &  0.046  0.036 &  0.000  0.000 &   0.000   0.000 & 0.960 1.212 &  0.109  0.111 & 0.970 0.957 &  5.83 \\
       68 & 18.867 & ...    & ...  & 1&  0& 0&95&0& +1  +1 & 0.833 0.974 & 0.762 0.816 & 0.823 0.816 &  1.415  1.228 &  1.521  0.901 &   2.282   0.787 & 0.757 0.797 &  0.549  0.567 & 1.021 1.016 &  3.14 \\
       69 & 18.989 & ...    & ...  & 1&  1& 0&31&0& +2  +6 & 0.744 0.666 & 0.788 0.543 & 0.788 0.745 & -0.919 -0.962 &  0.000  0.000 &   0.000   0.000 & 0.804 0.590 &  0.476  0.317 & 0.932 0.919 &  3.00 \\
       70 & 19.443 & ...    & ...  & 1&  1& 0&31&0& +1 +13 & 0.730 0.551 & 0.815 0.511 & 0.700 0.685 & -1.885 -1.435 &  0.000  0.000 &   0.000   0.000 & 0.691 0.541 &  0.867  0.539 & 0.895 0.876 &  3.23 \\
       71 & 19.640 & ...    & ...  & 1&  0& 1&95&0& +1 +12 & 0.718 0.590 & 0.725 0.502 & 0.721 0.665 &  0.002  0.002 &  0.000  0.000 &   0.000   0.000 & 0.704 0.549 &  0.206  0.128 & 0.959 0.940 &  3.23 \\
       72 & 20.154 & ...    & ...  & 1&  0& 0&95&0& -1  +6 & 0.727 0.596 & 0.582 0.609 & 0.596 0.678 & -1.403 -0.984 &  0.000  0.000 &   0.000   0.000 & 0.544 0.560 &  1.167  0.947 & 0.874 0.857 &  3.34 \\
\hline
\end{tabular}
\tablefoot{\fontsize{6.4pt}{0.90\baselineskip}\selectfont
The data in Tables~\ref{Tab_72dr3} and \ref{Tab_72nnflg} will be
combined and made available in electronic form at the CDS.
}
\end{table*}

\paragraph{No.07 = \object{Gaia DR3 3841458366321558656}}

The brightest of the eleven best candidates, concerning both apparent and 
absolute magnitude (see Table~\ref{Tab_72dr3} 
and Fig.~\ref{Fig_cmd72nnGG}), 
is \object{Gaia DR3 3841458366321558656}. For this moderately HPM star
\citetads{2013AJ....145...44Z} 
already measured a proper motion similar to that in DR3. After
\object{D6-2} and \object{D6-1}, its $RPlx$$\approx$7.4 represents the third 
largest value in the shortlist. With $Nper$$=$17 it shows a deficit of 
observations in the global comparison, where it also failed the
test of $IPDgofha$ (almost 60\% larger than the corresponding $q75$ value). 
Regarding all other parameters and flags there are no complaints.
The $vtan\_g$ of \object{Gaia DR3 3841458366321558656} of $\approx$1000\,km/s 
is nearly as high as with \object{D6-2}, but much more uncertain, mainly
because of a more than three times larger parallactic distance.
Similar to \object{D6-2}, \object{Gaia DR3 3841458366321558656} was already
selected as high-priority HVS candidate in DR2 (Sect.~\ref{subS_DR2})
and listed in
\citetads{2018RNAAS...2..211S}. 
It was also included in the DR2 HVS candidate list of
\citetads[][their Table 2]{2019ApJS..244....4D}, 
which otherwise was highly contaminated by spurious HPM objects
(Sects.~\ref{subS_DR2} and \ref{subS_falsepm}).
The DR3 parallax of \object{Gaia DR3 3841458366321558656} 
(0.325$\pm$0.044\,mas) remained stable but became more significant 
than in DR2 (0.334$\pm$0.063\,mas).
The small RV of $\approx$$-$10\,km/s measured by
\citetads{2018ApJS..238...16L} 
does not support an HVS status. They classified this star as
carbon-enhanced metal-poor (CEMP) candidate, in agreement with
\citetads{2023MNRAS.523.4049L} 
who used DR3 BP/RP spectra.

\paragraph{No.62 = \object{Gaia DR3 3730485894679948928}}

The faint ($G$$\approx$18.4\,mag) 
HPM star \object{Gaia DR3 3730485894679948928} was already known 
before {\it Gaia} as \object{LSPM J1257+0715}
\citepads{2005AJ....129.1483L}. 
With $RPlx$$\approx$5.0, it just entered the high-priority class in
this study. With $Nper$$=$17, it lies slightly below the local and global
median values, and in one proper motion component ($pmRA$) the error
exceeds the global comparison value by a few percent only. The
heliocentric tangential velocity of this highest Galactic latitude 
($GLAT$$\approx$$+$70$\degr$) star in the shortlist is $\approx$1000\,km/s, 
whereas the Galactocentric one is only above 800\,km/s.
Here, the DR3 parallax (0.898$\pm$0.179\,mas) is more significant but
10\% smaller than in DR2 (0.995$\pm$0.236\,mas), wherein this 
candidate would have been selected with low priority only.

\paragraph{No.29 = \object{Gaia DR3 3488721051716722048}}

This is one more candidate with $vtan\_g$$\approx$800\,km/s, but
with a smaller $RPlx$$\approx$4.3 corresponding to our low-priority class.
The relatively small proper motion of \object{Gaia DR3 3488721051716722048}
was known before {\it Gaia}, e.g. from
\citetads{2013AJ....145...44Z}. 
Except for its $Nper$$=$18, which is slightly below the global median of
17th magnitude stars, this candidate passed all
other local and global parameter tests. 
The parallax in DR3 (0.316$\pm$0.076\,mas) decreased by 16\%
compared to DR2 (0.375$\pm$0.114\,mas), where its priority would
be lower.

\paragraph{No.32 = \object{Gaia DR3 1451652599056932480}}

Another HVS candidate with high priority in DR3
($RPlx$$\approx$5.6) is \object{Gaia DR3 1451652599056932480}, for which
a similar proper motion as measured by {\it Gaia} was already provided by
\citetads{2003AJ....125..984M}. 
For this 17th magnitude star,
$Nper$$=$19 is smaller than the median in the local comparison, while
all other tests did not hint at any problems. Compared to DR2, the
parallax and proper motion components are similar, but the DR3 parallax
(0.453$\pm$0.080\,mas) has no higher precision than the DR2 one
(0.494$\pm$0.083\,mas) so that $RPlx$ decreased from DR2 to DR3. 
This is unusual (!) and may indicate binarity. The $vtan\_g$ of this and
the remaining candidates in the shortlist is already 
less extreme ($<$750\,km/s).

\paragraph{No.67 = \object{Gaia DR3 6884930019709368576}}

Our faintest ($G$$\approx$18.8\,mag) 
high-priority ($RPlx$$\approx$5.8) 
candidate is the HPM star \object{Gaia DR3 6884930019709368576}. 
Although not included in classical HPM catalogues,
its HPM was known before {\it Gaia}
\citepads{2003AJ....125..984M}. 
Here, $Nper$$=$17 falls below the median in the global comparison.
Both locally and globally
it has up to 20\% larger $e\_pmDE$ compared to $q75$ values. 
In our close NN search (Sect.~\ref{subS_owncnn}),
an equally faint ($G$$\approx$18.7\,mag) object was found at a separation
of $\approx$3\,arcsec, which was considered as non-critical.
With a parallactic distance of $\approx$860\,pc, similar to that of
\object{D6-2}, this is the second candidate in our shortlist that
is located within 1\,kpc. The parallax in DR3 (1.164$\pm$0.200\,mas) became
larger and more significant than in DR2 (1.086$\pm$0.280\,mas),
where this object would be assigned low priority only.

\paragraph{No.52 = \object{Gaia DR3 4715594303954641920}}  

This candidate
\citepads[not well measured in][]{2003AJ....125..984M} 
appears similar to \object{D6-3} 
in several aspects. In particular, one can see this with respect to its
low priority ($RPlx$$\approx$4.5) but excellent local and global test 
results (except for $IPDgofha$, which is just a few percent above the
local $q75$ value). It is also 
similarly faint ($G$$\approx$17.9\,mag)
and was even more frequently observed ($Nper$$=$29). However, 
with its colour of $G$$-$$RP$$=$0.69\,mag it does not fall in 
the CMD region of the D$^6$ objects, and the 
Galactocentric tangential velocity of \object{Gaia DR3 4715594303954641920}
is with $\approx$730\,km/s only moderately extreme. Its DR3 
parallax (0.394$\pm$0.088\,mas) is 11\% smaller than in 
DR2 (0.442$\pm$0.115\,mas). In a DR2 HVS selection it would have 
lower priority. 

\paragraph{No.10 = \object{Gaia DR3 4294774301679435776}}

The second brightest ($G$$\approx$16.0\,mag) of our 11 candidates,
\object{Gaia DR3 4294774301679435776}, is of high priority
($RPlx$$\approx$7.0). But in the local
comparison, $epsi$ and $sepsi$ turned out to lie 15\% and 50\%,
respectively above the corresponding $q75$ level. Note that this
object is located right in the Galactic plane ($GLAT$$\approx$$-$7$\degr$),
what is causing great concern.
Two objects within 5\,arcsec found in Sect.~\ref{subS_owncnn} also indicated 
image crowding but were (with their magnitudes $G$$>$17\,mag)
not considered as critical
close NNs to \object{Gaia DR3 4294774301679435776}. Rather different 
proper motions were measured before {\it Gaia} by
\citetads{2003AJ....125..984M} 
and
\citetads{2013AJ....145...44Z}, 
respectively.
Its parallax in DR3 (0.308$\pm$0.044\,mas) became enormous 29\% (!) larger
than measured in DR2 (0.239$\pm$0.055\,mas), where it would have been
selected with low priority only.

\paragraph{No.42 = \object{Gaia DR3 6620452702488805504}}

The last object in our shortlist is the low-priority ($RPlx$$\approx$4.3)
candidate \object{Gaia DR3 6620452702488805504}, for which
\citetads{2003AJ....125..984M} 
already found a proper motion almost equal to that in DR3. Similar to 
another low-priority candidate (\object{Gaia DR3 3488721051716722048}) in
the shortlist, it passed all parameter tests, except for the global
comparison of $Nper$, where it shows a deficit of observations.
The DR3 parallax (0.403$\pm$0.093\,mas) is 14\% larger than the very
uncertain value measured in DR2 (0.355$\pm$0.162\,mas), where it would 
not have been selected even with low priority.

\section{Summary and conclusions}
\label{Sect_concl}

This study draws attention to still uncertain {\it Gaia} parallaxes 
of HVS candidates, selected by high tangential velocities. 
First, a lower limit of 500\,km/s was used for computed 
Galactocentric tangential velocities (Sect.~\ref{subS_vtan_g}), 
and candidates with parallactic distances $<$10\,kpc 
and moderately HPMs $>$20\,mas/yr were selected. Similar 
numbers of about 1500 high-priority HVS candidates ($RPlx$$>$5) were found 
in {\it Gaia} DR2 (Sect.~\ref{subS_DR2}; 
Figs.~\ref{Fig_dr2pm}--\ref{Fig_dr2vtangPM}) and 
DR3 (Sect.~\ref{subS_DR3}; 
Figs.~\ref{Fig_dr3pm}--\ref{Fig_dr3vtangPM}).
But these two samples of candidates had less than
25\% in common, and the overlap further reduced with higher velocities.

It was shown that in high stellar density regions
in the GC and Galactic plane spurious HPMs can lead to false HVS 
candidates (Fig.~\ref{Fig_fpm}). 
This effect, which is stronger for small $Nper$, has been reduced 
from DR2 to DR3. Interestingly, in a recent comparison of proper motions
in crowded fields,
\citetads{2023A&A...677A.185L} 
found that DR3 proper motion uncertainties are underestimated by up to a 
factor of 4 in dense Galactic bulge fields. Their study was also
motivated by the search for HVSs among outliers in the proper motion
distribution.
One recommendation is to increase the lower limits of $Nper$ used for 
all objects in DR2 (6) and DR3 (10) (see 
Figs.~\ref{Fig_dr2NperG} and \ref{Fig_dr3NperG}) considerably 
with any astrometric selection of HVS candidates, in particular in 
crowded regions. 

In general, the HPMs of DR3-selected HVS candidates are in
good agreement with previous DR2 data and highly-significant. 
Therefore, 
their formal tangential velocity errors (Fig.~\ref{Fig_DR3evtanG})
are dominated by the parallax
errors. The high-priority HVS candidates ($RPlx$$>$5) selected in DR2
showed a trend towards larger parallaxes measured in 
DR3 (Figs.~\ref{Fig_dr2dr3plx} and \ref{Fig_dPlxG}). Consequently,
their tangential velocities became lower (Fig.~\ref{Fig_dr2dr3vtang}) 
and absolute magnitudes
fainter. Speculatively, this trend will continue for DR3-selected 
high-priority HVS candidates in DR4. 

The HVS selection in DR3 was extended to low-priority (3$<$$RPlx$$<$5) 
candidates. There are selection effects of the parallaxes and proper 
motions (Fig.~\ref{Fig_DR3astmG}). Both 
become larger at the faint end. This effect begins for high-priority 
HVS candidates already at brighter magnitudes than for low-priority ones.

With DR3, the RVs of bright stars, including one third of our DR3 
high-priority HVS candidates but only few low-priority ones, 
became available. For high- and low-priority candidates, the mean RV errors 
are 20 and 30 times smaller than the mean heliocentric tangential 
velocity errors. All RVs are within $\pm$450\,km/s, i.e.
smaller than the used lower limit for $vtan\_g$ (Fig.~\ref{Fig_RVvtang}). 
For $>$70\% of all candidates, the precise 
RVs are more than five times smaller than the uncertain heliocentric tangential 
velocities (Fig.~\ref{Fig_RVvtanh}). 
This leads to the conclusion that the DR3 HVS sample is strongly affected by 
underestimated parallaxes. 

In the second part of this study, the 72 nearest (parallactic 
distance $<$4\,kpc) and most extreme ($vtan\_g$$>$700\,km/s) HVS candidates
in DR3 were studied in detail. 
In the CMD (cf. Figs.~\ref{Fig_CMDextr} and \ref{Fig_cmd72nnGG}), 
these 11 high-priority 
and 61 low-priority candidates mainly occupy the MS, with many objects 
scattered towards redder $G$$-$$RP$ colours. A small group of four candidates
is located between the MS and the WD region. 
As expected, no blue massive HVS candidates were found 
within our relatively small distance limit.
The majority of apparently
too red candidates have close NNs, found in an own search
in DR3 (Sect.~\ref{subS_owncnn}; Fig.~\ref{Fig_nn16}), 
or DR3 flags and IPD parameters hinting at binarity. Ten most 
important astrometric quality parameters of each target were compared with
those of all objects of similar magnitude locally and HPM stars of equal 
magnitude globally. Median $Nper$ and 0.75 quantiles of nine other parameters 
were determined for the comparison objects and used as lower and upper critical 
limits, respectively. This is illustrated 
in Figs.~\ref{Fig_ePlxGnnGG}--\ref{Fig_NperGnnGG}.
The most decisive parameters, for which candidates 
failed the tests, were $RUWE$, $gofAL$, $sepsi$, and $epsi$ in the local
comparison and $Nper$, $RUWE$, and $gofAL$ in the global comparison
(Table~\ref{Tab_quali}). 
Combining all test results, and using $RPlx$ as the most important criterion,
none of the nearest extreme HVS candidates was fully convincing.

Local and global parameter tests complemented each other. An
interesting case was found with \object{D6-1}
\citepads{2018ApJ...865...15S}, 
where our global comparison went without any problems, probably
because of a very high $g\Delta$$=$+9, indicating many observing epochs
for this object and this sky region.
In the local comparison three parameters failed the test, i.e.
\object{D6-1} appears in comparison to other objects with many epochs
not so well measured. Deviations become visible the more often
an object is observed. Therefore, an alternative global test could involve
comparison objects not only of equal magnitude, but also of equal $Nper$.

Eleven candidates were selected with relaxed criteria: To include
the best of low-priority candidates, a minimum of $RPlx$$>$4 was used.
The best of high-priority candidates were selected by allowing for
$Nper$ values $\geqq$$-$2 compared to the local and global median
and by increasing the allowed upper limits of nine other astrometric 
parameters to 60\% above the 0.75 quantiles of comparison objects
The 'relaxed candidates' are discussed individually in
Sect.~\ref{subS_best} with respect to their problematic parameters, previous
proper motion measurements (generally in agreement with DR3), partly
low Galactic latitudes causing concern, previous DR2 parallaxes, 
external RVs mostly not supporting their HVS status, and other references.

The shortlist of eleven relaxed candidates, sorted by decreasing $vtan\_g$,
is led by three D$^6$ objects,
the low-priority candidate \object{D6-3}, and the
two high-priority candidates \object{D6-1} and \object{D6-2}
\citepads[all from][]{2018ApJ...865...15S}, 
followed by the bright high-priority 
candidate \object{Gaia DR2 3841458366321558656}
\citepads[from][]{2018RNAAS...2..211S}. 
Another DR2 candidate in the latter research 
note, \object{Gaia DR2 6097052289696317952} labelled in most figures,
and object D located next to the D$^6$ objects 
in Figs.~\ref{Fig_dr3cmd}, \ref{Fig_CMDextr} and \ref{Fig_cmd72nnGG} 
(see also Sects.~\ref{subS_DR3} and \ref{subS_DR3magsel})
failed in many of the parameter tests.
All test results are tabulated so that the reader can decide on an
own alternative selection of best candidates.

\begin{acknowledgements}
This work analyses results from the European Space Agency (ESA) space 
mission {\it Gaia}. {\it Gaia} data are being processed by the {\it Gaia} 
Data Processing and Analysis Consortium (DPAC). Funding for the DPAC is 
provided by national institutions, in particular the institutions 
participating in the {\it Gaia} MultiLateral Agreement (MLA). 
The {\it Gaia} mission website is https://www.cosmos.esa.int/gaia. 
The {\it Gaia} archive website is https://archives.esac.esa.int/gaia.
I have extensively used SIMBAD and VizieR at the CDS/Strasbourg
and would like to thank the CDS staff for their valuable work.
I also thank the anonymous referee for a helpful report.
\end{acknowledgements}

\bibliographystyle{aa}
\bibliography{hvsDR2DR3_lit}

\end{document}